\documentclass[apj]{emulateapj}
\def\gsim{\ \raise 3pt \hbox{$\rangle$} \kern -8.5pt \raise -2pt \hbox{$\sim$}\ }
\usepackage{graphicx,natbib,color,amsmath,subfigure}
\usepackage{ulem}
\newcommand{\blank}[1]{}
\usepackage{color}
\newcommand{\gfR}{} 

\newcommand{\dg}{} 
\newcommand{\gf}{} 
\newcommand{\gff}{} 

\begin{document}
\title{Probing Dynamics of Electron Acceleration with Radio and X-ray Spectroscopy, Imaging, and Timing in the 2002 Apr 11 Solar Flare
}
\author{Gregory D. Fleishman\altaffilmark{1,2}, Eduard P. Kontar\altaffilmark{3}, Gelu M. Nita\altaffilmark{1}, Dale E. Gary\altaffilmark{1}}
\altaffiltext{1}{Center For Solar-Terrestrial Research, New Jersey Institute of Technology, Newark, NJ 07102}
\altaffiltext{2}{Ioffe Physico-Technical Institute, St. Petersburg 194021, Russia}
\altaffiltext{3}{Department of Physics and Astronomy, University of
Glasgow, G12 8QQ, United Kingdom}

\begin{abstract}

Based on detailed analysis of radio and X-ray observations of a flare on 2002 April 11 {\gff augmented by realistic 3D modeling}, we have identified a radio emission component produced directly at the flare acceleration region. This acceleration region radio component has distinctly different (i) spectrum, (ii) light curves, (iii) spatial location, and, thus, (iv) physical parameters from those of the separately identified, trapped {\gfR or precipitating} electron components. To derive evolution of physical parameters of the radio sources we apply forward fitting of the radio spectrum time sequence with the gyrosynchrotron source function with 5 to 6 free parameters. At the stage when the contribution from the acceleration region dominates the radio spectrum, the X-ray- and radio-derived electron energy spectral indices agree well with each other. During this time the maximum energy of the accelerated electron spectrum displays a monotonic increase with time from $\sim300$~keV to $\sim2$~MeV over roughly one minute duration indicative of an acceleration process in the form of growth of the power-law tail; the fast electron residence time in the \textit{acceleration region} is about $2-4$~s, {\dg which} is much longer than the time of flight and {\dg so} requires a strong diffusion mode \textit{there} {\dg to inhibit free-streaming propagation}. The acceleration region has a relatively strong magnetic field, $B\sim120$~G, and a low thermal density, $n_e\lesssim2\cdot10^9$~cm$^{-3}$. These acceleration region properties are consistent with a stochastic acceleration mechanism.

\end{abstract}

\keywords{Sun: flares---acceleration of particles---turbulence---diffusion---Sun: magnetic fields---Sun: radio radiation}

\section{Introduction}


Probing the acceleration region of solar flares is known to be a highly nontrivial task \citep{Bastian_etal_2007, Xu_etal_2008, Holman_2012} since the direct emission from the acceleration site is often weaker than other competing emissions, e.~g., soft X-rays (SXRs) from thermal plasma and microwave continuum from a magnetically trapped fast electron component. Recently, we identified and reported a cold, tenuous flare \citep{Fl_etal_2011}, which displayed neither hot coronal plasma nor a magnetically trapped population of fast electrons in a coronal loop. This {\dg rare but} favorable combination of flare properties allowed us, for the first time, to firmly identify, {\gf spectrally and spatially,} the radio emission component produced directly at the acceleration region and then derive physical parameters of the acceleration site and accelerated electron population.

Although detection of even a single acceleration region is \blank{very} important for improving our understanding of where and how the flare electrons are accelerated, it might seem discouraging because such 'clean' cases without plasma heating and electron trapping are extremely rare. In contrast, in a typical case, the flare plasma heating is an essential component, vividly seen in coronal SXR sources, while in the microwave range three source contributions \citep[separate regions of acceleration, trapping, and precipitation, e.g.,][]{Aschwanden_1998} \blank{are} {\gf may be} present. To study acceleration processes, it is necessary to distinguish the acceleration region contribution in the presence of {\dg these other two} competing emissions. An ideal way to make this distinction, at least for cases when the acceleration is spatially displaced from the other components, is through the use of high spatial and spectral resolution microwave imaging spectroscopy, which is not yet routinely available. In the meantime, however, we are restricted to selecting rare cases where the direct emission from the acceleration region can be separated spectrally and/or temporally from the competing contribution in the spatially integrated dynamic spectrum.

Theoretical consideration of radio emission expected from the acceleration regions in flares \citep{Li_Fl_2009, Park_Fl_2010} suggests that the corresponding contribution should peak at a few GHz and have a reasonably narrow spectrum. This is, in particular, the case of the radio spectrum from the acceleration region of the cold, tenuous flare \citep{Fl_etal_2011}; however, having a spectral peak at this decimeter range does not guarantee that the emission originates from the acceleration region, so some additional evidence is needed. In this paper we present the study of a flare {\dg in which} a certain fraction of the radio emission {\gf (at a somewhat narrow spectral range during a limited time)} can be confidently attributed to the acceleration region, even though the competing contributions are strong and overall comparable with {\gf or even dominating} that from the acceleration region. In particular, we will demonstrate that the {\gfR radio} detection of the acceleration region in the given event is favored by stronger magnetic field at the acceleration site compared with the coronal `trapping site', where the fast electron accumulation occurs {\gfR at a later stage due to the well-known effect of magnetic trapping}. Thus, the gyrosynchrotron (GS) emission from the acceleration site dominates until the number of the magnetically trapped electrons rises above the level needed to dominate over the acceleration region contribution.

\begin{figure}\centering
\includegraphics[width=0.95\columnwidth]{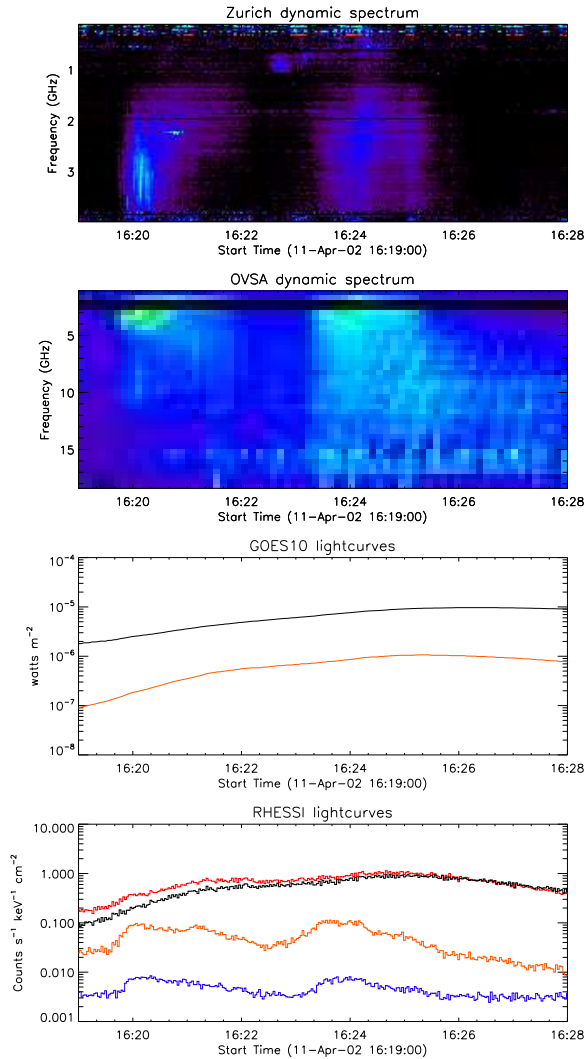}\\
\caption{\label{fig_11_apr_2002_over} Overview of  April 11, 2002 flare: Phoenix-2 and OVSA dynamic spectra, top panels. GOES (3\hspace{0.1cm}s) lightcurves
as measured by GOES-10 spacecraft. 
RHESSI (2 second bins) lightcurves  (bottom panel)
in: 3-9\hspace{0.1cm}keV (black), 9-15\hspace{0.1cm}keV (red), 15-30\hspace{0.1cm}keV (orange), 30-100\hspace{0.1cm}keV (blue).
}
\end{figure}

In section 2 we present the observations from Reuven Ramaty High Energy Solar Spectroscopic Imager \citep[RHESSI,][]{lin2002} in the X-ray range, Owens Valley Solar Array (OVSA) at 1--18\hspace{0.1cm}GHz, and Phoenix-2 spectrometer at 0.1--4\hspace{0.1cm}GHz \citep{1999SoPh..187..335M}. Context SXR observations made by GOES-10 and SOHO/MDI measurements of the photospheric magnetic field are utilized as well {\gff along with a new 3D modeling tool, \verb"GX_Simulator", which we have developed and recently included into Solar Software}\footnote{\url{www.lmsal.com/solarsoft/ssw\_packages\_info.html}}.
In section 3 we present spectral fitting of the radio and hard X-ray spectra using 5 to 6 physical parameters, and show that the acceleration region can be separately identified.
We {\gff  outline a possible 3D geometry} in section 4 and conclude with a discussion of the results in terms of a stochastic acceleration mechanism.

\section{Observations}

\subsection{X-ray Imaging and Spectroscopy} 

\begin{figure}\centering
\includegraphics[width=0.8\columnwidth]{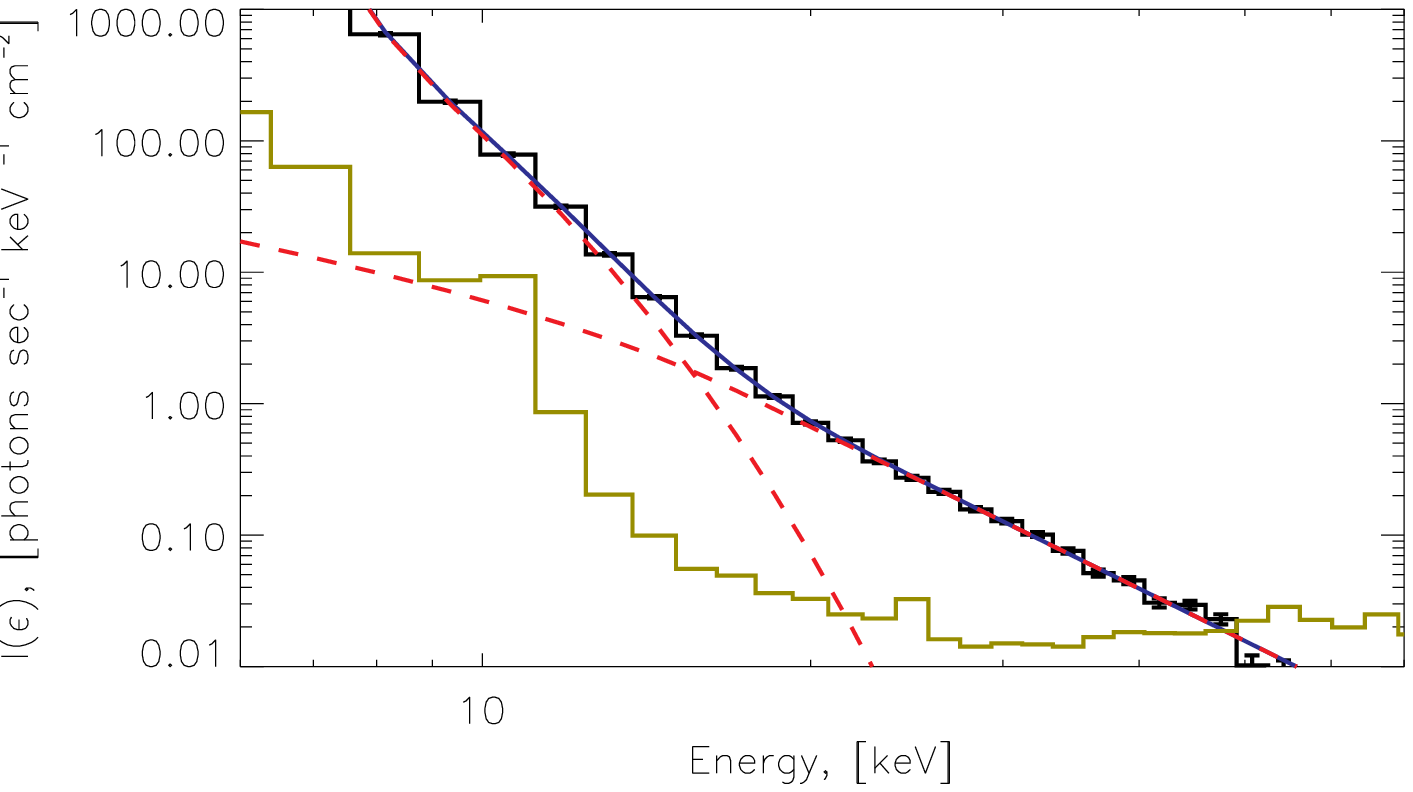}\\
\includegraphics[width=0.8\columnwidth]{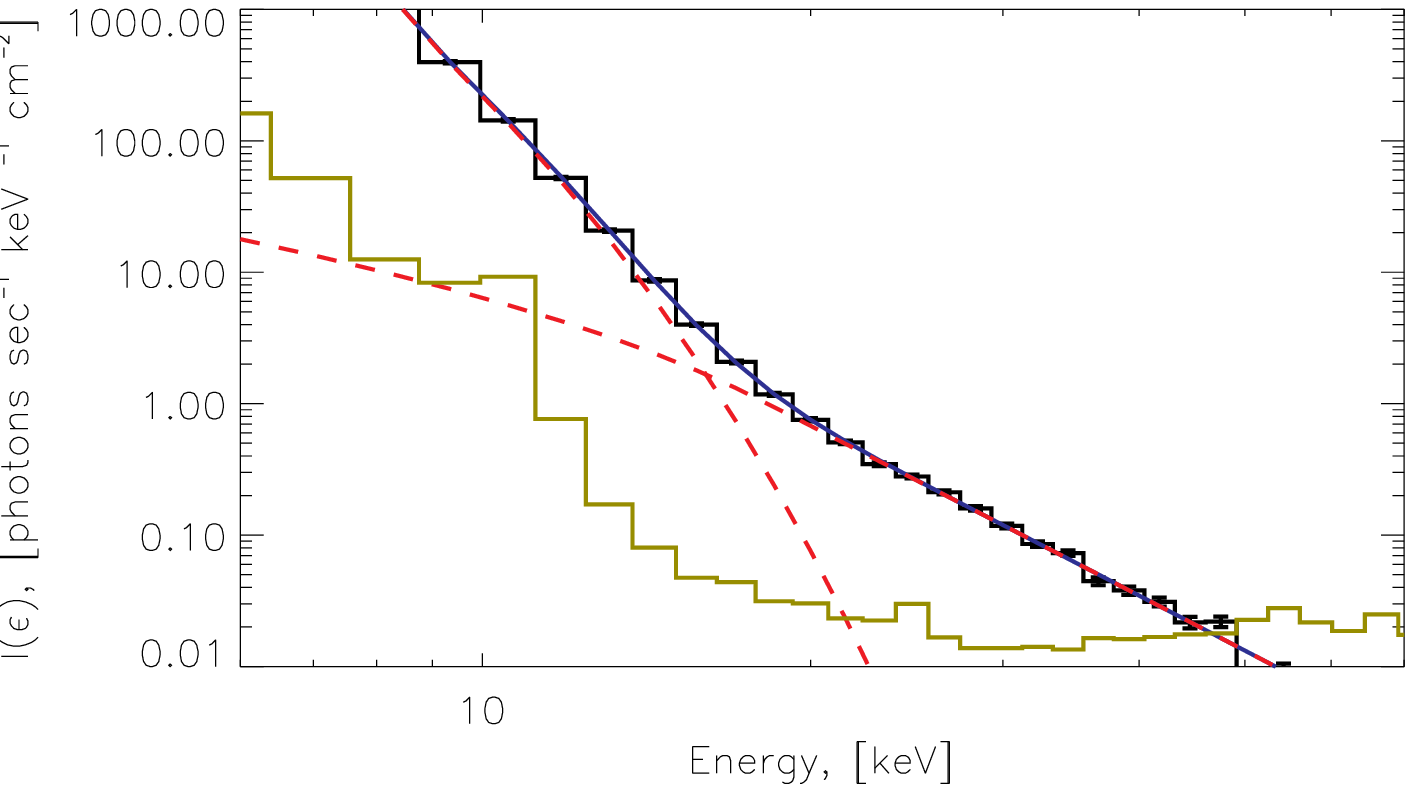}\\
\caption{\label{fig:peak_spectr} RHESSI spatially integrated X-ray spectrum above 6 keV
for two HXR peaks of the event:  $16:19:43-16:22:01$~UT - top panel and $16:23:07-16:25:13$ UT - lower panel.
In both panels, brown line indicates background, red-dashed lines thermal and non-thermal components, blue-line is the
total fit (thermal plus non-thermal), and black and brown histograms X-ray data for the signal and the background
respectively.}
\end{figure}

The flare demonstrates two hard X-ray peaks as evident from Figure~\ref{fig_11_apr_2002_over}.
The spatially integrated RHESSI X-ray spectrum (Figure~\ref{fig:peak_spectr})
over the duration of each peak indicates the presence of both thermal and non-thermal components
as often seen in X-ray spectra (e.g. see Kontar et al. 2011, as a review). To achieve better count rate statistics
and energy resolution,  we summed all front segment detectors but the detectors 2 and 7 were avoided
(see Smith et al. 2002 for the details). Spectral analysis was done using OSPEX \citep{Schwartz_etal2002}
with systematic errors set to 0.02\%.  As the flare appeared on the solar disk
($x\simeq 500''$, $y\simeq -180''$ ; heliocentric angle $\simeq 56^o$ ), albedo correction \citep{kontar2006}
was applied assuming isotropic emission (minimum correction). The spectrum was fitted with a standard thermal plus non-thermal thick-target model (Figure \ref{fig:peak_spectr}), with the isotropic
bremsstrahlung cross-section following \citet{haug1997}.

\begin{figure}\centering
\includegraphics[width=0.85\columnwidth]{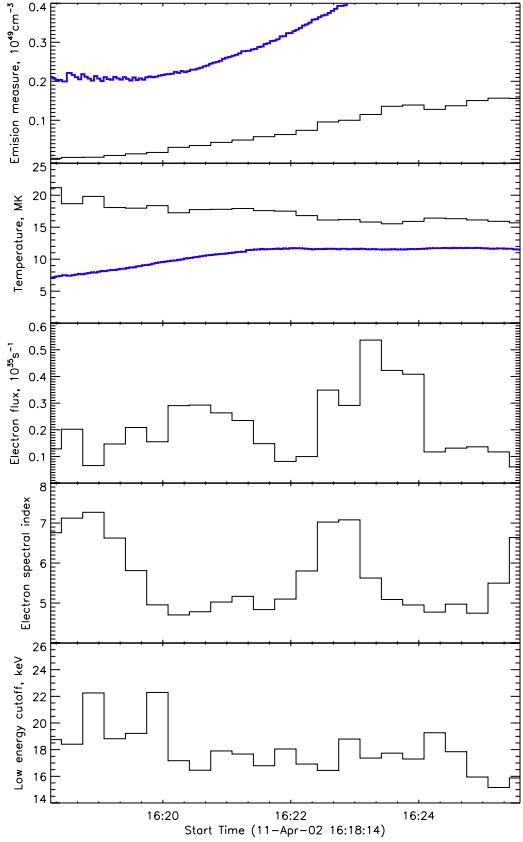}\\
\caption{\label{fig:fit_tt}Spectral fit parameters from the thermal plus thick-target X-ray model 
for 20 second time intervals of RHESSI data (black curves) as well as emission measure and temperature as obtained from  GOES-10 data (blue curves).
}
\end{figure}

The temporal variation of plasma parameters obtained from the {HXR} fit with time is presented in Figure \ref{fig:fit_tt}.
\blank{Both the number of accelerated electrons and the spectral index} {\gf The derived accelerated electron spectral evolution} demonstrates typical {\dg soft-hard-soft}
behavior and {\dg shows} two clear peaks in hard X-rays, well visible in $30-100$~keV lightcurves (Figure~\ref{fig_11_apr_2002_over}).
The emission measure of the thermal plasma increases from $\lesssim10^{47}$~cm$^{-3}$
to $\sim10^{48}$~cm$^{-3}$, while the temperature of the plasma stays rather constant, ranging from $\sim 20$ to $\sim 15$~MK
over the course of the flare. The emission measure and the temperature observed by GOES-10 {\dg are} somewhat {\gf higher and} lower,
in the range{\gf s above $2\cdot10^{48}$~cm$^{-3}$ and} between 12 and 10~MK{\gf, respectively}, but {\dg show} a similar temporal behavior.

\begin{figure}\centering
\includegraphics[width=0.32\columnwidth]{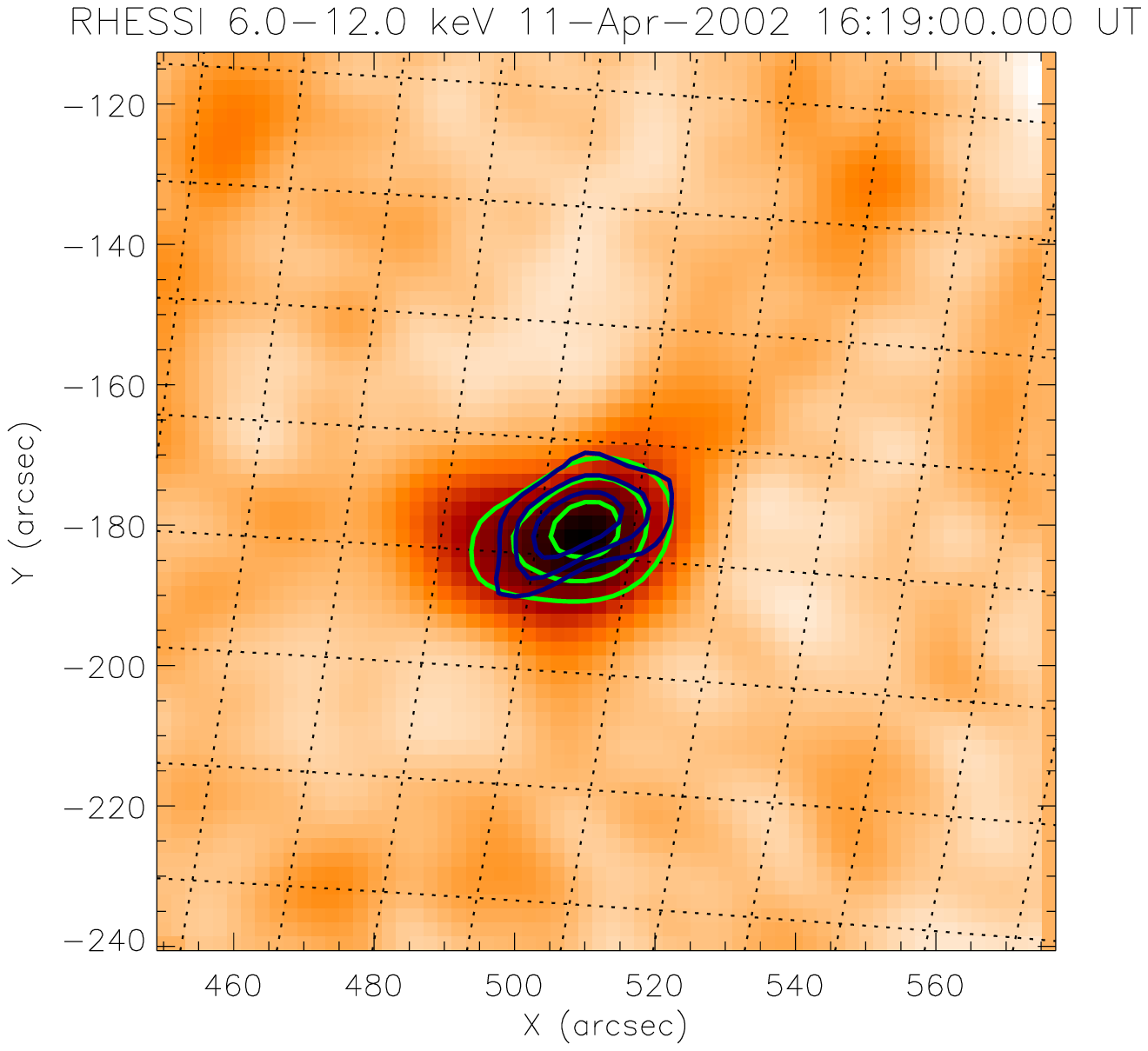}
\includegraphics[width=0.32\columnwidth]{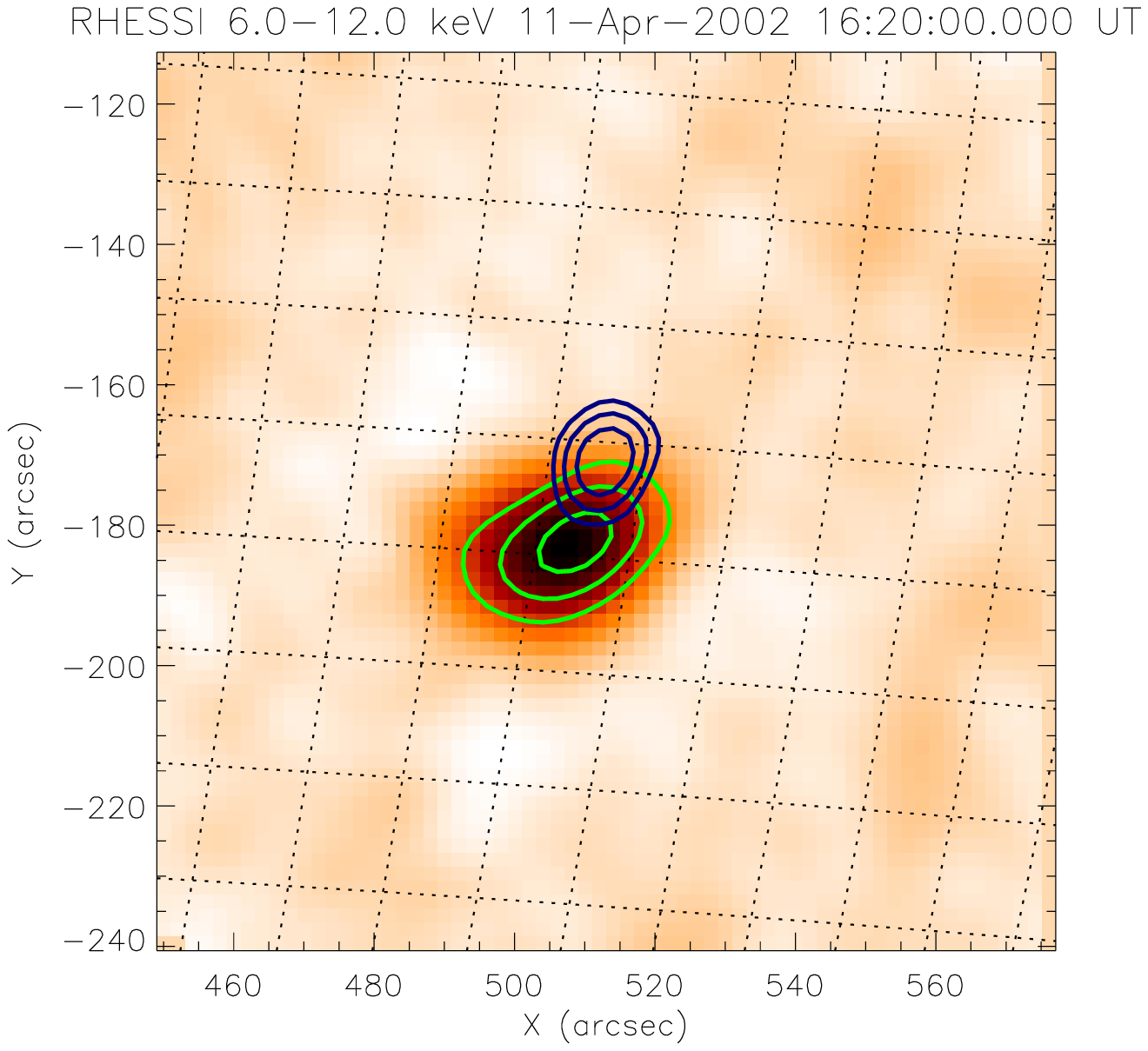}
\includegraphics[width=0.32\columnwidth]{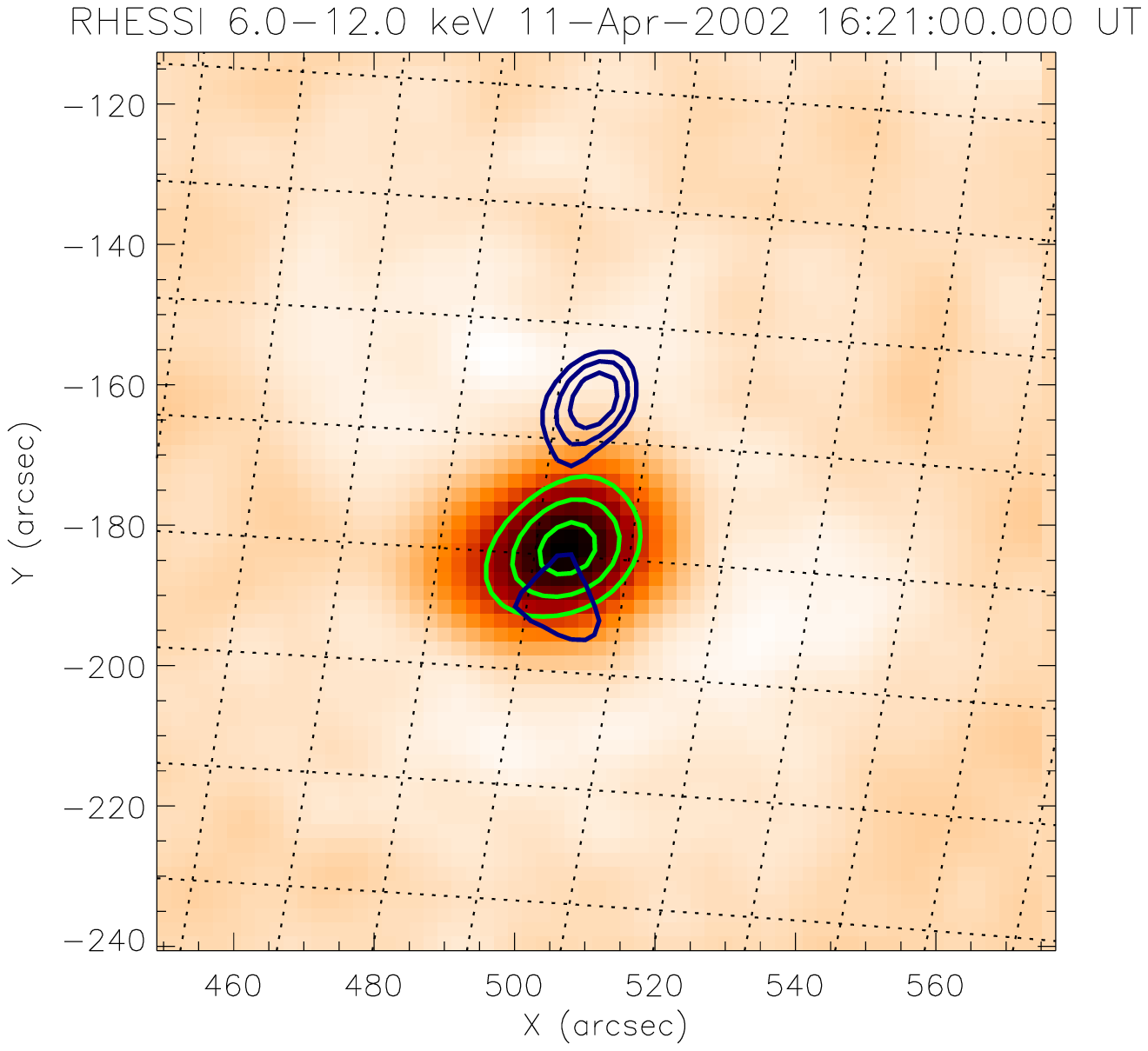}
\includegraphics[width=0.32\columnwidth]{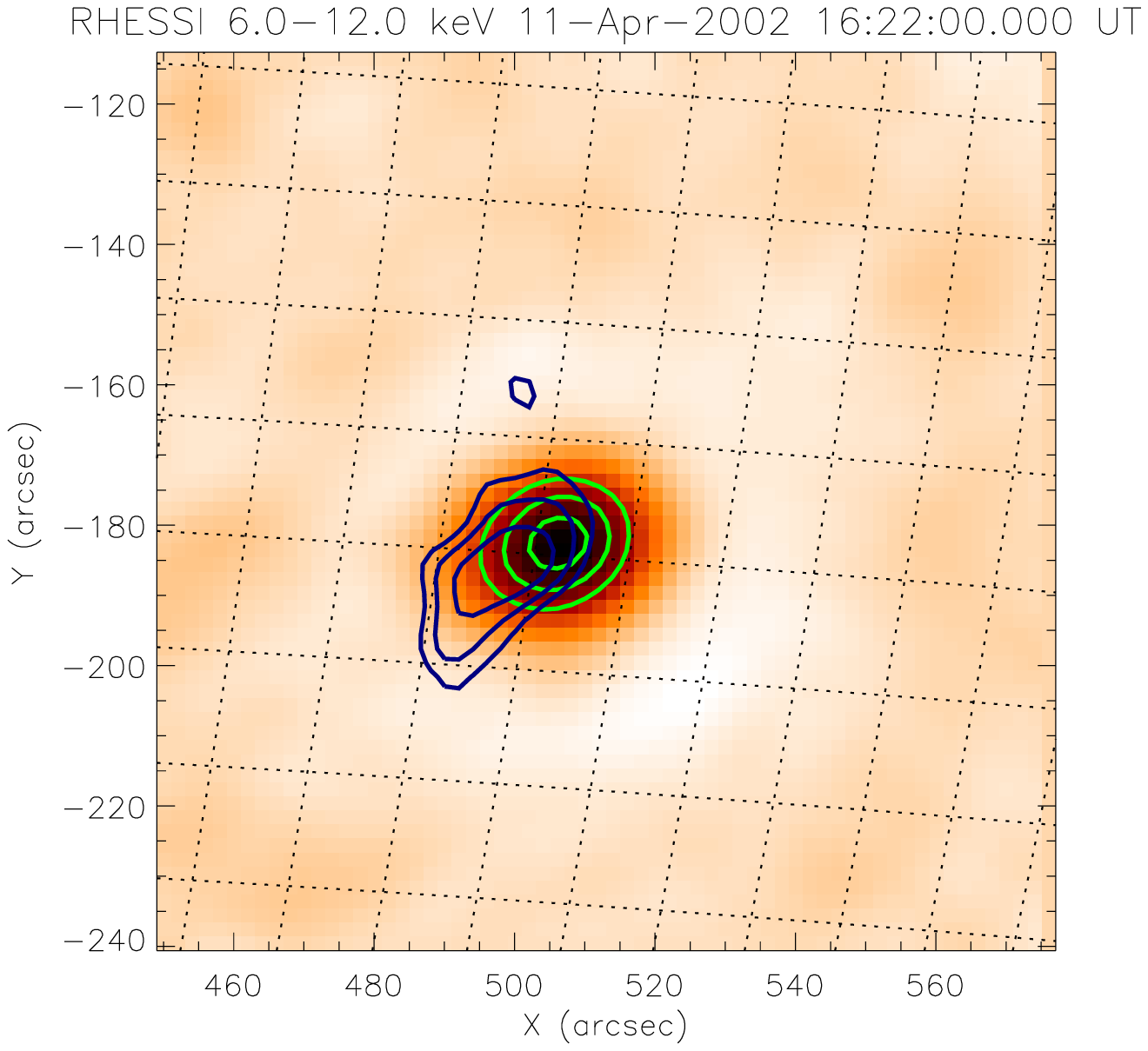}
\includegraphics[width=0.32\columnwidth]{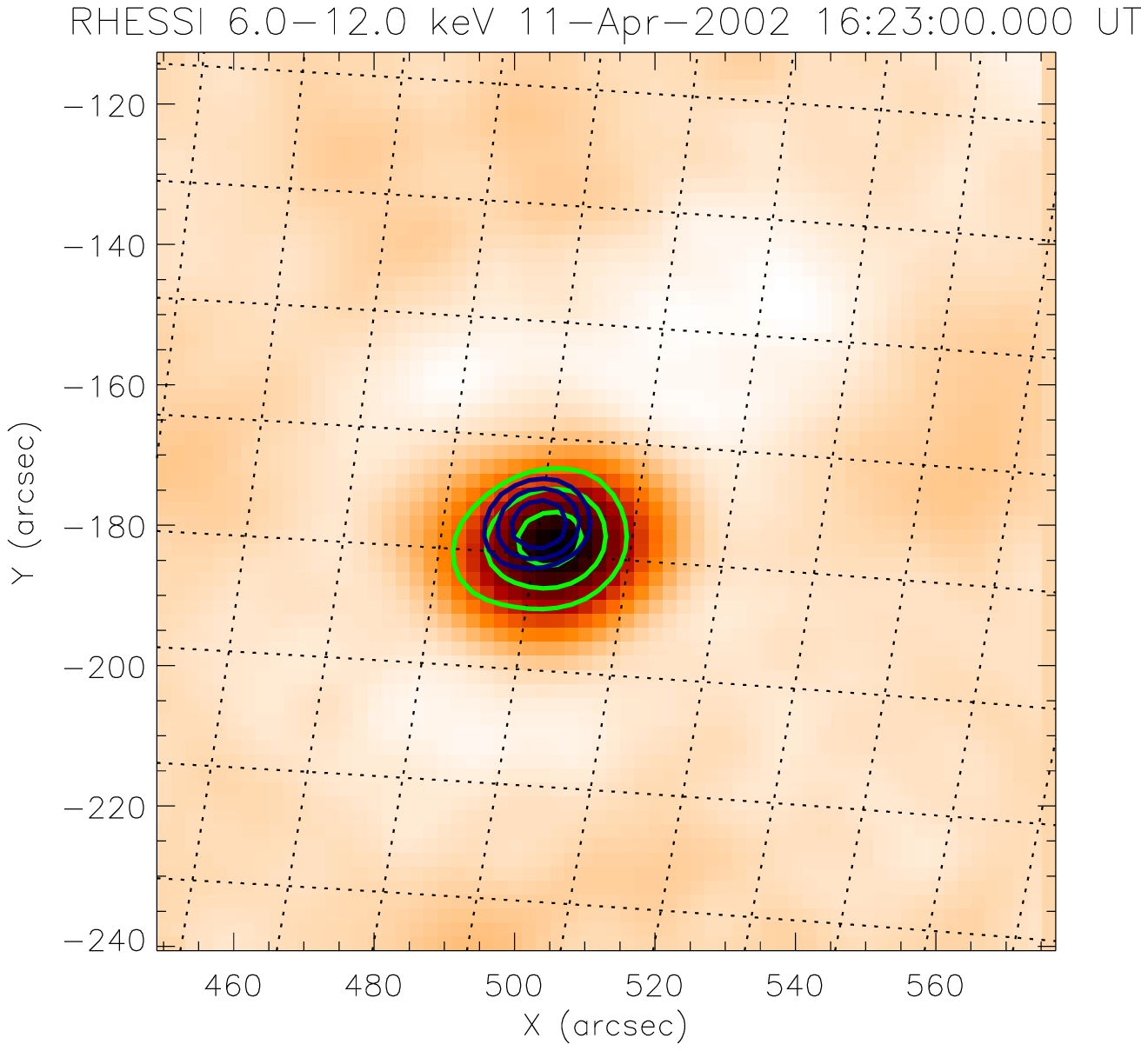}
\includegraphics[width=0.32\columnwidth]{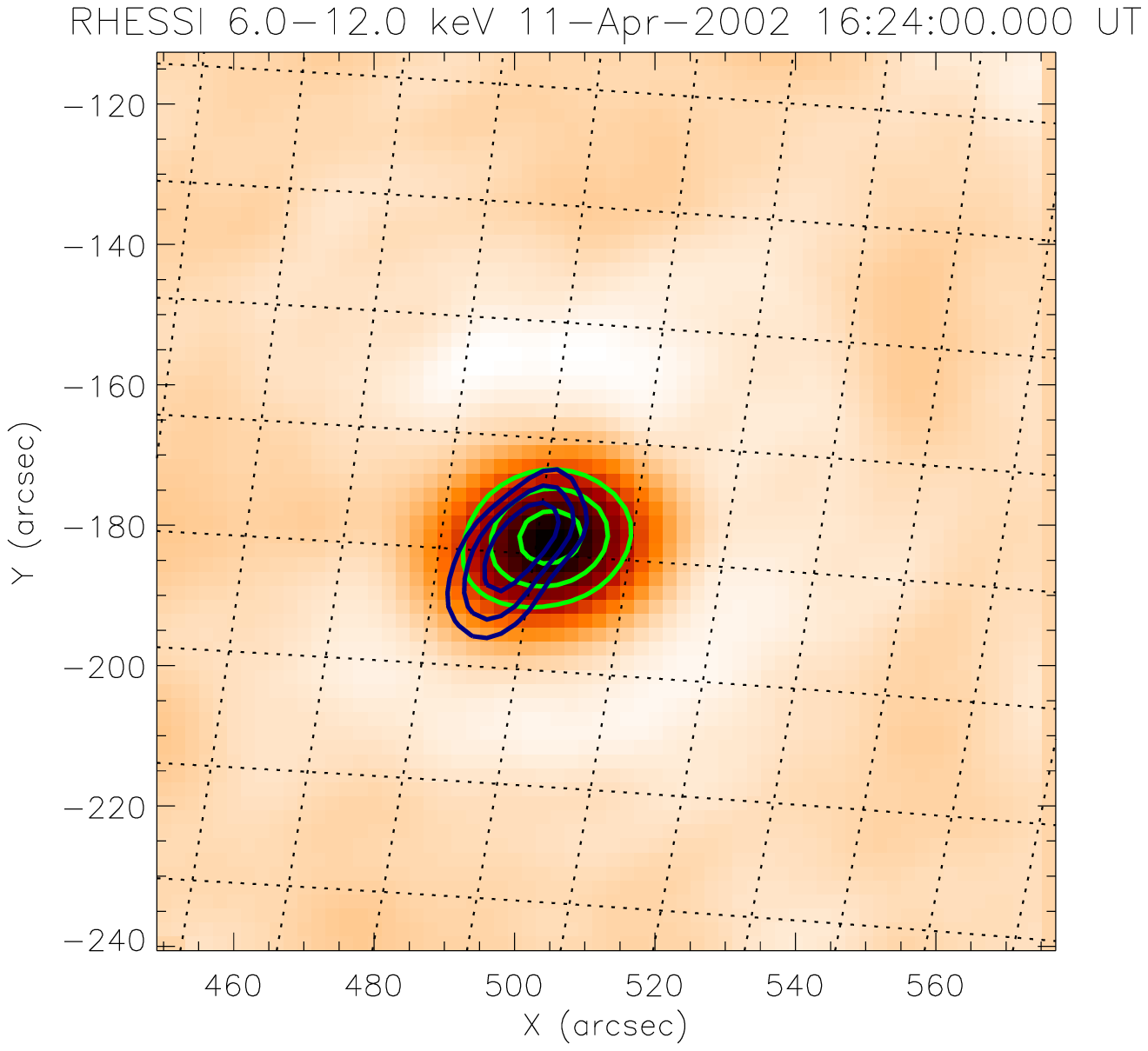}
\caption{\label{fig:Ximage} Spatial distribution of X-ray emission from April 11, 2002 flare
in various energy ranges. Background image $6-12$~keV, $12-20$~keV green (contours are at 50,~70,~90~\% levels), and $20-40$~keV dark blue
(contours are at 70,~80,~90~\% levels). The X-ray images are reconstructed using clean algorithm.
}
\end{figure}

\begin{figure}\centering
\includegraphics[width=0.99\columnwidth]{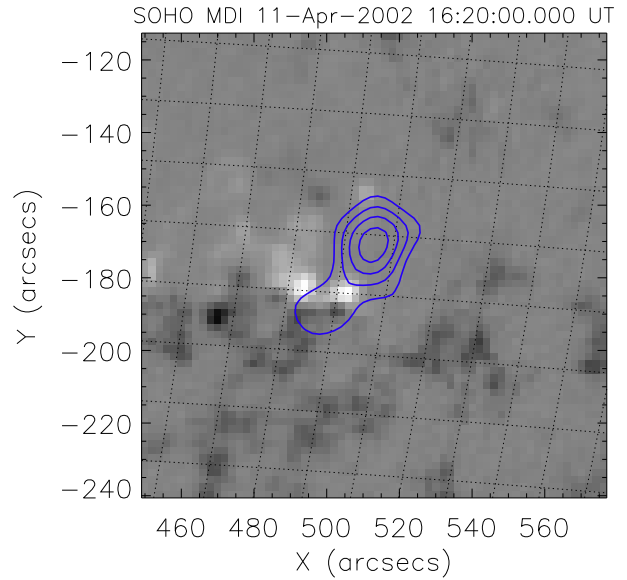}
\caption{\label{fig:XoverB_image} {\gf 20--40~keV} HXR contours  {\gfR [45; 60; 75; 90\%]} {\gf for 16:20--16:21~UT} superimposed on the SOHO/MDI line
of sight magnetic map. {\gf It is clearly seen that the HXR source is projected to the photosphere region free from any noticeable magnetic field enhancement, which supports the idea that the HXR source is coronal, rather than a footpoint.}}
\end{figure}

RHESSI imaging  (Hurford et al. 2002) reveals a spatially well-defined soft X-ray  (6-12 keV and 12-20 keV)
source (Figure \ref{fig:Ximage}), visible over the entire time of the flare, whose location gradually moves eastward
by roughly $10''$ over the flare duration. HXR sources imaged in the 20-40~keV range have higher temporal variability,
which coincides spatially with the SXR source at the troughs of X-ray light curves at $16:19$~UT and $16:23$~UT but deviates from this location at the peaks, when
the electron acceleration rate is the highest.
{\gf Based on SXR-to-HXR spatial relationship alone, one could interpret the event in terms of the standard flare picture, i.e. a looptop SXR source and two uneven HXR footpoints. This interpretation, however, is supported neither by the magnetic field data} {shown in Figure \ref{fig:XoverB_image} {\gff nor by the observation-based 3D modeling given below in Fig.~\ref{fig_3D_set}}. We} \blank{It is possible to} argue {based on analysis of the microwave data, which we will show later,} that the {\gfR main} HXR source remains coronal most of the time, \blank{as it is projected on a photospheric region free from any noticeable} {but support for this can be seen in the fact that the 20-40~keV source in Fig.~\ref{fig:XoverB_image} does not overlie any concentration of} magnetic field, \blank{Figure \ref{fig:XoverB_image}, which} that could \blank{have corresponded} {correspond} to footpoints \blank{as} often seen
in flares \citep[e.g.][]{Kosugi1992yohkoh,emslie2003,Kontar_etal2010, BattagliaKontar2011}.
\blank{Interestingly,} {\gf The only time period when the {\gfR main} HXR source possibly could be identified with a footpoint is late in the first peak}  16:21{\gf -- 16:22}~UT when the 20-40~keV source is {\gf more strongly} displaced to the north
from the SXR source, while later the HXR source moves {\gf back} slightly to the south. {\gfR A weaker HXR source, seen as a south-east extension of the main HXR source in Figures~\ref{fig:XoverB_image} and \ref{fig_OVSA_Images}, left, could correspond to a footpoint; alternatively, it can correspond to {\gff another part of the same magnetic structure or to} a lower, more compact loop.  }

\subsection{OVSA imaging}

We employ OVSA frequency synthesis imaging around the spectral peak frequency, 2.6--4.2~GHz, while no reliable imaging was possible at the higher frequencies, $\sim 10$~GHz due to the low flux levels. Given a limited OVSA spatial resolution at low frequencies, the radio sources were unresolved, although useful information on the source location at different stages of the burst has been obtained.

\begin{figure}\centering
\includegraphics[width=0.45\columnwidth]{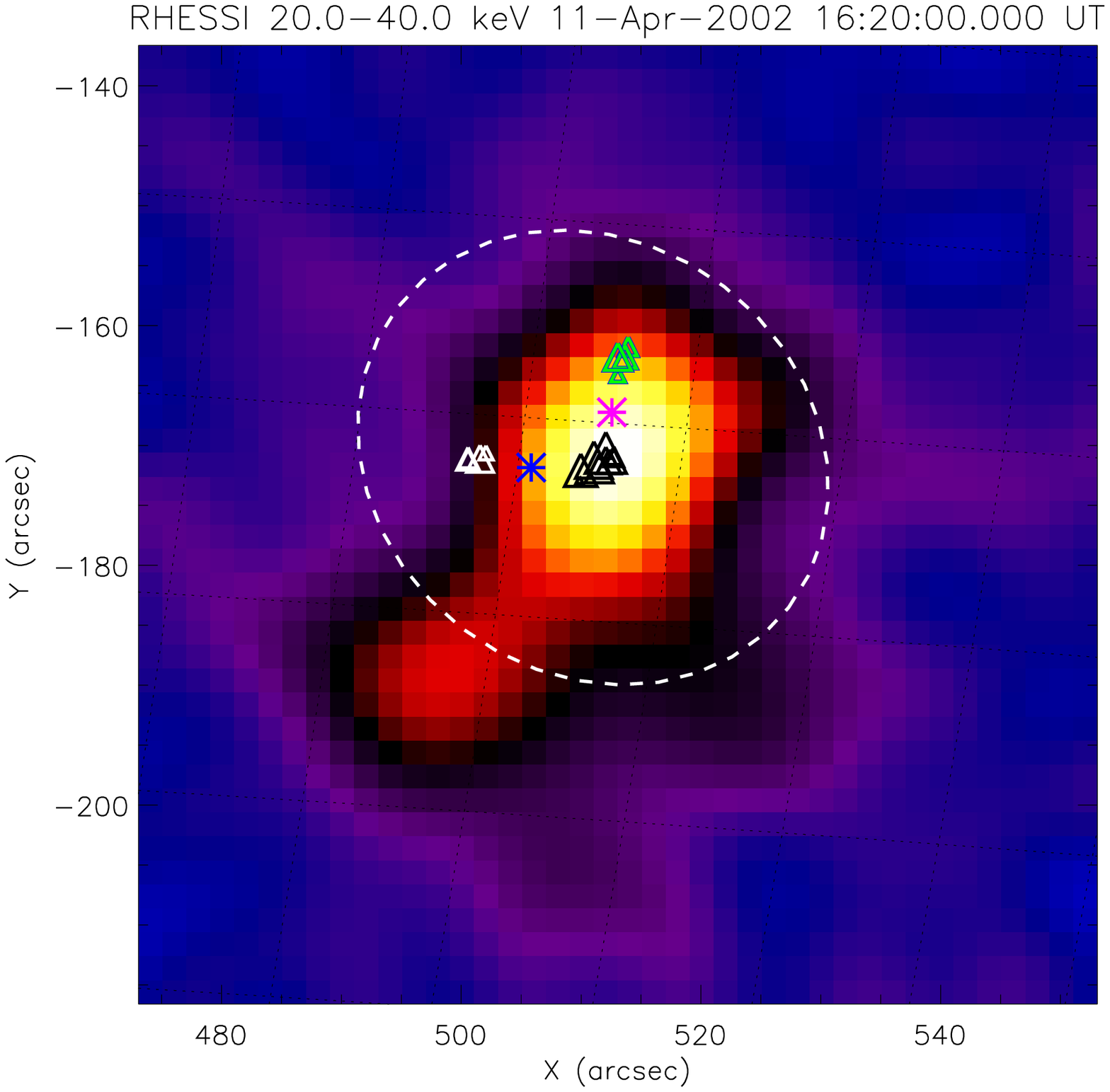} 
\includegraphics[width=0.450\columnwidth]{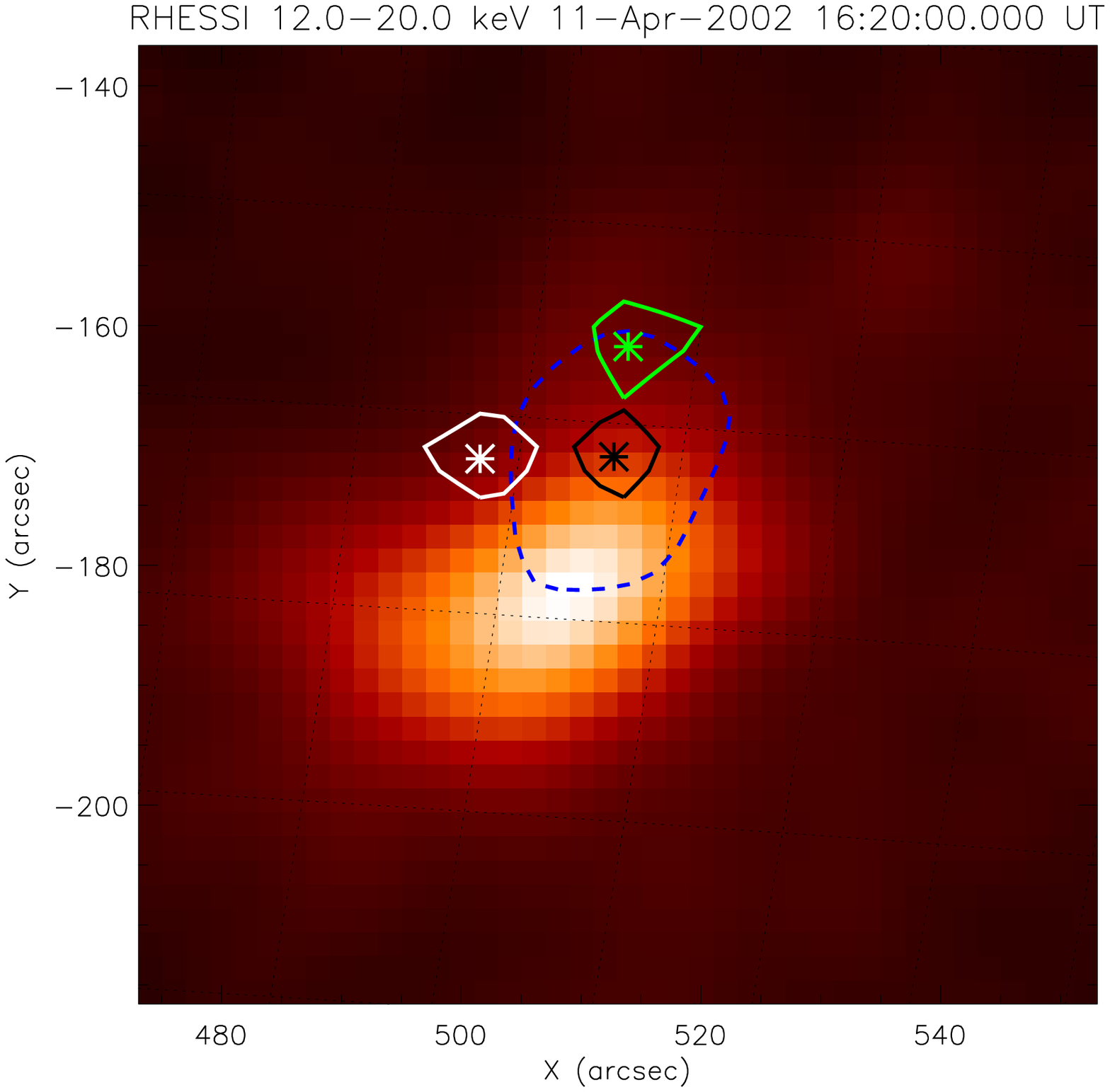} 
\caption{\label{fig_OVSA_Images} Left: Evolution of the spatial brightness peak of the radio emission at 2.6--4.2~GHz from April 11, 2002 flare.
Background: HXR image at $20-40$~keV at 16:20:00--16:21:00~UT. {\gf Symbols are OVSA image centroid positions separated by 8~s time interval (OVSA temporal resolution for imaging):
green triangles are for snapshots from 16:19:55 to 16:20:15~UT, pink asterisk for 16:20:20~UT, black triangles from 16:20:25 to 16:22:00~UT, white triangles from 16:23:40 to 16:24:20~UT, and the blue asterisk is for a late decay phase of 16:25:08~UT; larger triangles correspond to later snapshots within each group. The sequence of the contours clearly indicates that the radio source  is located at the northern part of the HXR image and stays there during the entire impulsive phase, then moves southward to exactly match the HXR centroid position and stays there the entire decay phase of the first peak, then it moves westward during the second peak, and finally returns back to HXR centroid location at the late decay phase. The accuracy with which the contours with the same color coincide with each other indicates the accuracy of the position measurements. The synthesized bean is shown by the dashed white oval.}
Right: {\dg centroids and 95\% contours of the three dominant source (corresponding to the green, black, and white sources on the left panel) positions} superimposed on the SXR image; blue dashed contour shows the HXR source at 60\% level.}
\end{figure}

Figure~\ref{fig_OVSA_Images} displays the  {\gf centroid} location of the radio emission during different stages of the event superimposed on the HXR and SXR images plotted for time interval 16:20:00-16:21:00~UT. A few important conclusions can be made based on the presented spatial relationships. First, we note a clear spatial evolution of the radio source, which is initially (during the impulsive peak of low-frequency microwave emission at 2.6--4.2~GHz) displaced by roughly $10''$ from the accompanying HXR source with even greater displacement, by roughly $20''$, from the accompanying SXR source. Upon transition to the decay phase, the radio brightness peak shifts by $\sim10''$ to exactly match the brightness center of the HXR source, which remains displaced $10''$ from the SXR source. Finally, during the second temporal peak of the flare around 16:24~UT, the radio brightness peak is displaced by $10''$ to the east, after which it eventually returns back to the location of the {\gfR radio} source {\gfR during the decay phase} of the first flare peak.

\begin{figure}\centering
\includegraphics[width=0.4\columnwidth,angle=90]{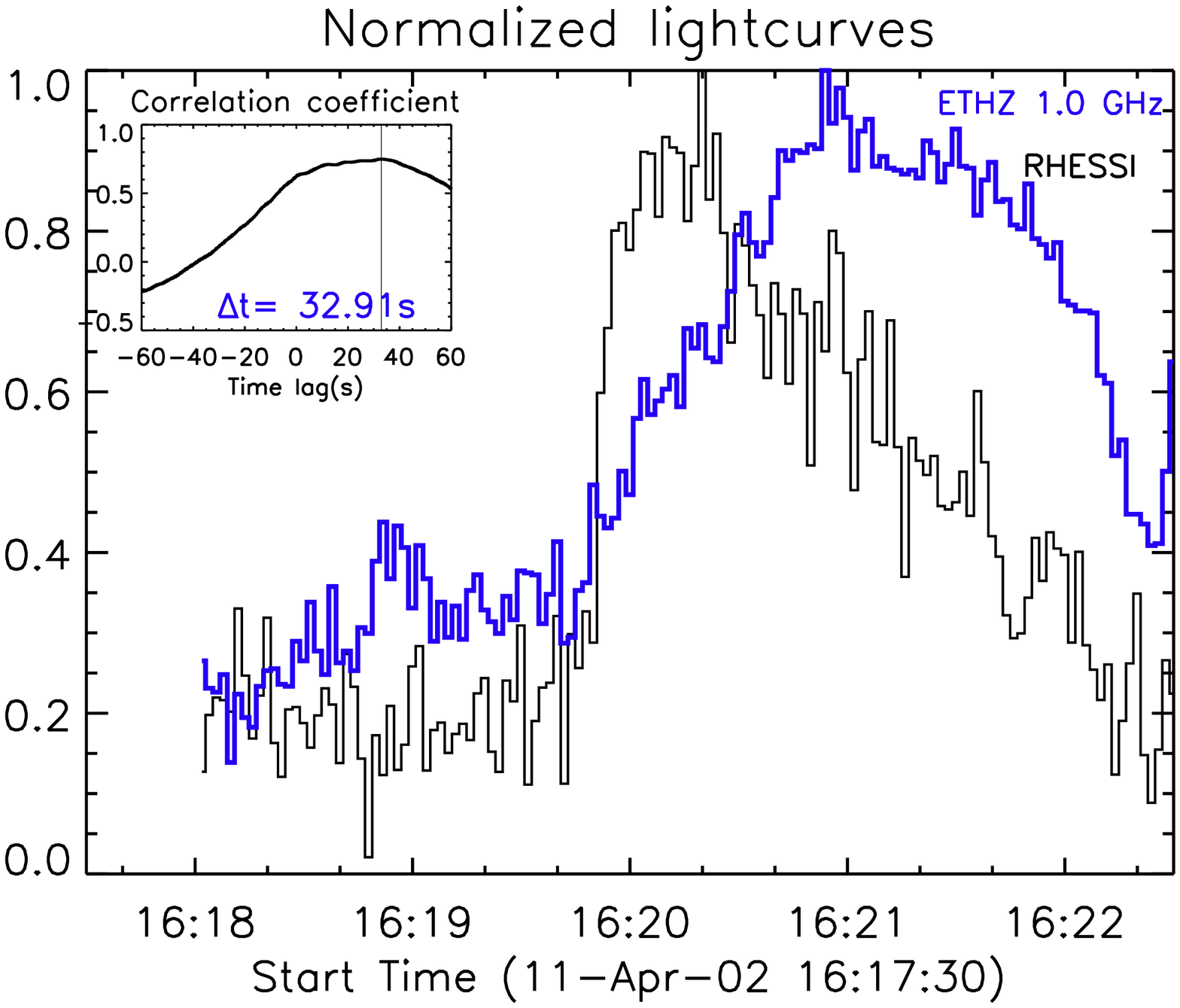}
\includegraphics[width=0.4\columnwidth,angle=90]{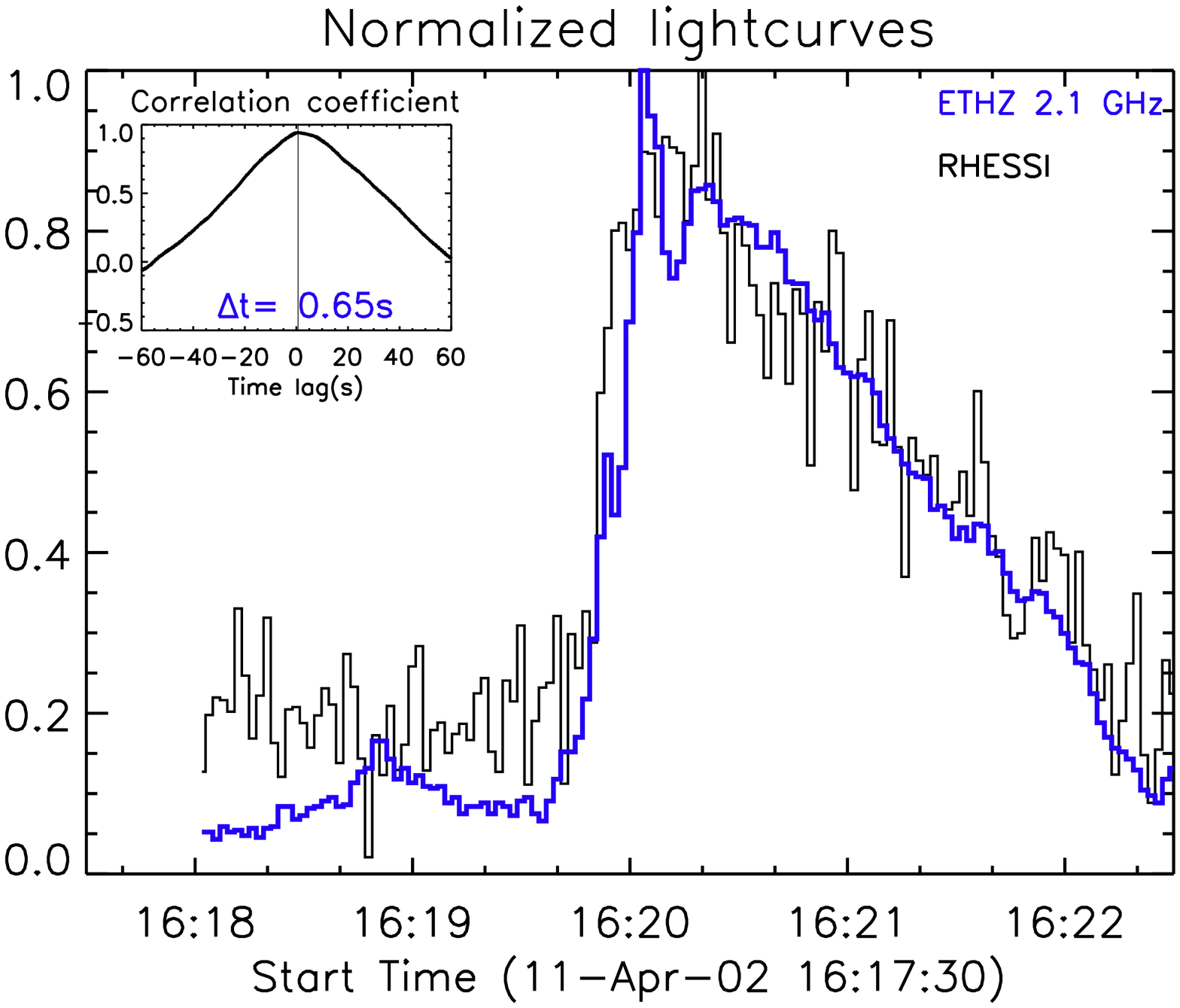}
\includegraphics[width=0.38\columnwidth,angle=90]{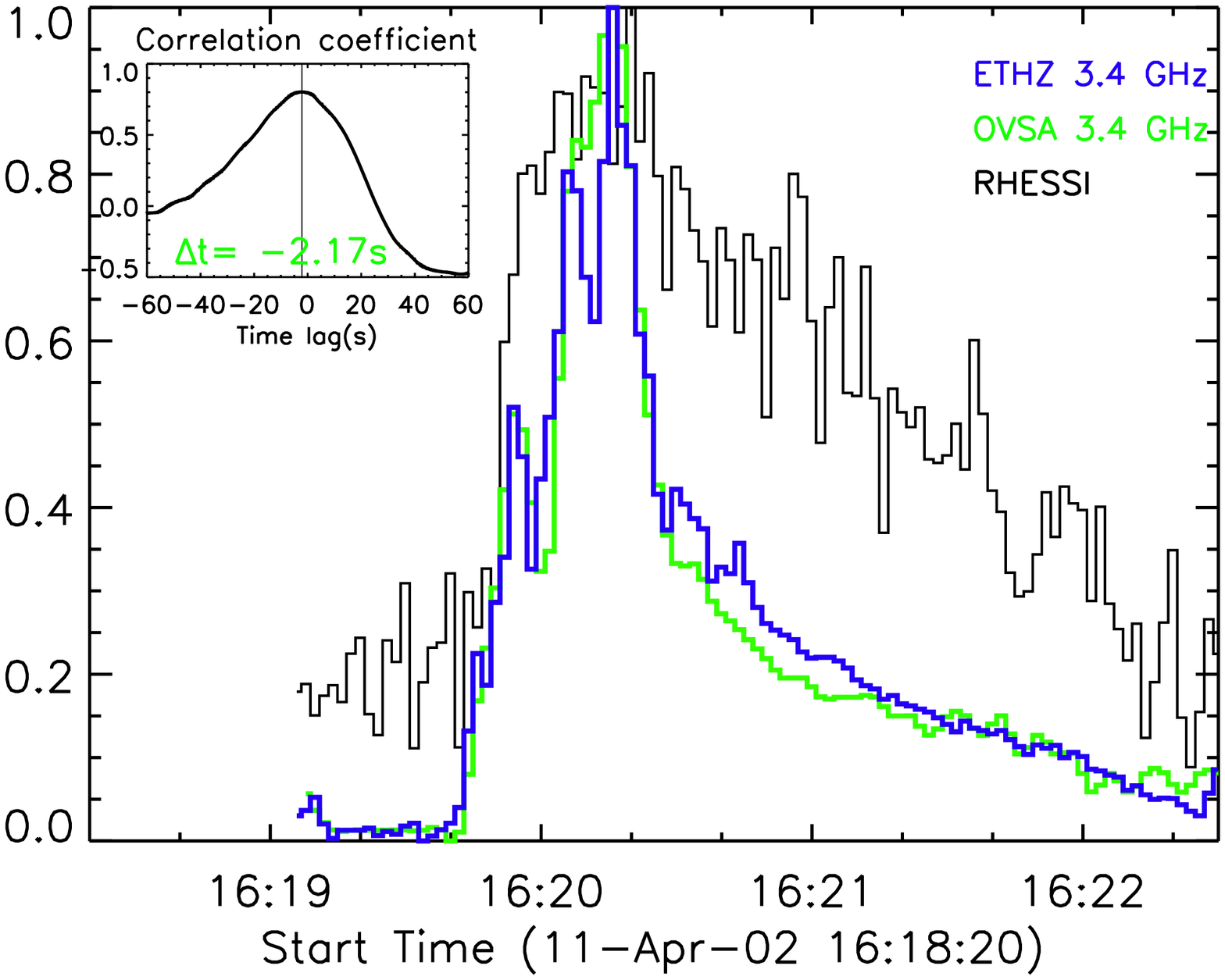}
\includegraphics[width=0.38\columnwidth,angle=90]{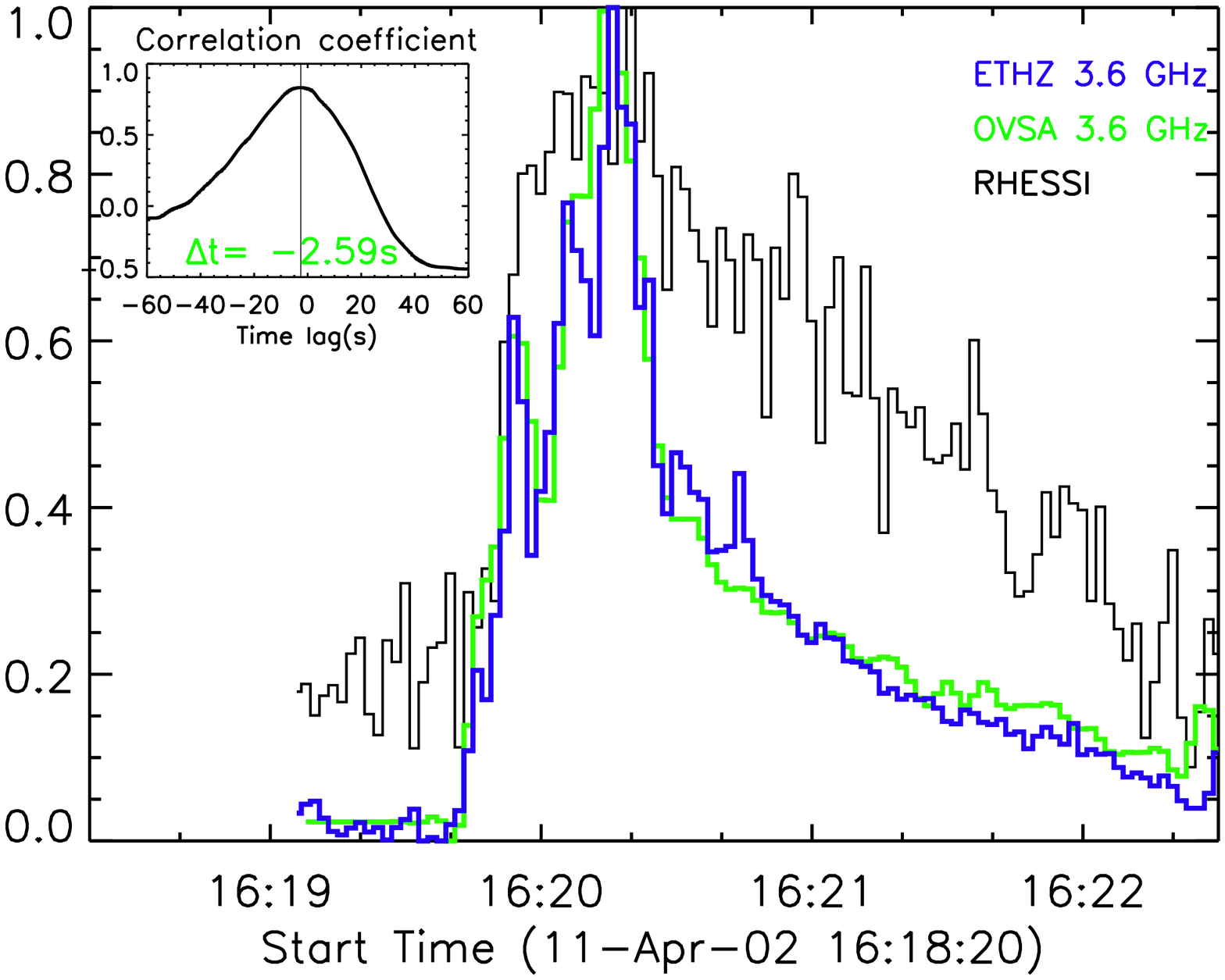}
\includegraphics[width=0.38\columnwidth,angle=90]{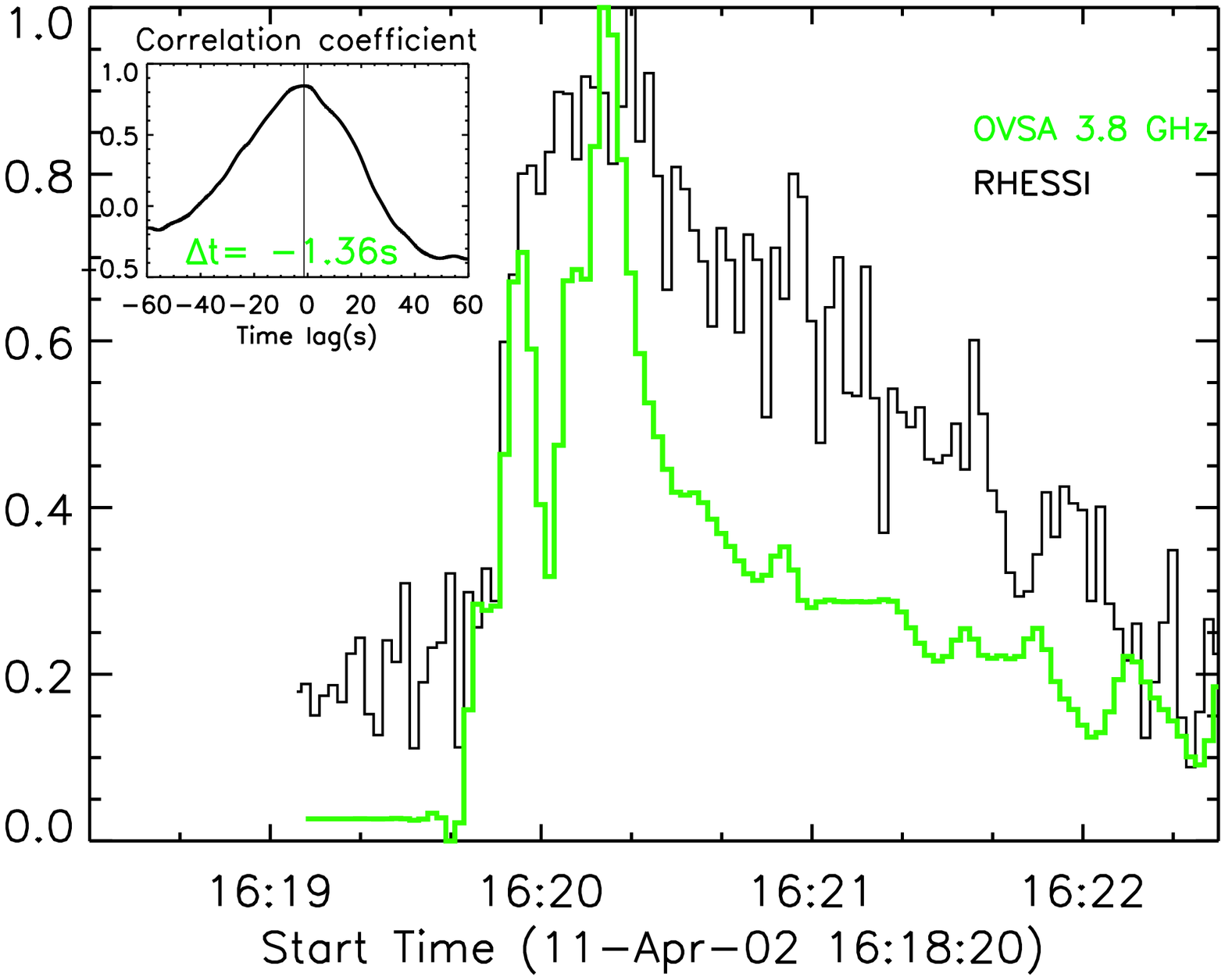}
\includegraphics[width=0.38\columnwidth,angle=90]{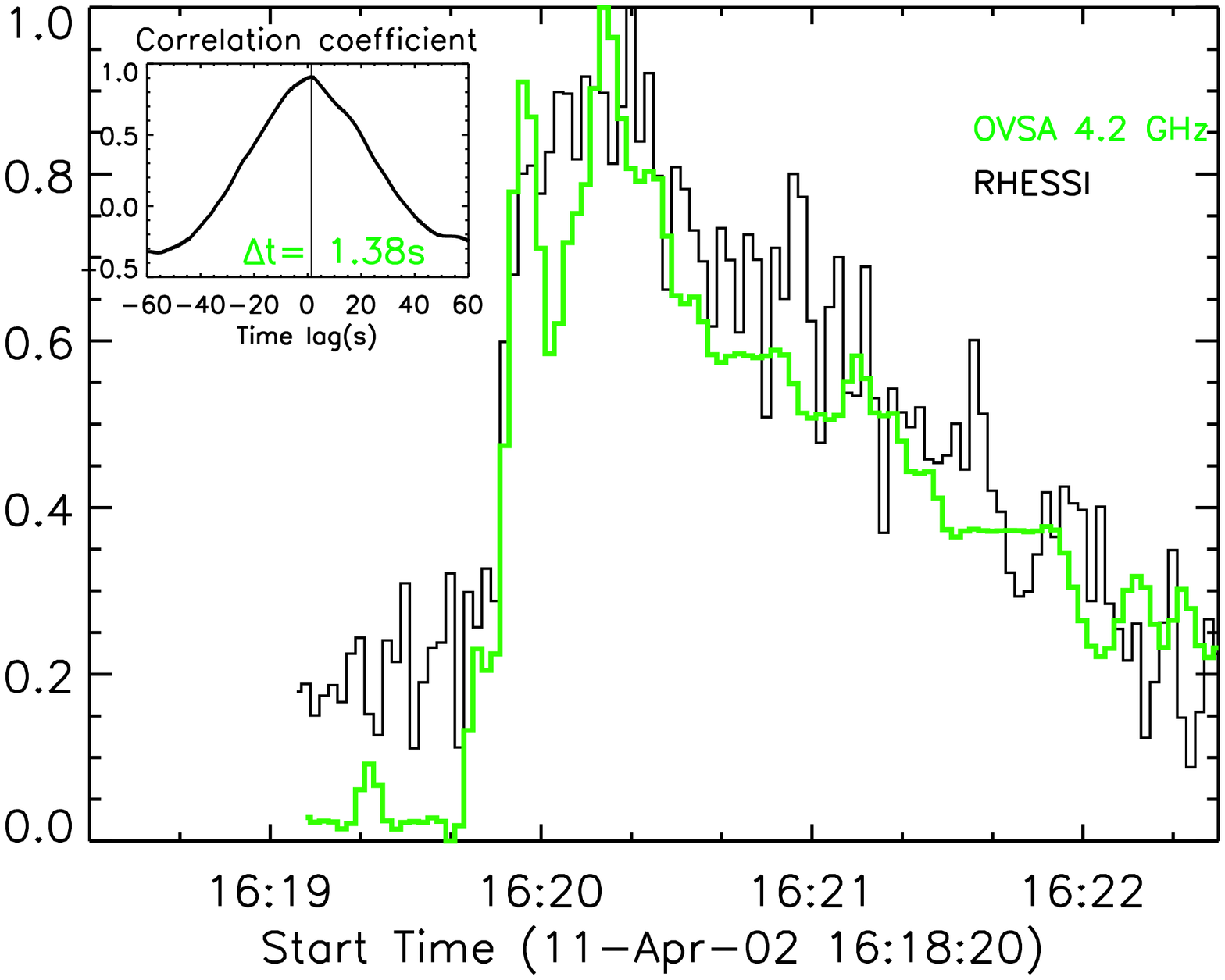}
\includegraphics[width=0.38\columnwidth,angle=90]{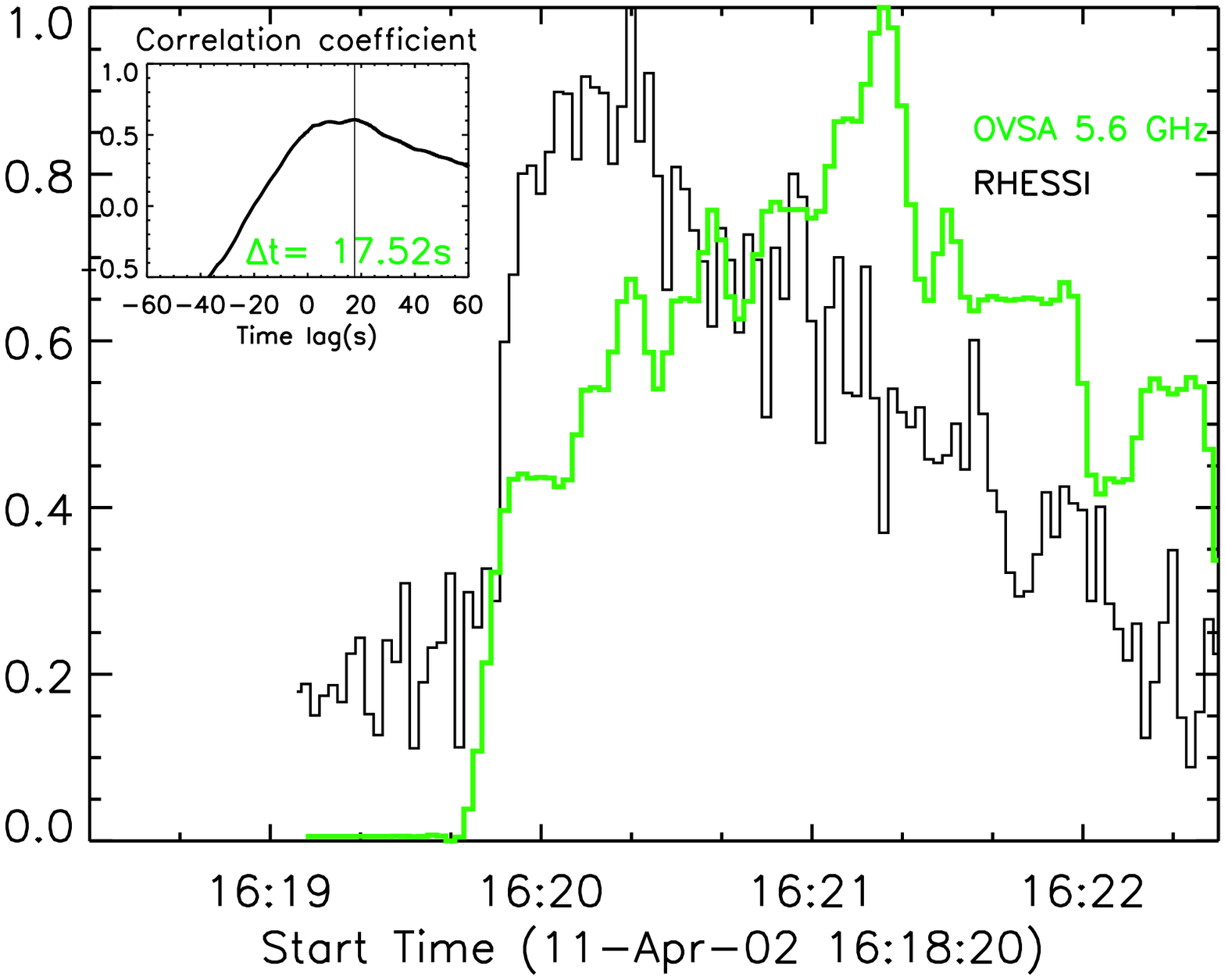}
\includegraphics[width=0.38\columnwidth,angle=90]{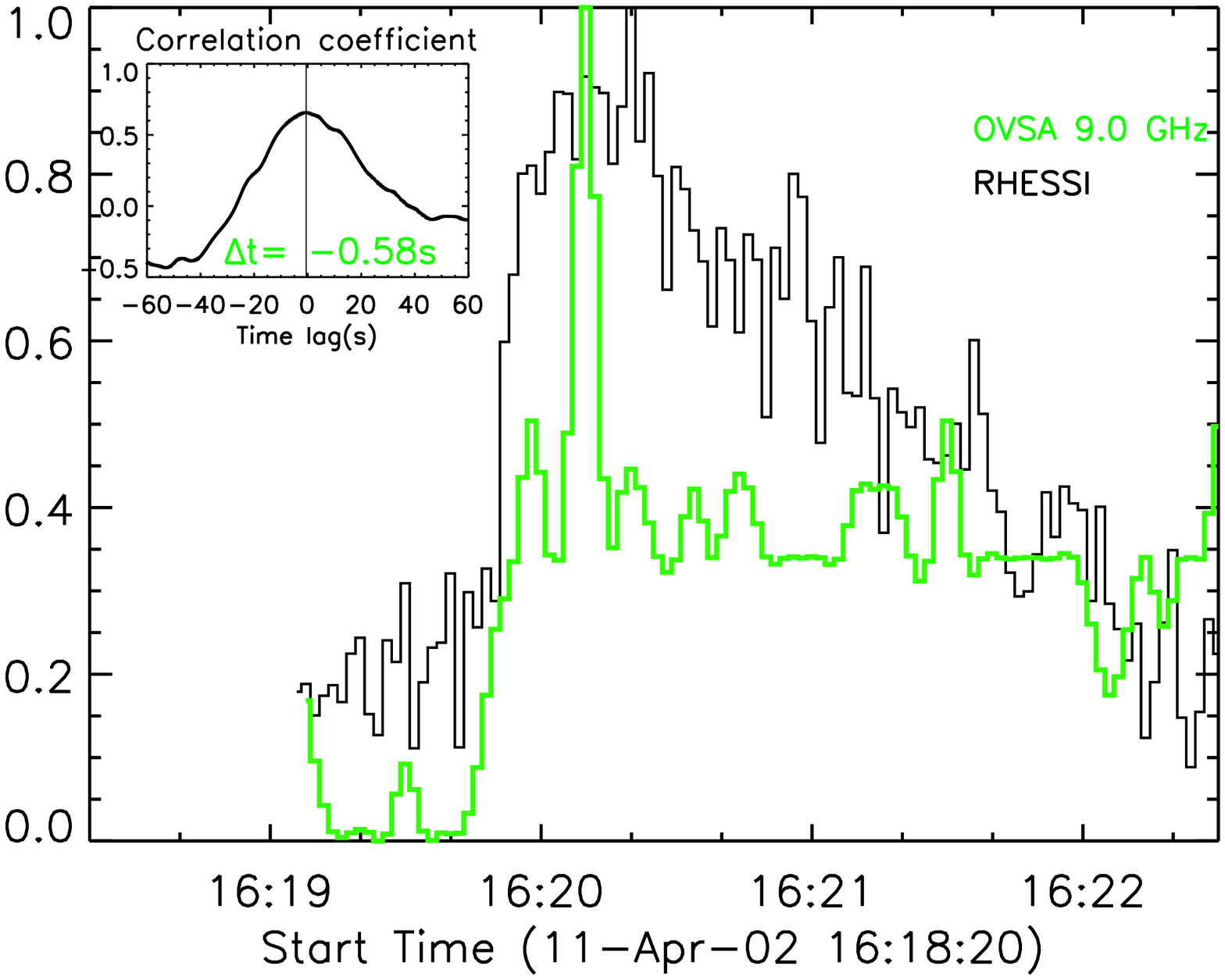}
\caption{\label{fig:Rtiming_main} Radio to HXR timing with Phoenix-2 and OVSA at different frequencies. Black thin curve, same throughout all the panels: HXR (20--40~keV) light curve with 2~s resolution normalized to 1;  blue thick curves: resampled (to the RHESSI 2~s resolution) Phoenix-2 light curves at different frequencies indicated at the panels; green curves: spline interpolated OVSA light curves from original 4~s resolution to the RHESSI 2~s resolution at different frequencies indicated at the panels. The insets: corresponding Radio-to-HXR lag correlations in which the time delay is printed by either blue (for Phoenix-2 data) or green (for OVSA data). The light curves at other frequencies are given in Figs.~\ref{fig:Rtiming}---\ref{fig:Rtiming_4} in the end of the file.}
\end{figure}

\subsection{Radio to X-ray Timing}

Study of the emission light curves including cross-correlations and delays between various microwave and X-ray channels is often highly helpful in identifying distinct emission components, which as noted in the introduction may be produced by electrons at the acceleration site, by magnetically trapped electrons, or by precipitating electrons. Figure~\ref{fig:Rtiming_main} displays temporal relationships between radio light curves recorded by Phoenix-2 and OVSA on one hand and HXR light curves recorded by RHESSI  on the other hand. To perform the lag cross-correlation analysis between the radio and  HXR light curves, \blank{and} the Phoenix-2 light curves were resampled, while OVSA light curves spline interpolated to match 2~s RHESSI resolution. The displayed radio light curves are distinctly different from each other depending on the frequency and can be grouped as follows.

The low frequency radio light curves, $f<2$~GHz, display a significant time delay compared with the HXR light curve: up to half a minute as determined from lag cross-correlation and up to 45~s if the peak times are considered. Then, at higher frequencies, 2~GHz~$<f<$~5~GHz, there is no significant delay between the light curves; the delay varies within $\pm2$~s in this frequency range. {\dg However, the various radio light curves do show spectral evolution} within this range: they become more and more impulsive as frequency changes from 2 to 2.8~GHz, then the light curves stay roughly similar to each other at $2.8-3.8$~GHz, and then they again become less impulsive at $4-5$~GHz. At higher frequencies, $5.0-7.5$~GHz, the radio light curves are again significantly delayed compared with the HXR light curve similarly to the lowest frequency range. Finally, the highest frequency range is characterized by light curves with an impulsive peak coinciding with the HXR peak and with the impulsive radio peak at frequencies $2.8-3.8$~GHz, but followed by a more extended part of the light curves.

We conclude that the radio light curves consist of two essentially different components---impulsive and delayed---whose relative contributions depend strongly on the frequency.\footnote{No impulsive component {\gfR (acceleration region contribution)} is detected during the second peak of the flare occurring around 16:24:00~UT; not shown in Figure~\ref{fig:Rtiming_main}.} 
The delayed radio light curves behave exactly as expected for the microwave emission produced by a {\gf magnetically} trapped component of fast electrons
\citep[e.g.,][]{Melnikov_1994, Bastian_etal_1998, Kundu_etal_2001, Melnikov_2006}. 
{\dg In contrast, the impulsive radio light curves around the spectral peak frequency of 3.2~GHz reach maximum flux simultaneously with the HXR light curve and so cannot be ascribed} to the trapped component. Instead, their timing shows remarkable similarity to that found {\dg in the cold, tenuous flare \citep{Fl_etal_2011}, i.e.} microwave emission produced directly
from {\dg an} acceleration region at low frequencies and from precipitating electrons at high frequencies.
{\dg We suggest} that the impulsive and delayed emissions come from physically different sources, which is consistent with {\dg their} different spatial locations {\dg discussed in} the previous section. 

\section{Radio Spectral Fit}

\begin{figure}\centering
\includegraphics[width=0.9\columnwidth]{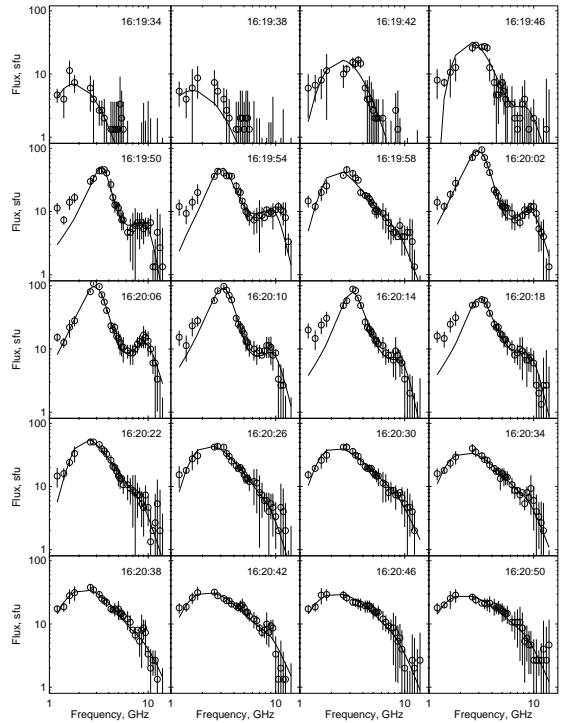}
\caption{\label{fig_OVSA_fit_Spectra_0} Radio  spectral fit as described in the text (solid curves) of the OVSA spectra (symbols with scatter) at consequent snapshots with 4~s cadence; Part I.
Other spectra with fits are given in Figs.~\ref{fig_OVSA_fit_Spectra_1}---\ref{fig_OVSA_fit_Spectra_4} below.}
\end{figure}

{\dg An alternative way of looking at spectral evolution, rather than \textit{frequency-dependent light curves}, is through \textit{time-dependent spectra}} during the burst. {\dg These are} shown in Figure~\ref{fig_OVSA_fit_Spectra_0}, which includes both the measured total power radio spectra (points) and spectral fits to be described below.  The observed spectral evolution is in marked contrast with the event of 2002 July 30 analyzed by \citet{Fl_etal_2011}, where almost no spectral evolution was present. Indeed, \citet{Fl_etal_2011} successfully fitted the entire sequence of the radio spectra in the 2002 July 30 event with a source model having only one free parameter---the instantaneous number of the accelerated electrons 
at the source. Judging from the significant spectral evolution in the event under study, such a simplified approach to the radio spectral fit cannot work. A multi-parameter fit similar to that developed by \citet{Fl_etal_2009} is needed in this case. However, \citet{Fl_etal_2009} applied their multi-parameter fit to a sequence of spatially resolved (model) spectra, while here we are forced to deal with the total power spectrum, in which source nonuniformity can be expected to play a major role \citep{Kuznetsov_etal_2011}.

Note that many of the spectra in Figure~\ref{fig_OVSA_fit_Spectra_0} show two spectral peaks. A similar two-component structure was present in all radio spectra in the 2002 July 30 event \citep{Fl_etal_2011}, which was successfully interpreted as a combination of a main (low-frequency) coronal source, {\gfR representing in fact the acceleration region in that event,} and a secondary (higher-frequency) source formed by the precipitating component of the fast electrons. Here we employ a similar two-{\gfR component spectral}  model, i.e., that the radio spectrum is formed in two uniform sources physically related to each other.
{\gf While   {\gfR many} of the radio spectra do not show two clearly separated spectral peaks, nevertheless, we fit two homogeneous spectral components at all times because they are needed to account for the source inhomogeneity, which broadens the spectrum to the extent that it cannot be fitted by a single-component, homogeneous GS spectrum.}

Specifically, we adopt {\gf the premise} that the bulk of low-frequency emission comes from a coronal source with unknown (free) parameters such as magnetic field $B_1$ and thermal density $n_e$, where a power-law  population of fast electrons is present with an unknown number density $n_r$, spectral index $\delta$, and high-energy cut-off $E_{\max}$, but with a fixed low-energy cut-off $E_{\min}=17$\blank{20}~keV, as shown in the lower panel of Figure~\ref{fig:fit_tt}.
{\gf The GS emission, especially its polarization, depends on the viewing angle and angular distribution of the fast electrons \citep[e.g.,][]{Fl_Meln_2003a}. Since no polarization data \blank{was} are available, we fix the viewing angle at a quasi transverse value, $\theta=80^\circ$ and adopt the simplest case of an isotropic electron distribution.
}

Based on the imaging data we fix the  {\gfR main} source size to be $25''$ {\dg in each dimension}
so the low-frequency source is fully characterized by these five free parameters. For the  {\gfR second} (`precipitating') source we begin with one free parameter, the magnetic field $B_2$, keeping $n_e$, $E_{\min}$, and $E_{\max}$ values the same as for the `coronal' source. The spectral index of the {\gf instantaneous distribution of} precipitating electrons is taken to differ by 1/2 from that of the main electron population, $\delta_{pr}=\delta+1/2$, to account for the time of flight effect, {\gf $\tau_{\rm TOF} \sim L/v \propto E^{-1/2}$, which assumes  {\gfR energy-independent escape and} a free propagation transport regime for the precipitating electrons}, and the size of the high-frequency source is corrected to conserve magnetic flux along the flaring magnetic flux tube.
{\gfR  Certainly the adopted precipitation model  may be oversimplified \citep[see, e.g.][]{van_den_Oord_1990, Su_etal_2011, Holman_2012}\, but it is sufficient for the spectral fitting and allows clear isolation of the low-frequency spectral component during the time when the double peak structure is clearly present in the radio spectrum.  Since our focus is on the low-frequency component, we forego more-detailed discussion of the precipitating source.}
The gyrosynchrotron source function is computed by the fast GS code developed by \citet{Fl_Kuzn_2010}.

\begin{figure}\centering
\includegraphics[width=0.93\columnwidth]{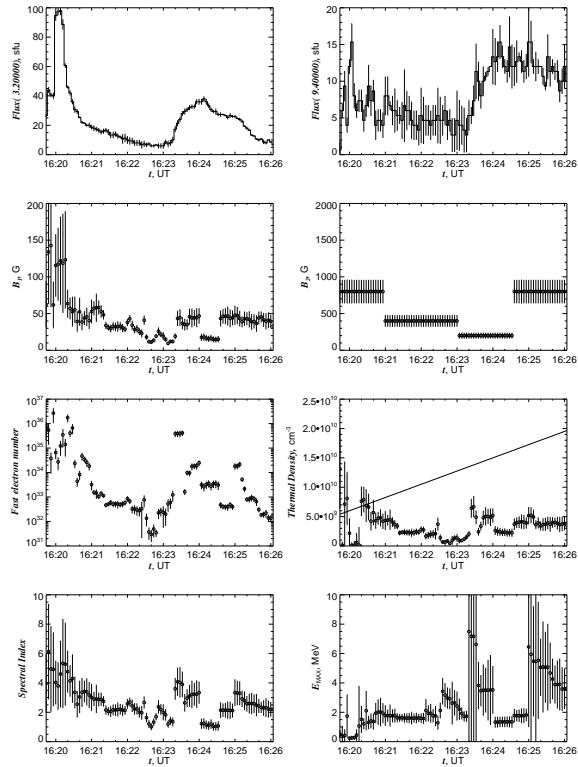}
\caption{\label{fig_OVSA_fit_parms} Radio source parameters {\gf as derived from} the OVSA spectral fit {\gf for five parameters of the low-frequency coronal source and adopted magnetic field value $B_2$ for the 'precipitating' source as described in the text}. A solid curve at the thermal plasma number density shows a number density {\gf evolution} of the SXR source derived from emission measure from the RHESSI fit. {\gf Two top panels show the radio light curves recorded at 3.2~GHz and 9.4~GHz given for the reference purpose.} 
}
\end{figure}

Although this six-parameter fitting succeeds throughout the entire flare duration, {\gfR there is a mismatch between the fit and observed spectrum at some time frames at the lowest frequencies. This is a well-known indication of source nonuniformity \citep{Nita_etal_2004, Altyntsev_etal_2008, Qiu_etal_2009}, which could have been accounted by adding a contribution from the electron accumulation site (see below); we avoid this complication, however, because the parameters of this third spectral component could not be reliably constrained from the fit.  }
The derived physical parameters show noticeable scatter, particularly in the magnetic field $B_2$, whose values clustered around several discrete levels near 800, 400, and 200~G during different flare stages\footnote{In fact, when $B_2\simeq200$~G, the contribution from this source is negligible, thus, a one-component GS fit works well during this period of time.}.
Accordingly, to improve stability of other derived parameters we fixed the $B_2$ (whose exact value is inessential to what is discussed below, but needed to obtain a good fit, {\gfR especially, for the spectra with two peaks}) to one of the above values based on the original six-parameter fit, for time ranges corresponding to peak 1 impulsive phase (800~G), peak 1 decay phase (400~G), peak 2 rise phase (200~G) and peak 2 decay phase (again 800~G), as is explicitly shown in Figure~\ref{fig_OVSA_fit_parms}.  We then repeated the fitting for the remaining five free parameters in the low-frequency (`coronal') source, and the same parameters with the above-stated adjustments in $\delta$ and source size for the high-frequency (`precipitation') source{\gf , therefore, the presence of the precipitation source does not add any additional free parameter to the model}.  The time evolution {\gf of the derived parameters of the coronal source and adopted magnetic field $B_2$ of the precipitation source} is shown in the bottom six panels of Figure~\ref{fig_OVSA_fit_parms}, compared with the observed 3.2 and 9.4~GHz microwave time profiles (upper panels).

The {\dg derived evolution of the} physical parameters deserves {\dg some discussion}. The magnetic field {\dg in} the low-frequency source is about 120~G during the impulsive phase of the radio burst, while it drops quickly to 30--50~G at the transition to the decay phase around 16:20:20~UT. Remarkably, this magnetic field change derived from the \textit{spectral fit}, {\gfR Figure~\ref{fig_OVSA_fit_parms},} happens at the very same time as the $10''$ shift of the \textit{spatial brightness peak}, {\gfR see Figure~\ref{fig_OVSA_Images}}. 
This implies that it makes sense to distinguish between these two {\dg spatially distinct} {\gf low-frequency} sources,  {\gfR which as we show below represent the very acceleration region (the early source, with $B\sim120$~G, producing the impulsive radio emission) and the  classical looptop radio source (the  later source, with $B\sim40$~G, producing the radio emission {\dg from magnetically trapped electrons} {\gf over} the decay phase) {\gff spatially coinciding with the HXR source}.}

\textbf{Acceleration region.} At the {\gf impulsive low-frequency} source, {\gfR $\sim$ 16:20:00--16:20:20~UT,} the thermal number density {\gff obtained from the radio fit} is somewhat low, {\gf $n_e \lesssim 2\cdot10^9$~cm$^{-3}$, implying the {\gfR radio} source is located in the corona, not at a chromospheric footpoint}, while the number of nonthermal electrons is consistent with the acceleration rate derived from the HXR data, $(1-3)\cdot10^{34}$~electron/s, see Figure~\ref{fig:fit_tt}, \blank{with the lifetime of} {\gf if they reside at the {\gff radio} source for} 2--4~s,  {\gfR which requires the strong diffusion transport mode}. The radio derived electron spectral index does not display any significant departure from the HXR derived electron spectral index  {\gfR during this time interval}, cf. Figure~\ref{fig:fit_tt}. All these properties are similar to those determined for the acceleration site in the 2002 July 30 event \citep{Fl_etal_2011}, {\dg from which we conclude} that we have here another instance of the acceleration region detection in a solar flare.   {\gfR The electrons accelerated at this source escape from there in roughly 3~s and then accumulate in another, 'trapping' source, which dominates the radio spectrum and spatial location after 16:20:20~UT. The acceleration, however, continues for a longer time:} we note that the maximum electron energy, $E_{\max}$, displays a monotonic increase from $\sim300$~keV to $\sim2$~MeV over this phase of the burst ($\sim$16:20:00--16:20:50~UT), which is reasonable to interpret as {\dg the} growing of a power-law `tail', i.e., the very process of the electron acceleration. Thus, the \blank{`first' source} {\gf impulsive low-frequency source dominating the radio emission over roughly 16:20:00--16:20:20~UT},  {\gfR  which produces fast electrons and  supplies them to the coronal trapping site  until at least 16:20:50~UT,} can confidently be identified with the acceleration region of the flare under study.

\begin{figure}\centering
\includegraphics[width=5.9cm,angle=90]{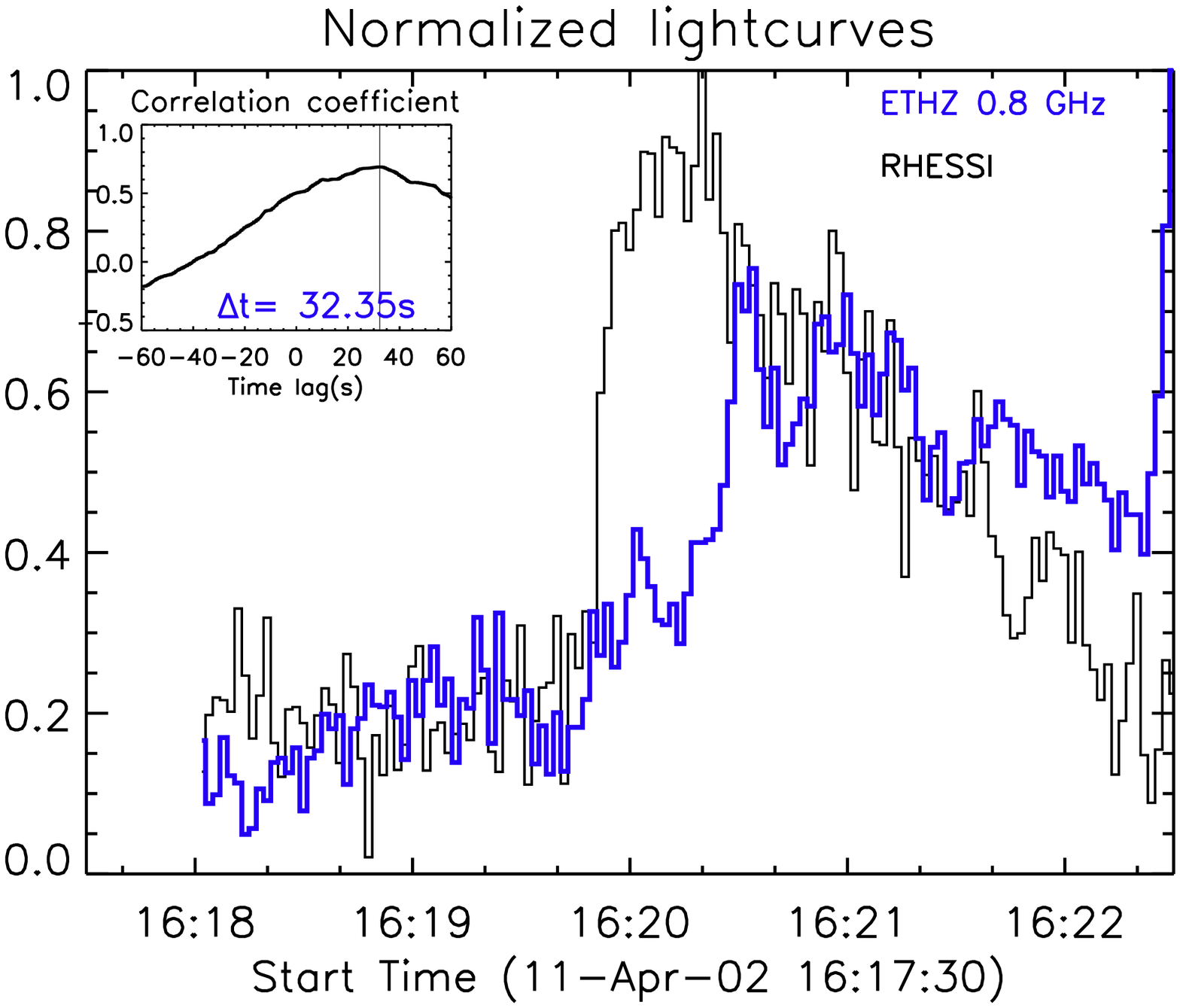}
\includegraphics[width=5.9cm,angle=90]{rhessi-radio_10.eps}
\includegraphics[width=5.5cm,angle=90]{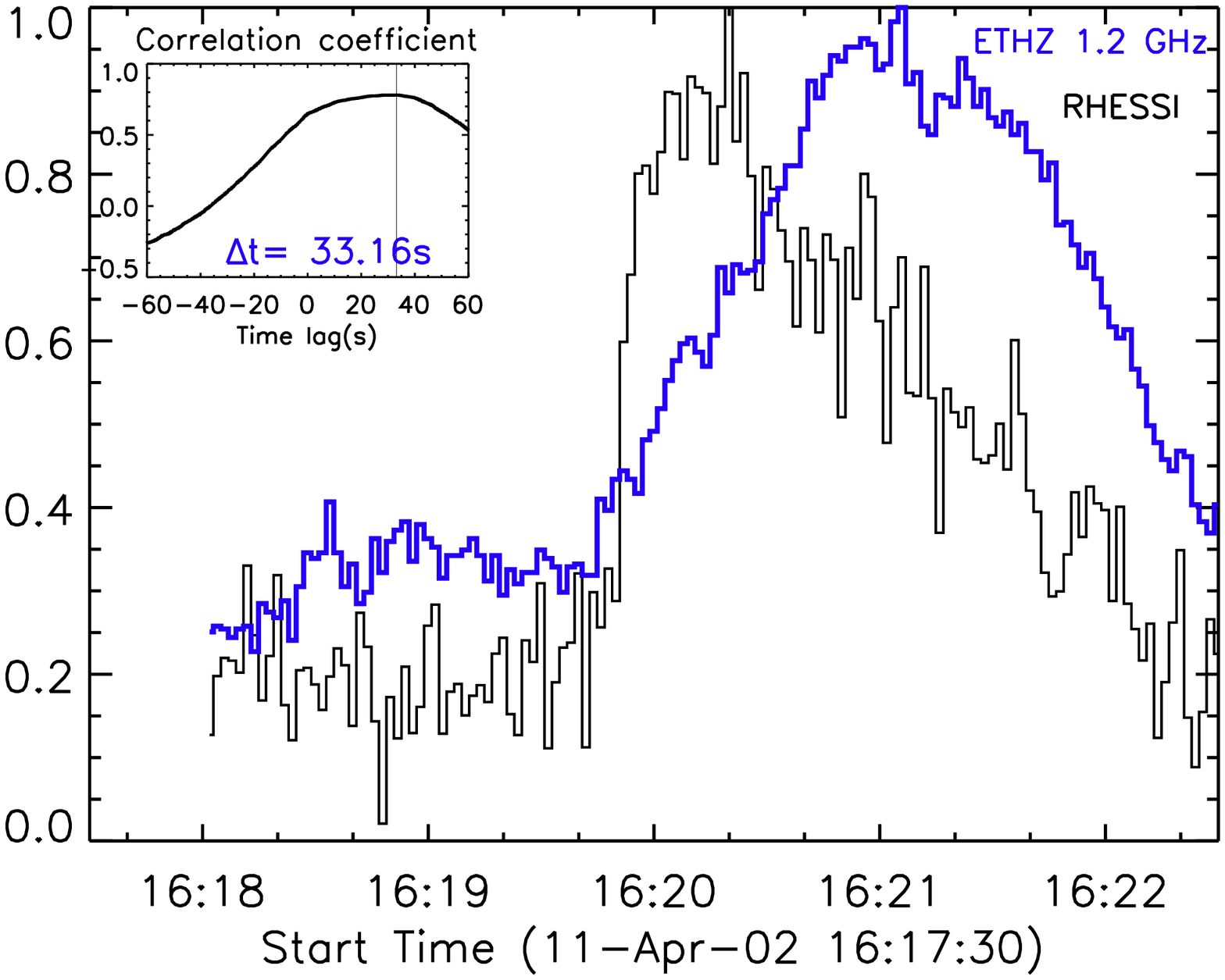}
\includegraphics[width=5.5cm,angle=90]{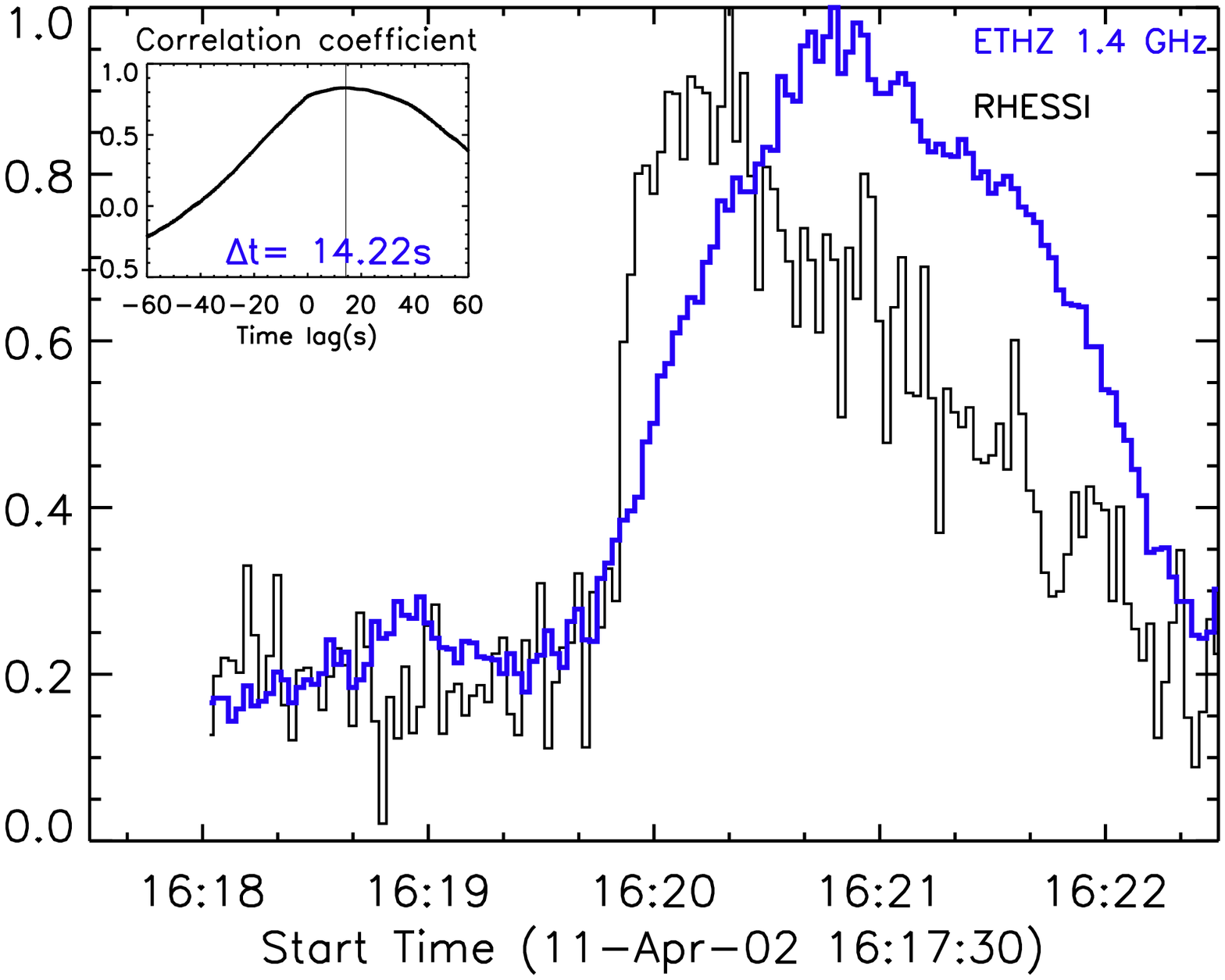}
\includegraphics[width=5.5cm,angle=90]{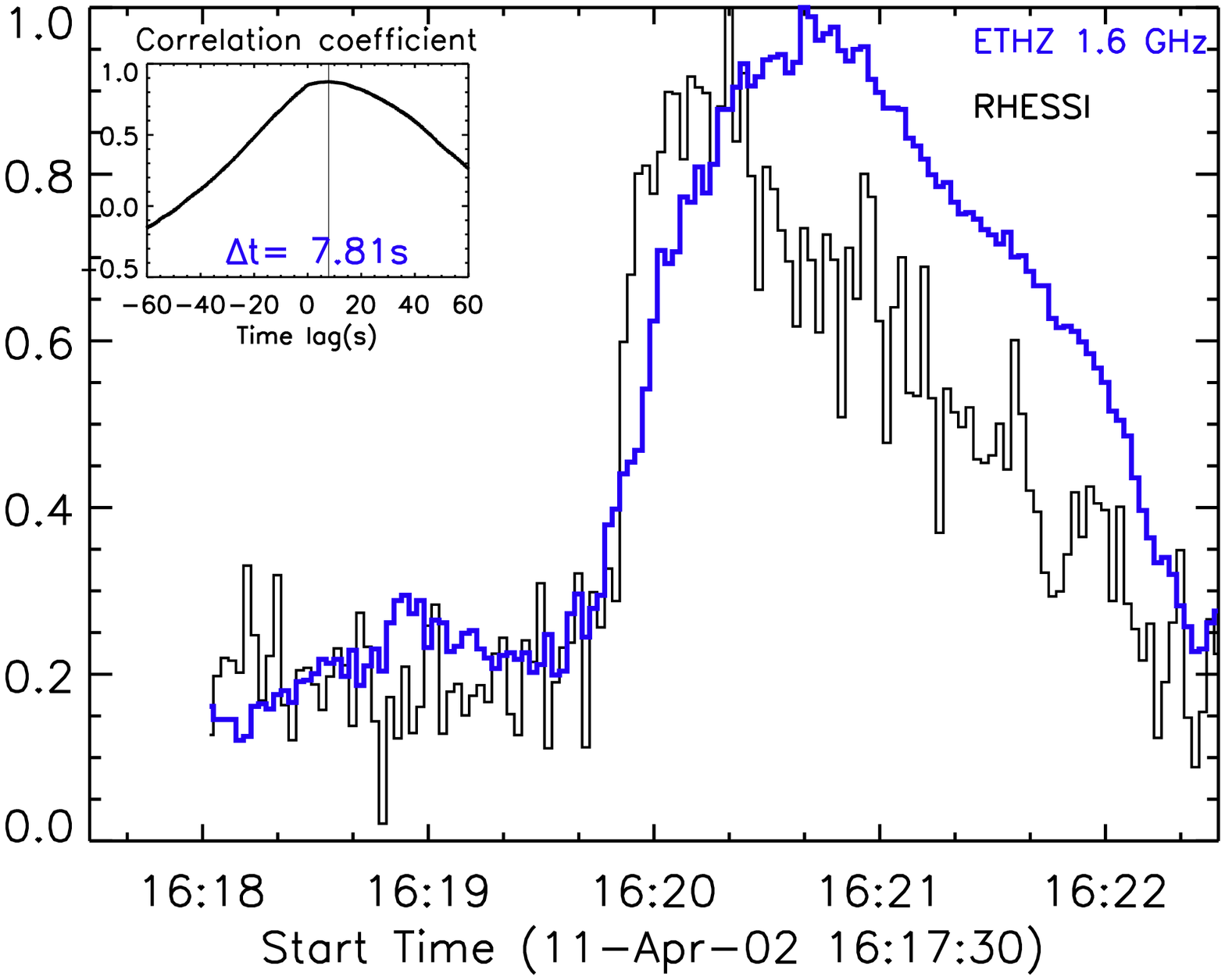}
\includegraphics[width=5.5cm,angle=90]{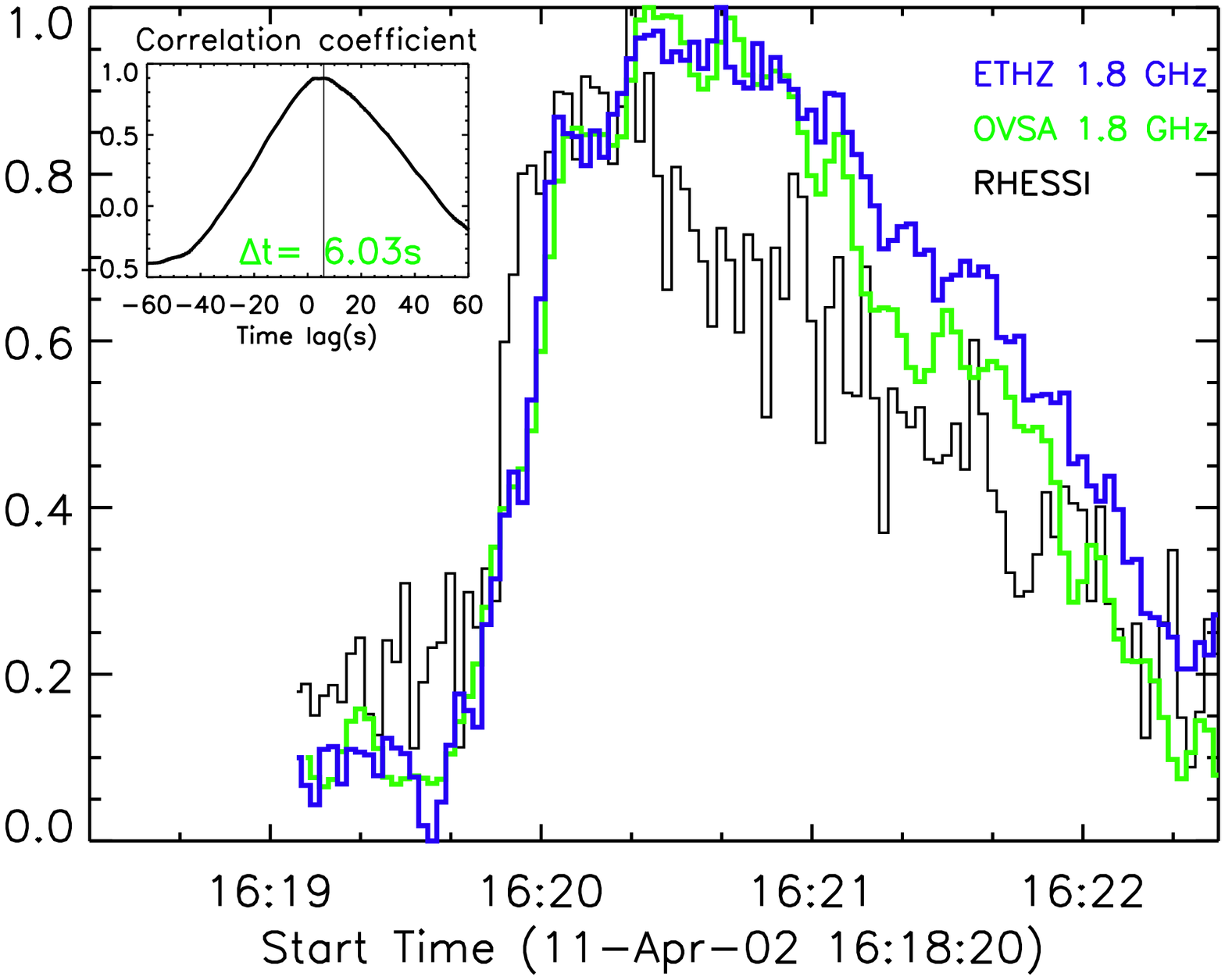}
\caption{\label{fig:Rtiming} Radio to HXR timing with Phoenix-2 and OVSA at different frequencies. Black thin curve, same throughout all the panels: HXR (20--40~keV) light curve with 2~s resolution normalized to 1;  blue thick curves: resampled (to the RHESSI 2~s resolution) Phoenix-2 light curves at different frequencies indicated at the panels; green curves: spline interpolated OVSA light curves from original 4~s resolution to the RHESSI 2~s resolution at different frequencies indicated at the panels. The insets: corresponding Radio-to-HXR lag correlations in which the time delay is printed by either blue (for Phoenix-2 data) or green (for OVSA data). 
}
\end{figure}

\textbf{Electron accumulation site.}
Transition to the  {\gfR gradual} decay  {\gfR phase}\footnote{Its main parameters, $B_1\sim 30-50$~G and $n_e \lesssim 5\cdot10^9$~cm$^{-3}$, clearly indicate its coronal location, although spatially distinct from the acceleration region showing a different \textit{coronal} location.} at about 16:20:20~UT  {\gfR manifests the stage when the trapping site has accumulated a sufficient number of fast electrons to dominate the radio spectrum. At this time} the derived number of accelerated electrons with $E\gtrsim20$~keV reaches a maximum of $\sim10^{36}$~electrons, which corresponds to the number density of the accelerated electrons of $n_r\sim2\cdot10^8$~cm$^{-3}$ for the adopted source volume. For the determined earlier acceleration rate of $(1-3)\cdot10^{34}$~electron/s, having {\dg a total} electron number {\dg of} $\sim10^{36}$ requires a highly efficient electron trapping with the trapping time longer than 30~s. Indeed, that long trapping time is fully confirmed by the measured delay between the HXR/impulsive radio light curves and 'non-impulsive' radio light curves, see Figure~\ref{fig:Rtiming}, some of which are delayed by almost one minute. This   {\gfR means} that  {\gfR in} this {\gfR electron accumulation site}  the low-frequency radio emission is dominated by the \textit{magnetically trapped} electron component,  {\gfR which is often seen in microwave bursts \citep[e.g.][]{Melnikov_1994, Meln_Magun_1998, Bastian_etal_1998, Lee_Gary_2000, Kundu_etal_2001, Melnikov_etal_2002, Bastian_2006_Nobe, Tzatzakis_etal_2008, Reznikova_etal_2009}. In contrast to the strong diffusion mode in the acceleration region established above, effective magnetic trapping implies a \textit{weak diffusion mode} at the accumulation site, mediated, perhaps, by Coulomb collisions.}

I{\gfR ndeed, i}t is well known \citep[e.g.,][]{Melrose_Brown_1976, Melnikov_1994, Bastian_etal_1998, Meln_Magun_1998, Lee_Gary_2000, Melnikov_2006} that Coulomb losses of the trapped electrons result in a flattening of the fast electron energy spectrum. {\gf This effect must be present and accounted for since the Coulomb loss time is about 10~s for 20~keV electrons, which is much shorter that the trapping time at the decay phase of the burst.} Indeed, the recovered evolution of the electron spectral index (Figure~\ref{fig_OVSA_fit_parms}$e$) displays a monotonic decrease over the entire decay phase, nicely in agreement with the theoretical expectation, while the maximum electron energy remains roughly constant at the level of $\sim2$~MeV implying that no acceleration to even higher energy is demanded by the radio spectrum fit.  {\gfR Although some of the details of the accumulation site interpretation are not unambiguously confirmed, e.g., some regimes of the wave scattering can result in a weak diffusion mode, \citep[see, e.g.,][]{Bespalov_etal_1987, Stepanov_Tsap_2002}, and still be compatible with efficient magnetic trapping, \textbf{our main point is that} this {\gff radio} source {\gff (coinciding with the HXR source)} is distinctly different from the acceleration region source, discussed above. }

In addition to the spectral index evolution, the fit results also suggest a modest decrease of the magnetic field and the thermal number density at the radio source. This behavior can be understood if we suppose, as is likely, that the flaring loop is nonuniform with height in such a way that higher field lines are linked with more tenuous thermal plasma (implying slower Coulomb loss rate) than more compact lower field lines. In this case the trapped fast electrons will live longer at the outer field lines, so the outer looptop regions (with lower magnetic field and smaller thermal density) will dominate the radio emission at the late decay phase. We note that this source remains at the same projected position during the entire decay phase as seen from the OVSA imaging. The derived thermal number density in the radio source is always below the thermal number density derived from {\dg the emission measure} determined from the RHESSI X-ray spectral fit, which is consistent with significant spatial displacement between these two sources.

The described fit results are highly stable vs. various modifications of the fitting process including varying weights of the data points and initial values of the fitting parameters used as input. In contrast, the fit is not so stable during the second flare peak around 16:24~UT, which is a direct consequence of broader observed radio spectra, indicative, perhaps, of a stronger role of source inhomogeneity at this second peak making the spectral fit not unique. Nevertheless, the overall range of the derived parameters is consistent with that for the first flare peak, although no direct contribution from the acceleration region is detectable here.

\section{Discussion and Conclusions}

We have demonstrated that the combination of X-ray and microwave imaging, spectroscopy, and timing may allow a firm detection of the solar flare acceleration site even when the thermal SXR emission and microwave contribution from a magnetically trapped electron population are strong. For the 2002 April 11 event discussed here, the direct radio emission from  the \textit{acceleration region} dominates {\gf the radio spectrum} during the early, impulsive phase of the radio emission at 2.6--4.2~GHz.

For a better understanding of the acceleration region and its relationship with other emission components it would be highly desirable to somehow constrain the 3D geometry and magnetic connectivity within the flaring region. The current state-of-the-art of magnetic field reconstruction implies using a nonlinear force-free field (NLFFF) extrapolation of vector magnetic field measurements at the photospheric level to the corona. In our case, however, no vector magnetogram is available, but the line-of-sight data only from SOHO/MDI. This makes the 3D modeling described below non-unique, and is only offered for illustrative purposes.

\begin{figure}\centering
\nonumber
a\centerline{aa\includegraphics[width=0.4\columnwidth]{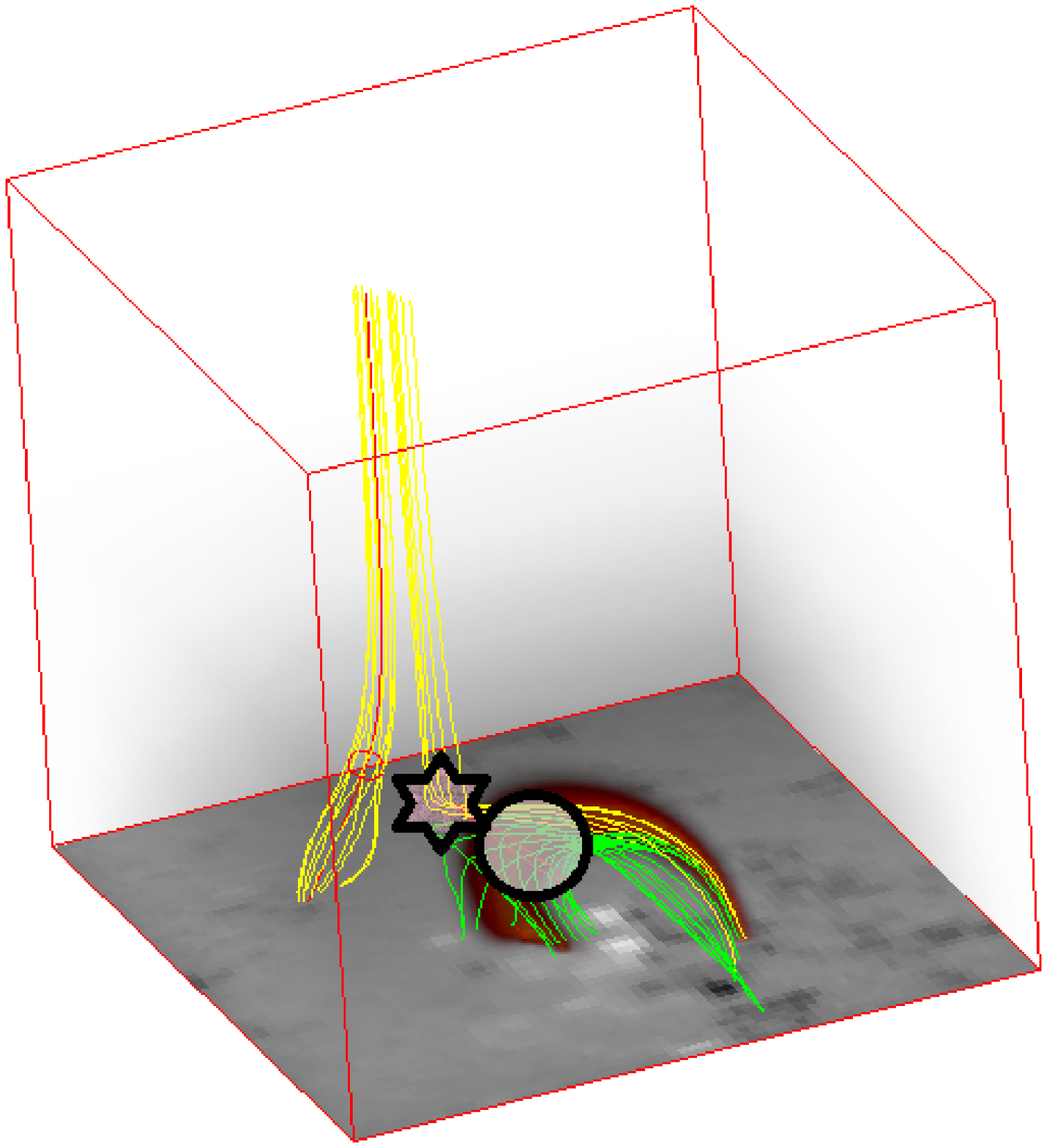} \qquad
d \includegraphics[width=0.4\columnwidth]{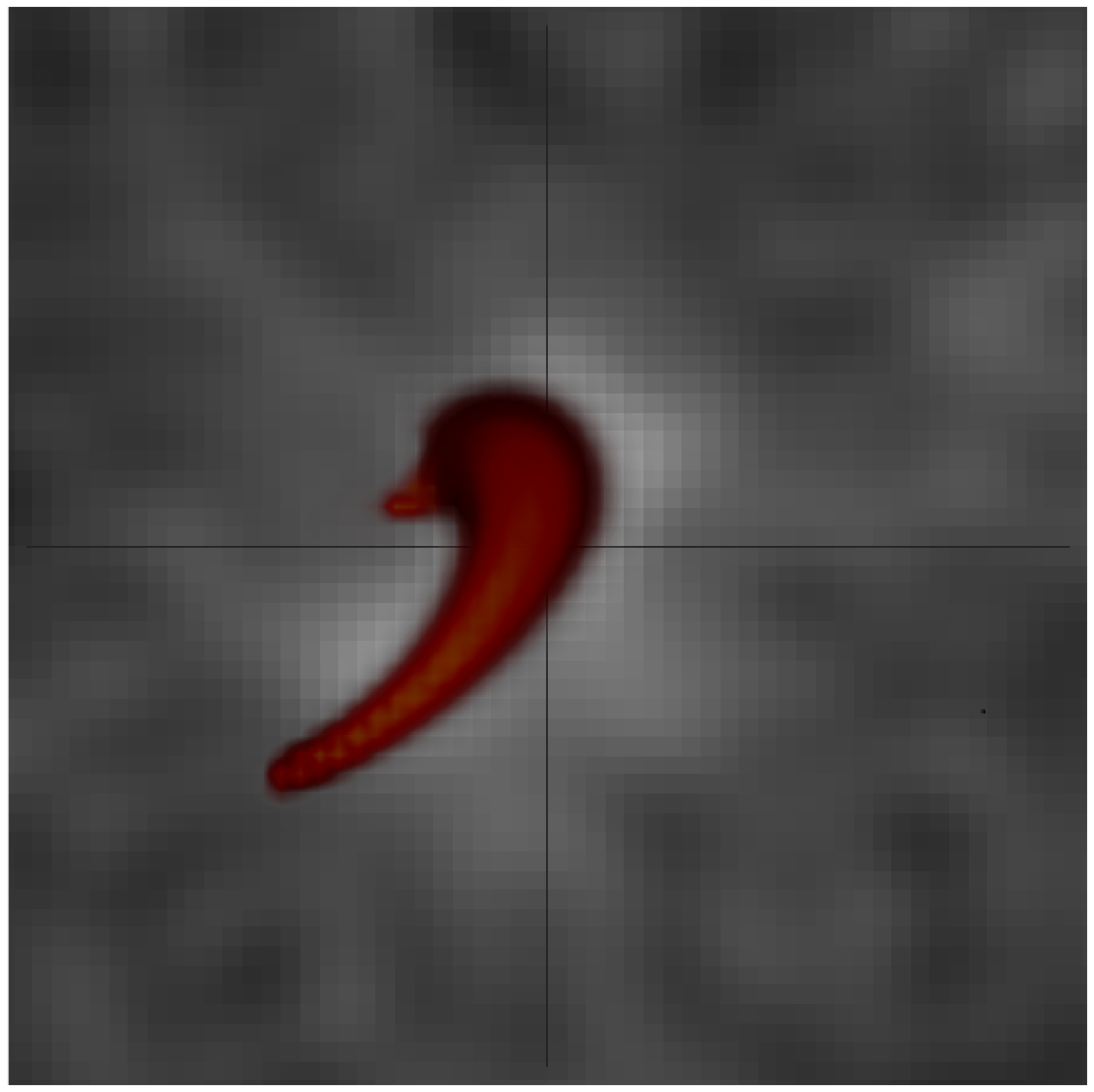}
}\\
\vspace{1cm}
b\centerline{\includegraphics[width=0.40\columnwidth]{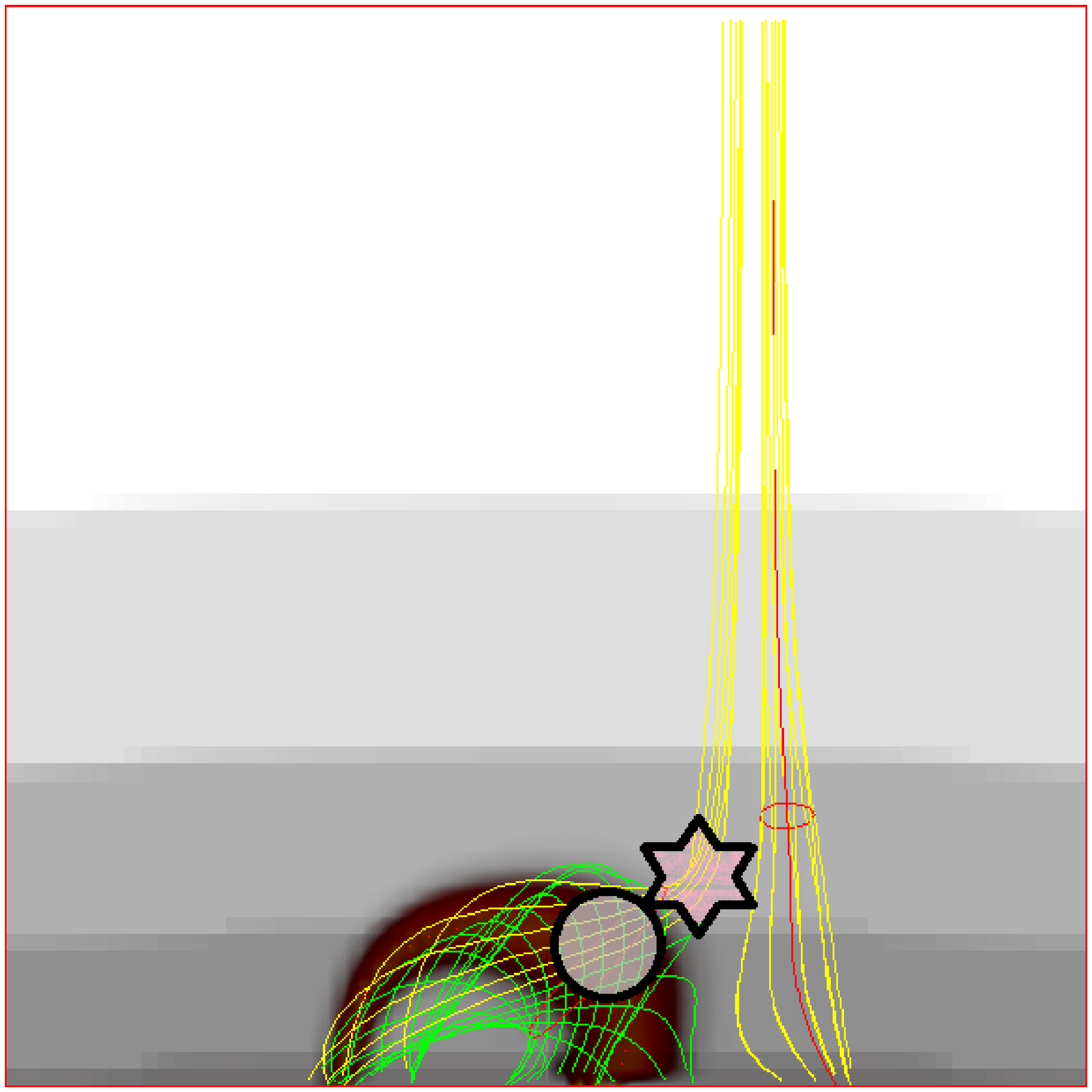}\qquad \ \
e \includegraphics[width=0.40\columnwidth]{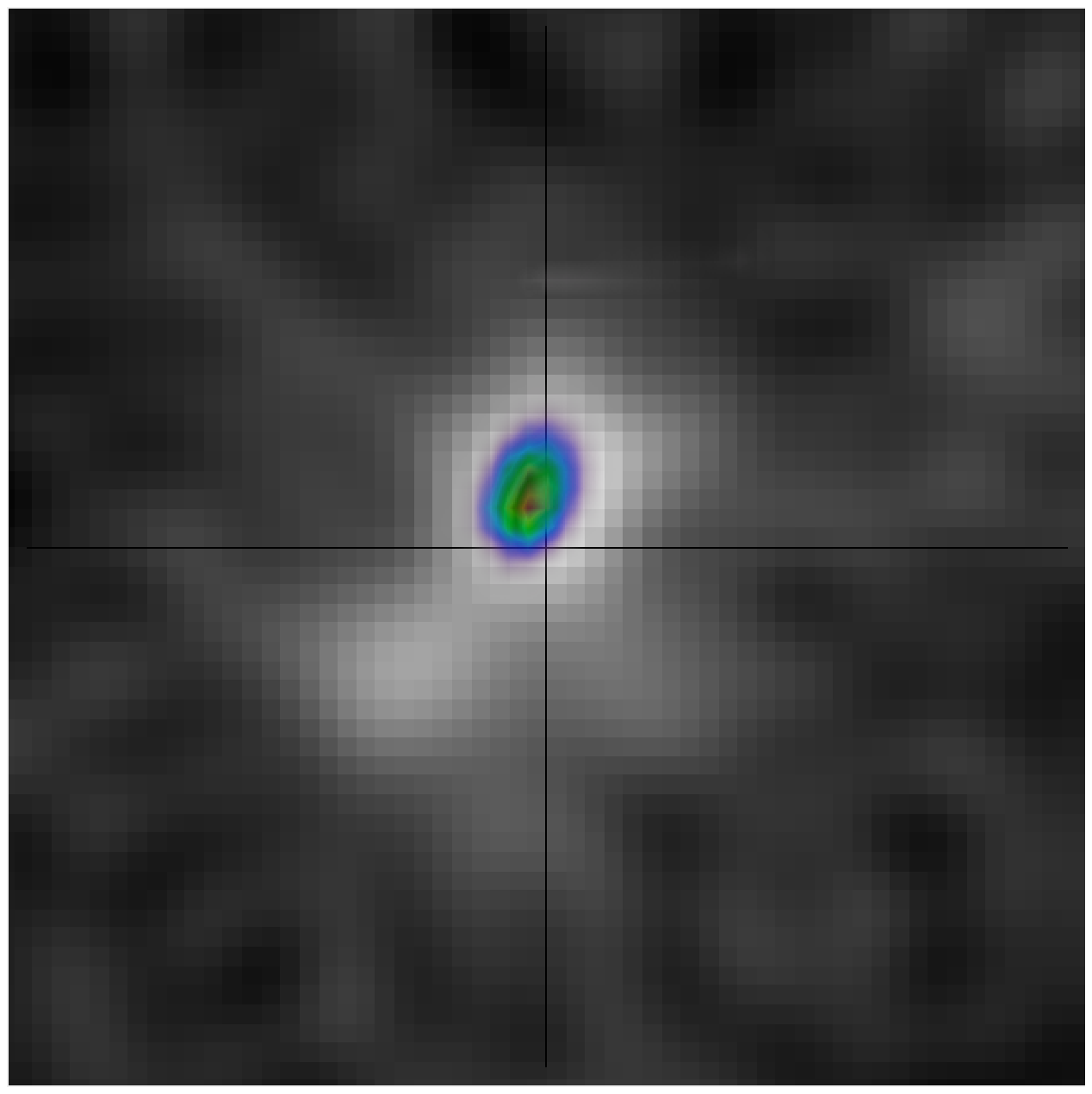}}\\
\vspace{1cm}
c\centerline{\includegraphics[width=0.4\columnwidth]{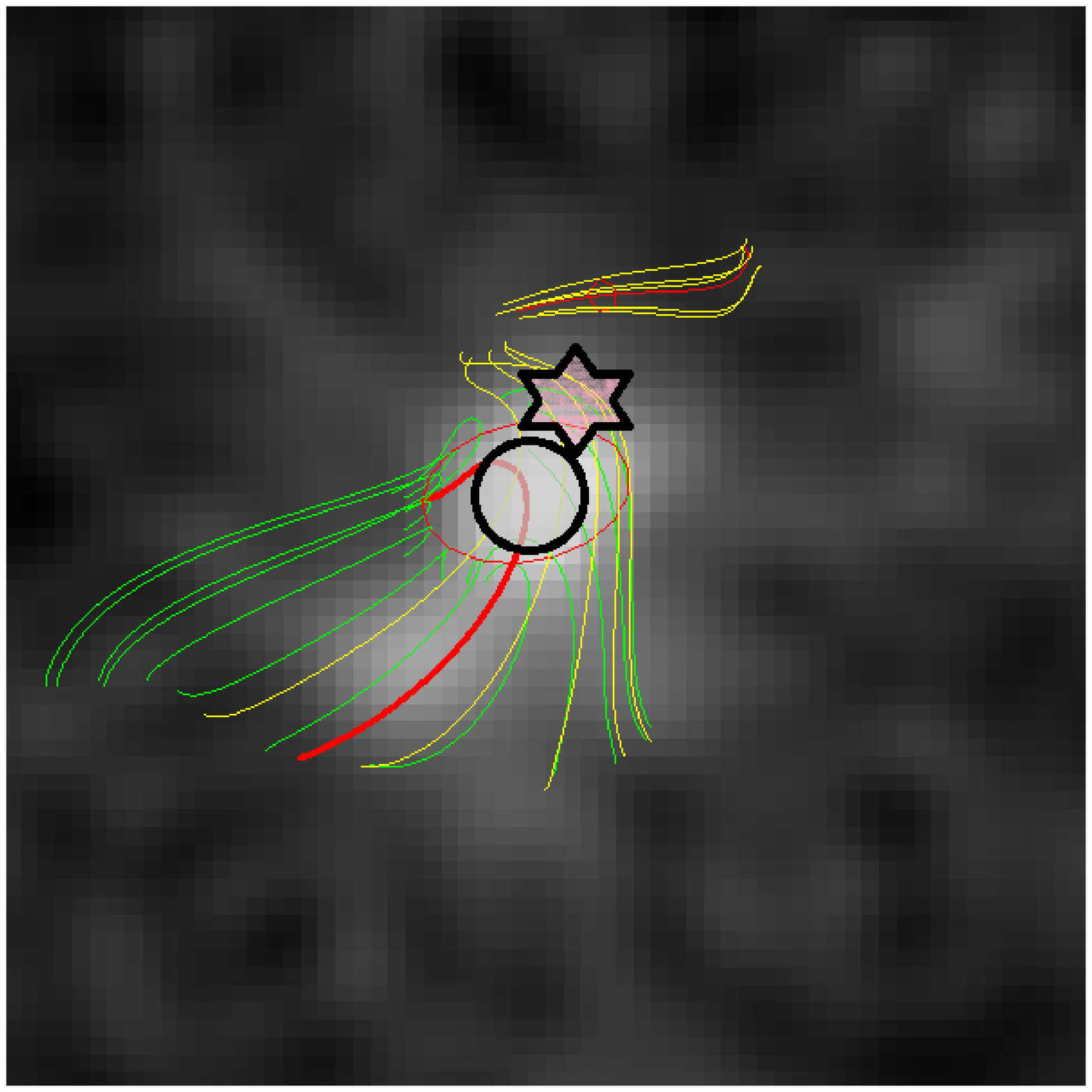}\qquad\ \
f \includegraphics[width=0.4\columnwidth, bb=89 180 519 609,clip]{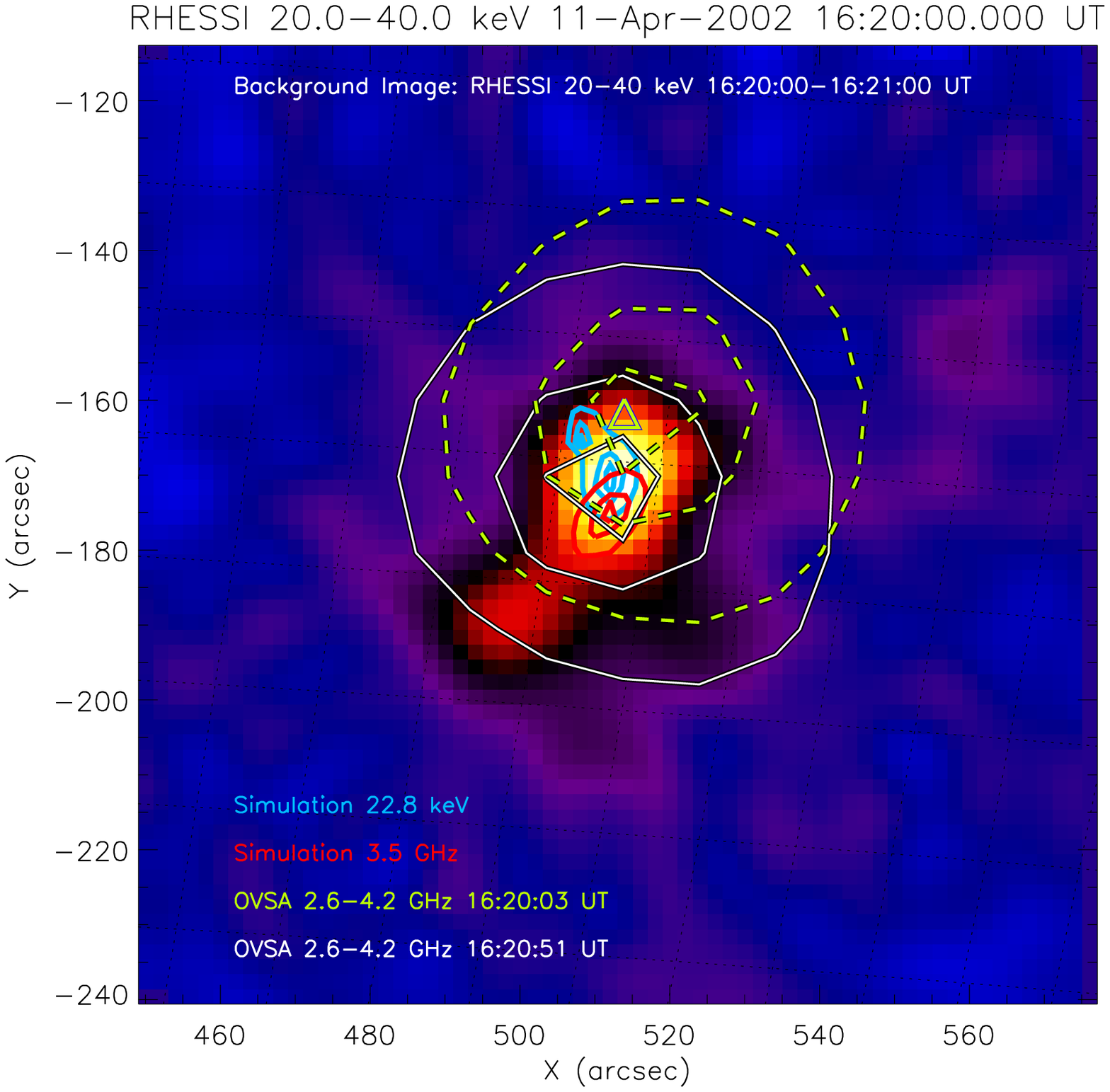}}
\caption{\label{fig_3D_set}
{\gff
3D model of the flaring region based on LFFF ($\alpha\approx -5.5\cdot10^{-10}$~cm$^{-1}$) extrapolation of the photospheric SOHO/MDI magnetogram.
($a$-$c$) Extrapolated 3D magnetic data cube visualized by two magnetic flux tubes (central field lines are red): one consists mainly of closed field lines (green) with a few outer open field lines (yellow), while the other one consists of open field lines only. The locations of the acceleration region (star symbol) and trapping source (circle) are shown. ($a$) perspective view with the line-of-sight magnetogram shown at the bottom boundary; the dark red structure represents the model spatial distribution of the thermal plasma. ($b$) side view of the structure; step-like grey stripes represent the assumed hydrostatic plasma distribution in the corona outside the flaring loop. ($c$) top view of the structure (thermal distribution not shown); the greyscale background shows the HXR image taken at 16:20:00--16:21:00~UT.
($d$) top view of the thermal plasma distribution, ($e$) nonthermal electron distribution, and ($f$) a combination of simulated and observed images. The same HXR image is used as a background in all these panels. The contours at the right bottom panel are for 30\%, 70\%, and 90\% of the corresponding peak value. Centroid of the acceleration region (the OVSA image at 16:20:03~UT), which defines the location of the star symbols in ($a$-$c$), is separately shown by the green triangle.
}
}
\end{figure}

To build a 3D model we use our open\footnote{\url{www.lmsal.com/solarsoft/ssw\_packages\_info.html}} modeling tool, %
\verb"GX_Simulator",
and start from the line-of-sight SOHO/MDI photospheric magnetic measurements at the region of interest. We then have the choice to apply either potential or linear (constant-$\alpha$) force-free field (LFFF) extrapolations to build a 3D magnetic data cube with which we can plot the field lines and form flux tube around some of them. Our tests with different values of force-free parameter $\alpha$ show that it is not possible to build a magnetic structure consistent with locations of the X-ray and radio sources for either the potential field ($\alpha=0$) or for most LFFF extrapolations. However, we succeeded to find the required connectivity for $\alpha\approx -5.5\cdot10^{-10}$~cm$^{-1}$. We use the corresponding magnetic data cube for our 3D illustrative model, summarized in Figure~\ref{fig_3D_set}. The left column of panels shows, respectively, the perspective, side, and top views of the magnetic structure in the flaring region. The locations of the acceleration region (star symbol) and trapping source (circle) are shown in relation to this structure.

It is interesting that the trapping source matches perfectly the top of the flaring loop as expected, while the acceleration region source is displaced compared  with the loop top toward a region where many of the magnetic field lines are open. Moreover, inspection of neighboring field lines implies that formation of a cusp nearby to the acceleration region is likely. Speculating further, one might conclude that the particle acceleration site coincides with or is adjacent to the magnetic reconnection site.
The presence of open magnetic field lines 
is also supported by lower-frequency radio observations including WIND detection of type III bursts indicating escaping energetic electrons.

To complete the 3D modeling we populated the flaring loop (i.e., the region of the closed field lines shown by green with the central field line shown by red) with appropriate nonuniform spatial distributions of the thermal plasma (red in panels $a$ and $d$) and nonthermal electrons (rainbow colors in panel $e$).  We then use this 3D model to compute the corresponding radio and X-ray images by solving the corresponding radiation transfer equations numerically. Results of this modeling in the form of simulated images at 22.8~keV and 3.5~GHz are overplotted on top of the observed radio and X-ray images. One can clearly see that the locations of the simulated images are consistent with each other and with observed locations of the HXR source and late-stage (trapped) radio source, while the simulated radio source from the trapped component is clearly displaced compared with radio location of the acceleration region.

The global parameters of the acceleration region are $B\sim120$~G and $V\sim6\cdot10^{27}$~cm$^{-3}$, which are, respectively, two times and ten times larger than reported in the earlier cold, tenuous (2002 July 30) flare \citep{Fl_etal_2011}. The residence time of the fast electrons at the acceleration region is $\sim3$~s, which is comparable to that in the cold, tenuous flare, and is much longer than the {\dg free-streaming time} through the acceleration region.  As noted in \citet{Fl_etal_2011}, this favors {\gf diffusive electron transport due to their scattering by turbulent waves and, thus,} a stochastic 
acceleration mechanism. {\gf The available data is, unfortunately, insufficient to firmly specify the version of stochastic acceleration mechanism \citep[see e.g.][for a recent review]{Petrosian_2012} operating in the event; however, it does favor those models predicting a roughly energy-independent diffusion time at the source, like in the cold flare event \citep{Fl_etal_2011}}

Properties of the accelerated electron components are somewhat different from those in the 2002 July 30 {\dg cold} flare. Firstly, in the 2002 April 11 event the accelerated electron spectrum is noticeably softer ($\delta\approx 5$) than in the cold flare ($\delta\approx 3.5$). Secondly, in the cold flare the accelerated electrons are detected at the energies above 6~keV, while in the April 11 event they are only seen above $\sim20$~keV; lower-energy X-ray emission is dominated by the thermal background. Thirdly, the acceleration efficiency is different: in the cold flare almost all available electrons were accelerated, while in the April 11 event even the peak instantaneous number density of the fast electrons ($n_r\sim2\cdot10^8$~cm$^{-3}$) does not exceed 10\% of the thermal electron density. Fourthly, in the April 11 event we clearly see a spectral evolution indicative of the growth of a power-law tail ($E_{\max}$ increases with time at the acceleration stage), whereas no spectral evolution was detected in the cold flare event, which implies a nearly instantaneous growth of the power-law tail. And finally, the released flare energy divides between the thermal and nonthermal components in remarkably different proportions in these two events.

Let us discuss from whence all these differences could originate. We have already concluded that the bulk acceleration mechanism is likely to be a stochastic/Fermi process with a relatively long residence time of the electrons controlled by their spatial diffusion on the turbulent magnetic irregularities at the acceleration region. For a diffusive Fermi acceleration process {\dg the shape of the particle energy spectrum} depends primarily on the ratio of two key parameters---the acceleration rate $\tau_a$ {\gfR (this is the time needed to establish the nonthermal particle spectrum, not to be interpreted as a duration of the acceleration process)} and the residence/diffusion time of the electrons $\tau_d$ at the acceleration region in such a way that the larger the $\tau_a/\tau_d$ ratio the steeper (softer) the accelerated electron spectrum, \citep[see, e.g.,][]{Hamilton_Petrosian_1992}. The residence times, $\tau_d \sim 3$~s, are comparable in the two events under comparison; the acceleration   {\gfR rates} are, however, different. Indeed, the acceleration time $\tau_a$ can be roughly estimated as the time needed for the power-law tail to grow, which is clearly shorter than the residence time, $\tau_a < 3$~s, in the cold flare (recall, no spectral evolution  {\gfR was} noted), while longer, $\tau_a > 3$~s, in the April 11 event ($E_{\max}$ increases with time). Thus, for other conditions being equal, the accelerated electron energy spectral index must be larger in the April 11 event in agreement with observations.

The acceleration efficiency and energy balance in the flare depend, in addition to the acceleration mechanism itself, on the process of electron extraction from the thermal pool and their injection into the main acceleration process. 
In the cold flare almost all available thermal electrons were injected and accelerated, although their consequent energy losses were insufficient to significantly heat the thermal plasma. In contrast, in the April 11 event, only a relatively minor fraction of the thermal electrons were accelerated, making the collisional heating of the thermal plasma even less efficient than in the cold flare case (given that other relevant physical parameters are similar in these two cases). Thus, the presence of a very hot flaring {\dg SXR} plasma with $T\sim20$~MK (which is present even before the flare impulsive phase) requires another heating mechanism distinct from the collisional plasma heating by accelerated electrons. This conclusion is further supported by the spatial displacement between the thermal SXR source and nonthermal coronal HXR and microwave sources.

Although the available data are insufficient to firmly identify the flare energization process in either event, or the mechanism of energy division between the thermal and nonthermal components, we can conclude that this process does show some resemblance to that controlling the balance between Joule heating and runaway electrons in a DC electric field. Indeed, suppose that there is a relatively weak sub-Dreicer electric field directed along the flaring loop magnetic field. This electric field will initiate an electric current, which will lose its energy {\dg through} Joule heating, while the fraction of the runaway electrons available for further stochastic acceleration will be relatively minor. In the case of a stronger electric field, e.g., comparable to the Dreicer field, the fraction of the runaway electrons becomes large, while the Joule heating is reduced so the plasma heating is modest. Even though it is a long way from these speculations to even a qualitative model, the analysis performed favors a flare picture in which electrons are first extracted from the thermal pool by a DC electric field (of yet unspecified origin) and then stochastically accelerated to form a power-law-like energy distribution. Therefore, a stochastic acceleration mechanism naturally containing a DC electric field is called for.

We have shown that observations of radio emission directly from the acceleration site provide important constraints on the acceleration mechanism in solar flares.  Despite the great differences between this flare and the cold flare, a similar acceleration mechanism, although operating in a somewhat different parameter regime, seems to be called for.  Future radio {\dg spectral imaging} observations that can better separate the acceleration site from the sites of trapping and precipitation are needed to investigate the flare acceleration mechanism(s) in more detail.

\acknowledgments
This work was supported in part by NSF grants
AGS-0961867, AST-0908344,  AGS-1250374 and NASA grants NNX10AF27G and NNX11AB49G to New Jersey
Institute of Technology and by the RFBR  grants 12-02-00173 and 12-02-00616.
This work was supported by a UK STFC rolling grant and the Leverhulme Trust, UK. This work also benefited from workshop support from the International Space Science Institute (ISSI).
Financial support by the European Commission through HESPE (FP7-SPACE-2010-263086)
(EPK) and the "Radiosun" (PEOPLE-2011-IRSES-295272) Networks
 is gratefully acknowledged.

\bibliographystyle{apj}
\bibliography{xray_refs,fleishman,ms_bib}



\begin{figure}\centering
\includegraphics[width=5.3cm,angle=90]{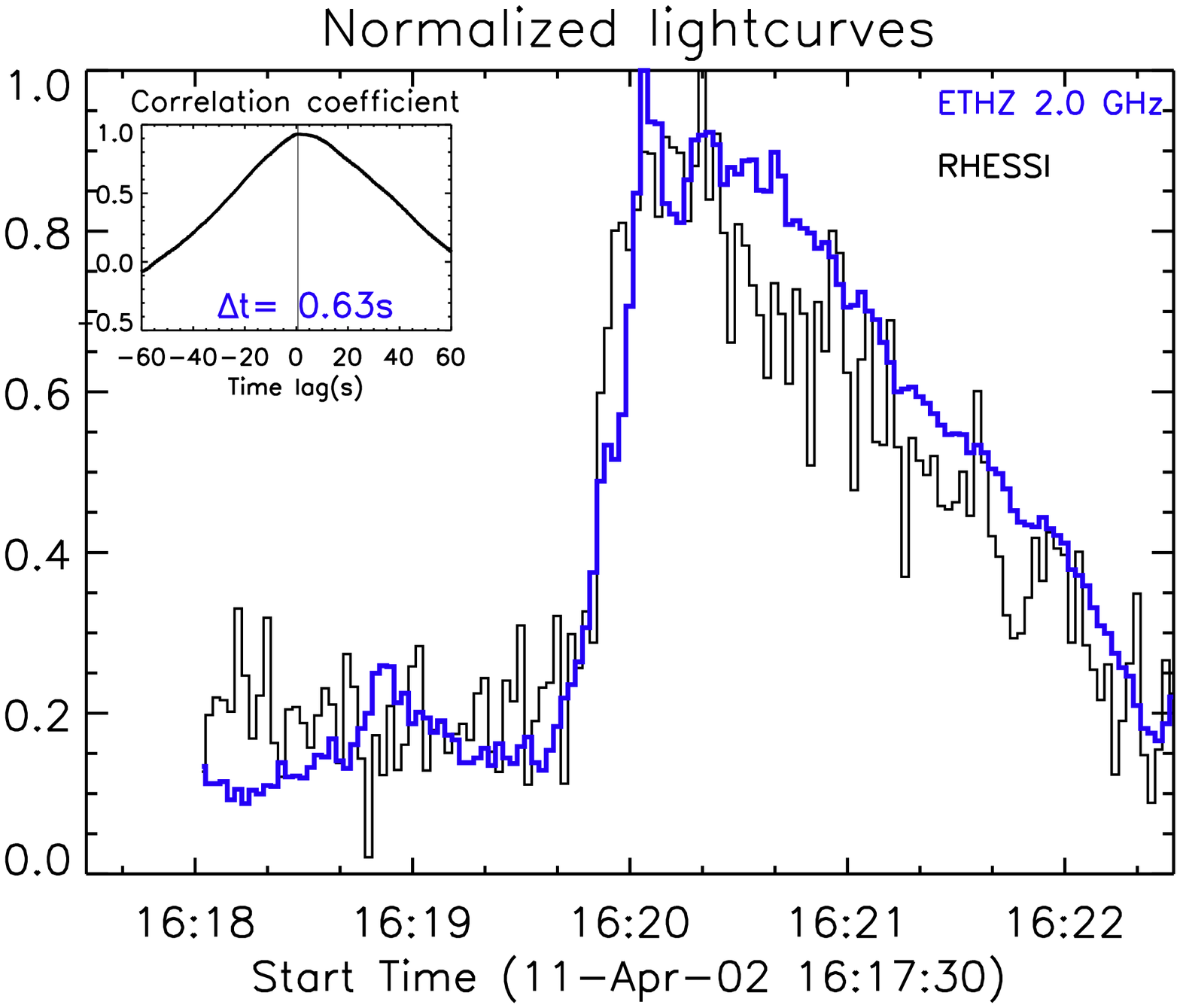}
\includegraphics[width=5.3cm,angle=90]{rhessi-radio_21.eps}
\includegraphics[width=5cm,angle=90]{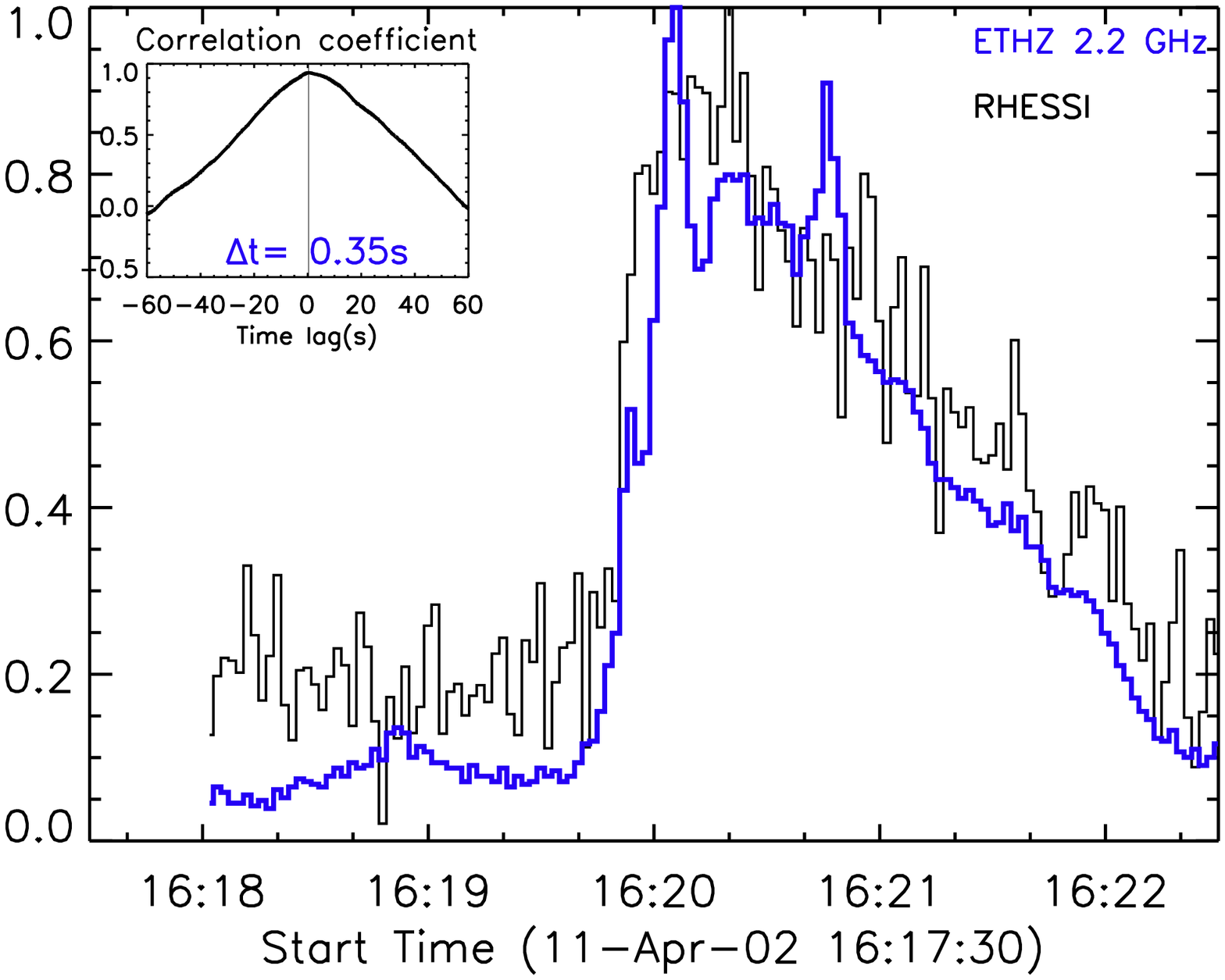}
\includegraphics[width=5cm,angle=90]{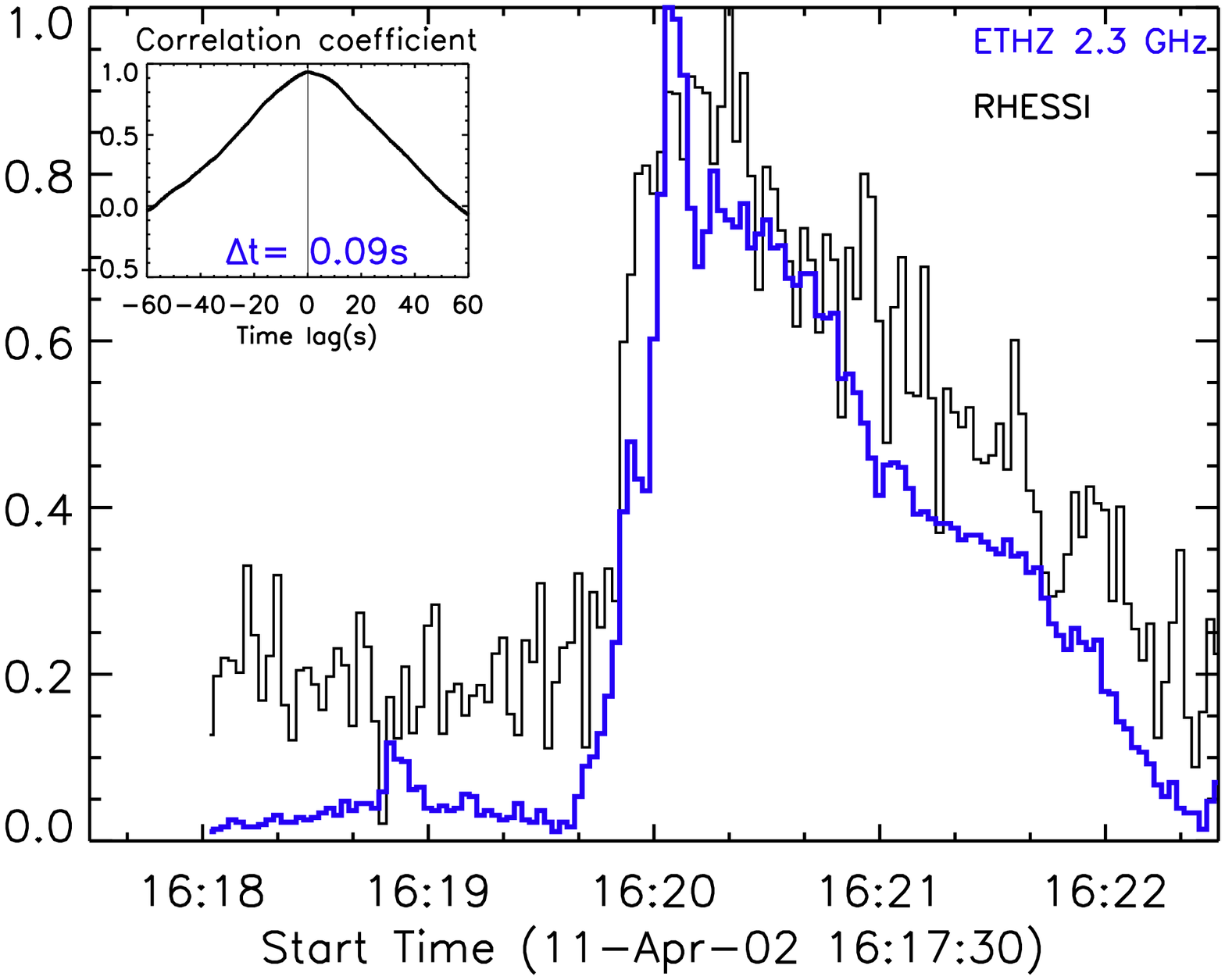}
\includegraphics[width=5cm,angle=90]{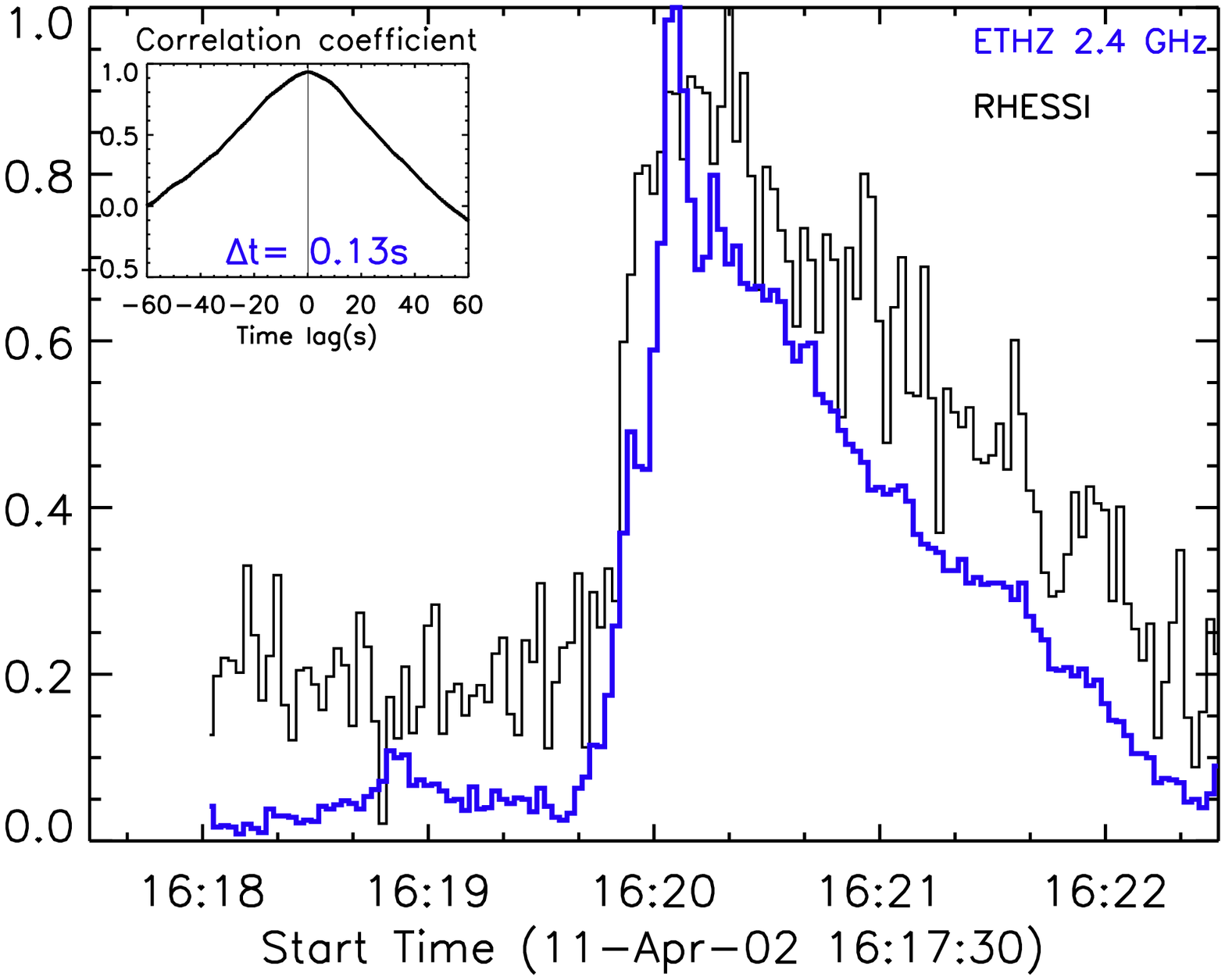}
\includegraphics[width=5cm,angle=90]{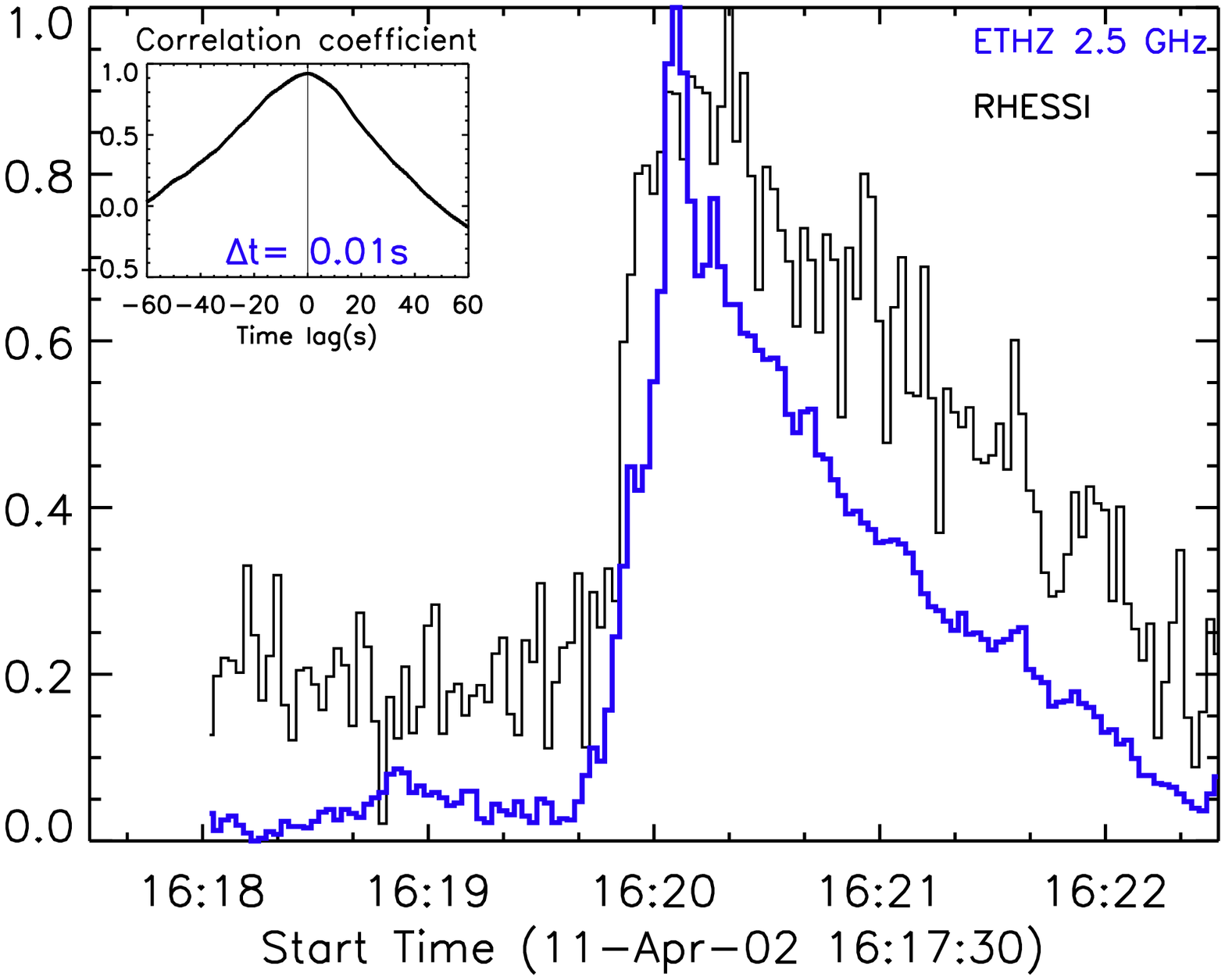}
\includegraphics[width=5cm,angle=90]{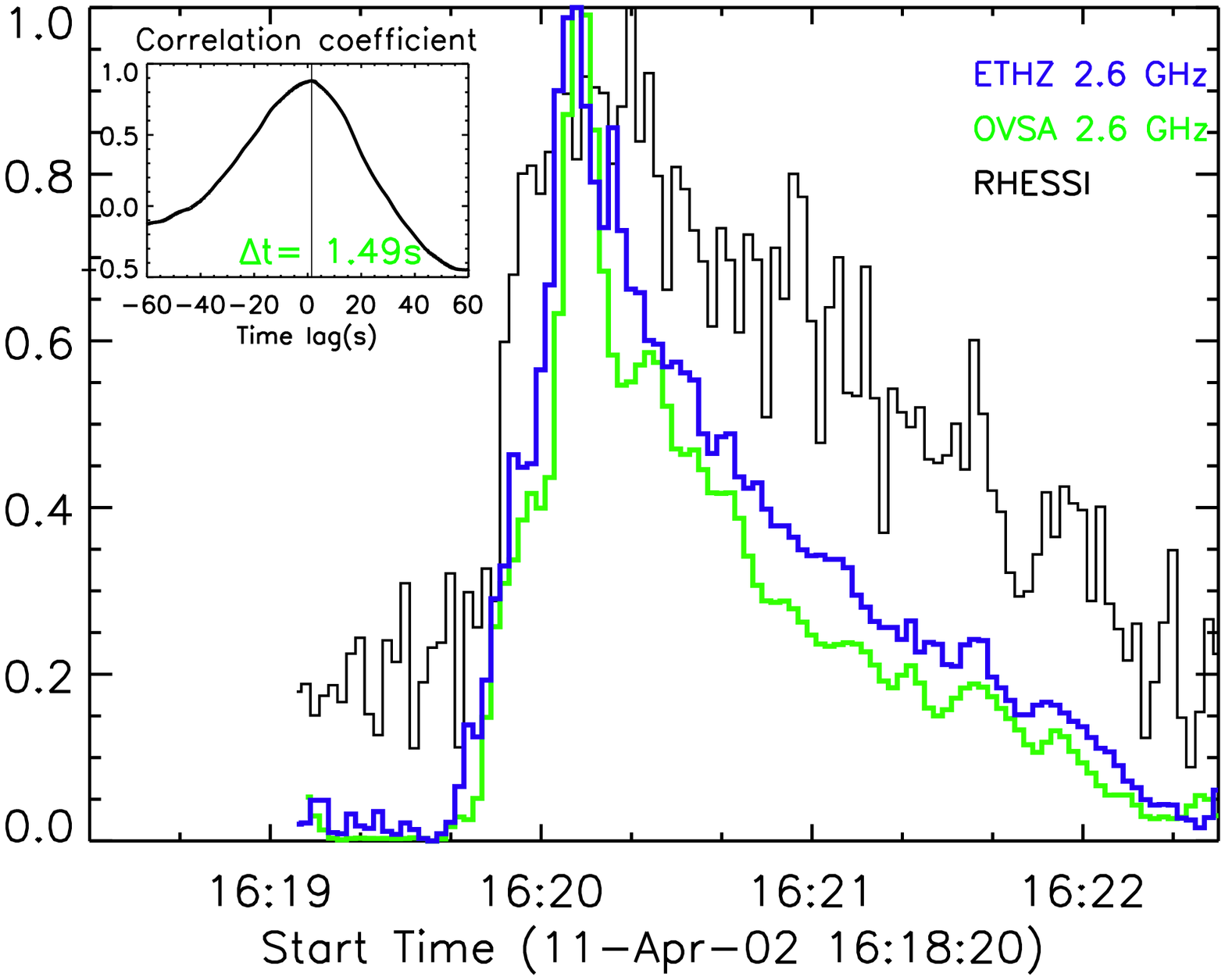}
\includegraphics[width=5cm,angle=90]{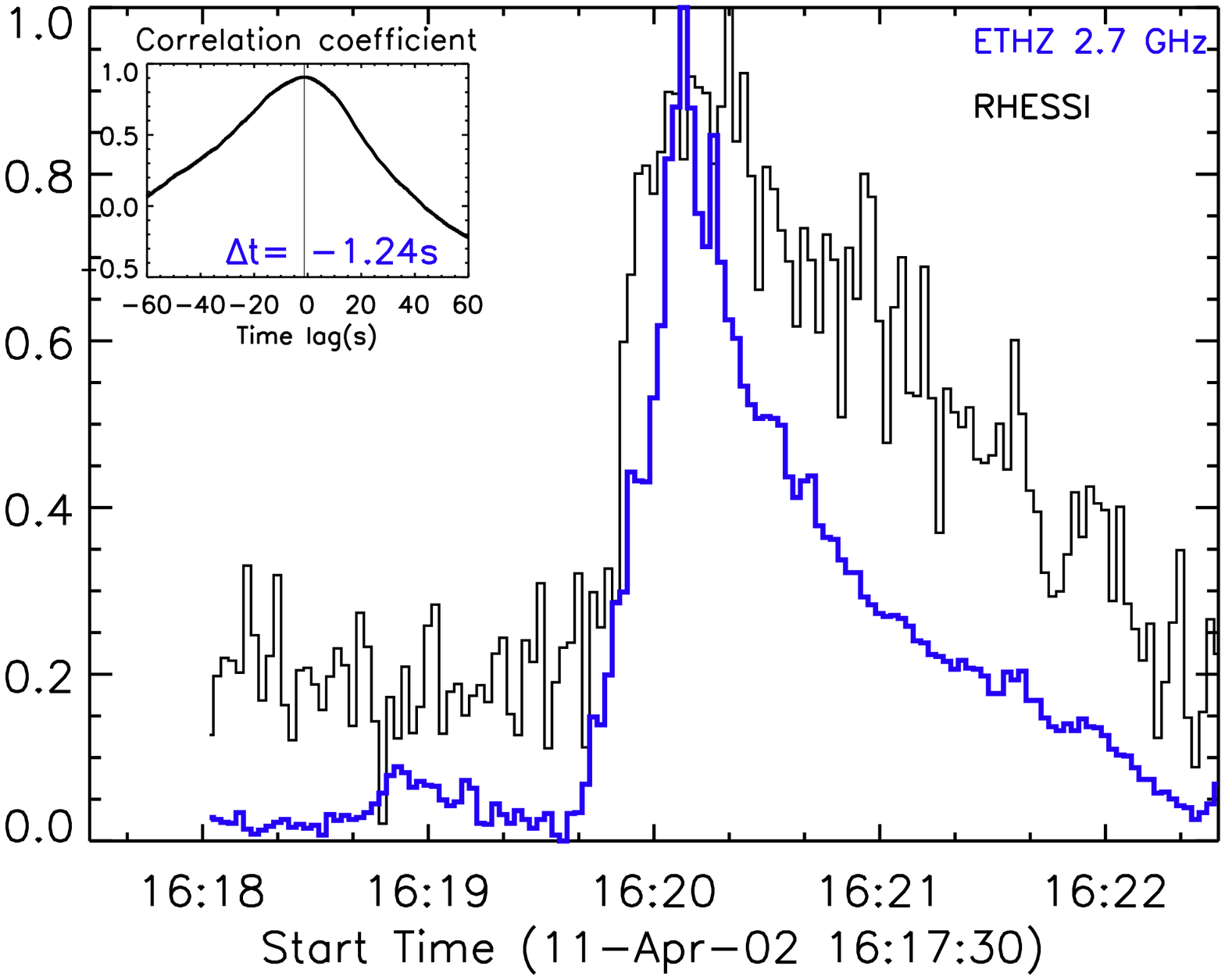}
\caption{\label{fig:Rtiming_1} Radio to HXR timing; continued.}
\end{figure}

\begin{figure}\centering
\includegraphics[width=5.3cm,angle=90]{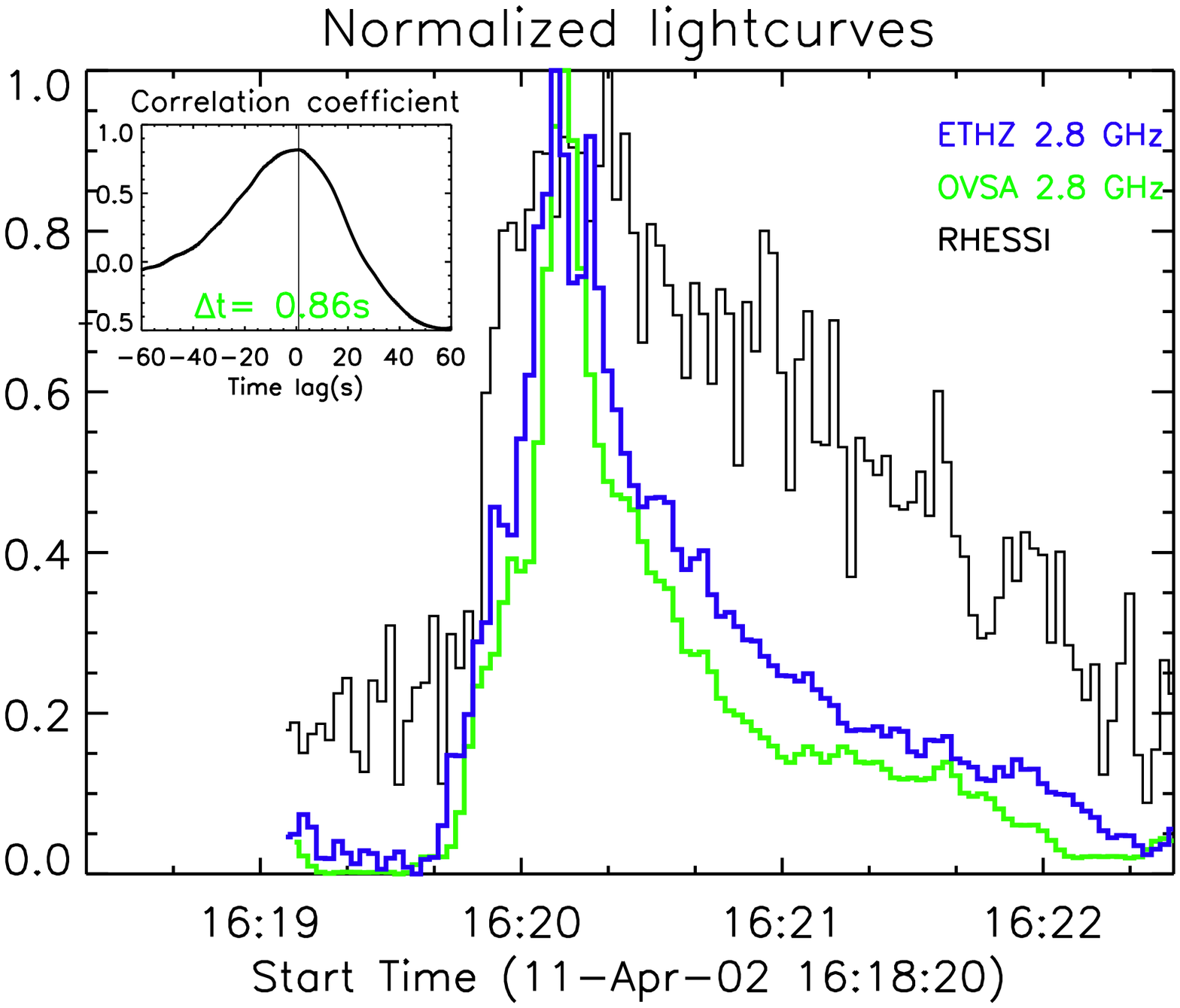}
\includegraphics[width=5.3cm,angle=90]{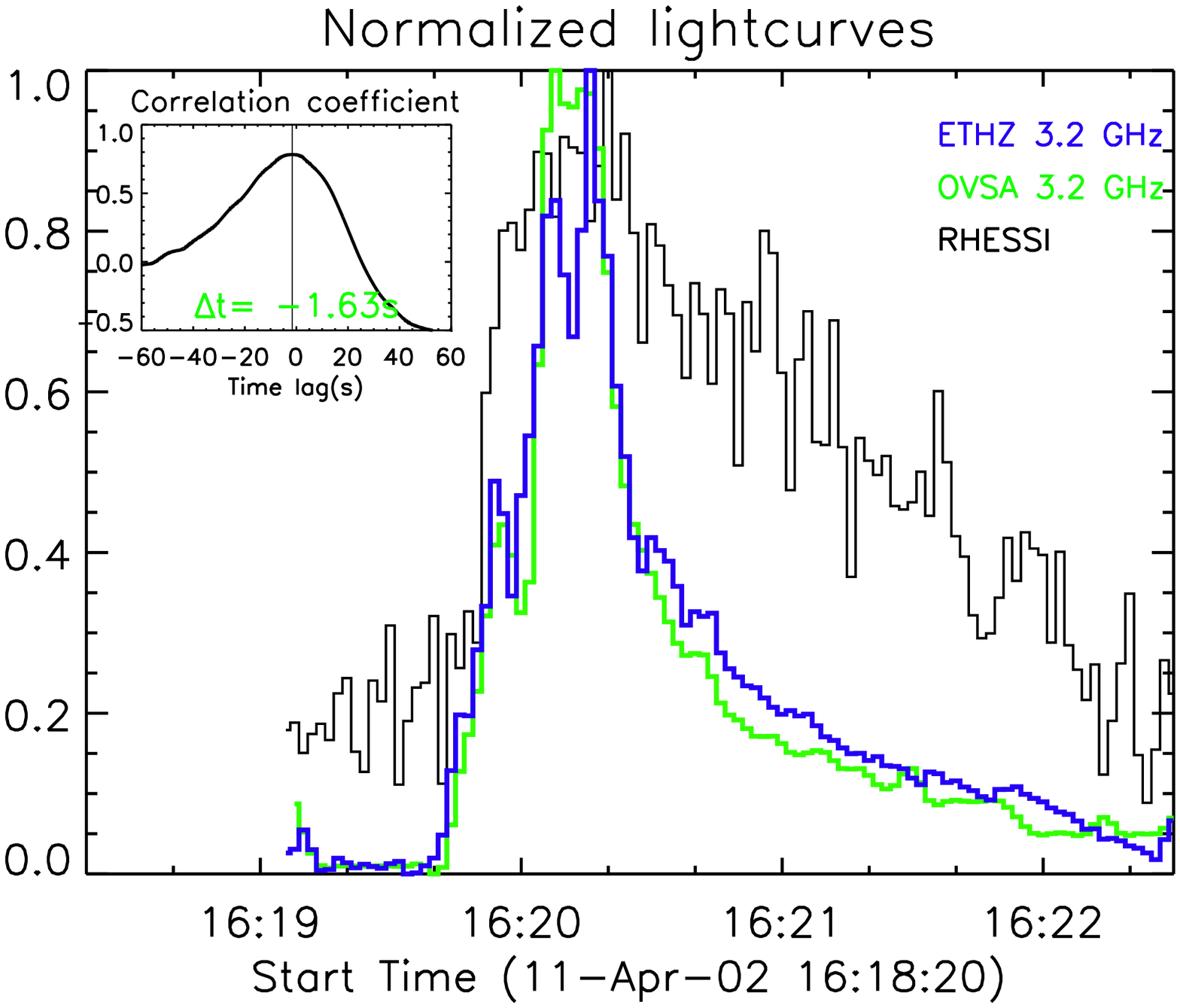}
\includegraphics[width=5cm,angle=90]{rhessi-radio_34.eps}
\includegraphics[width=5cm,angle=90]{rhessi-radio_36.eps}
\includegraphics[width=5cm,angle=90]{rhessi-radio_38.eps}
\includegraphics[width=5cm,angle=90]{rhessi-radio_42.eps}
\includegraphics[width=5cm,angle=90]{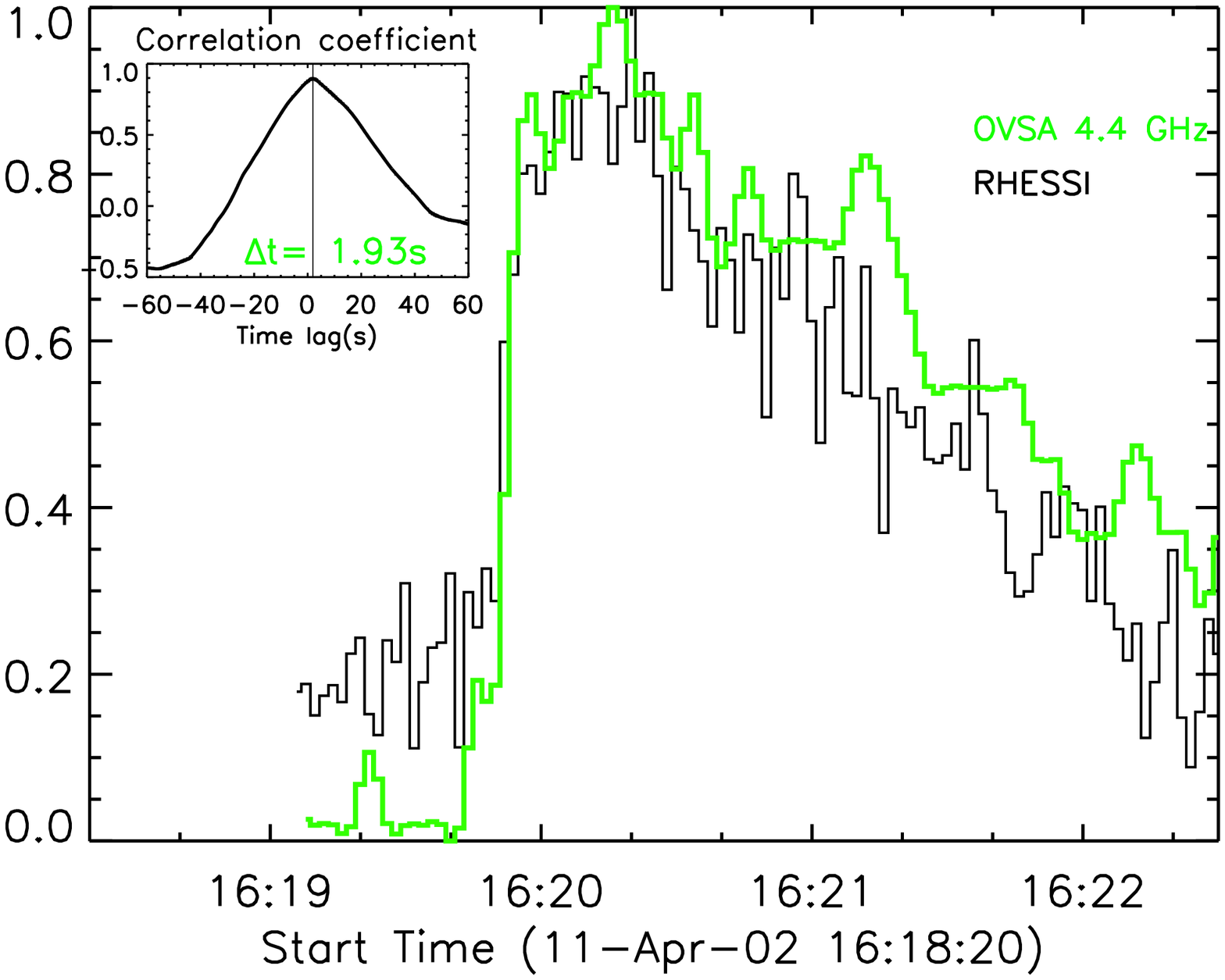}
\includegraphics[width=5cm,angle=90]{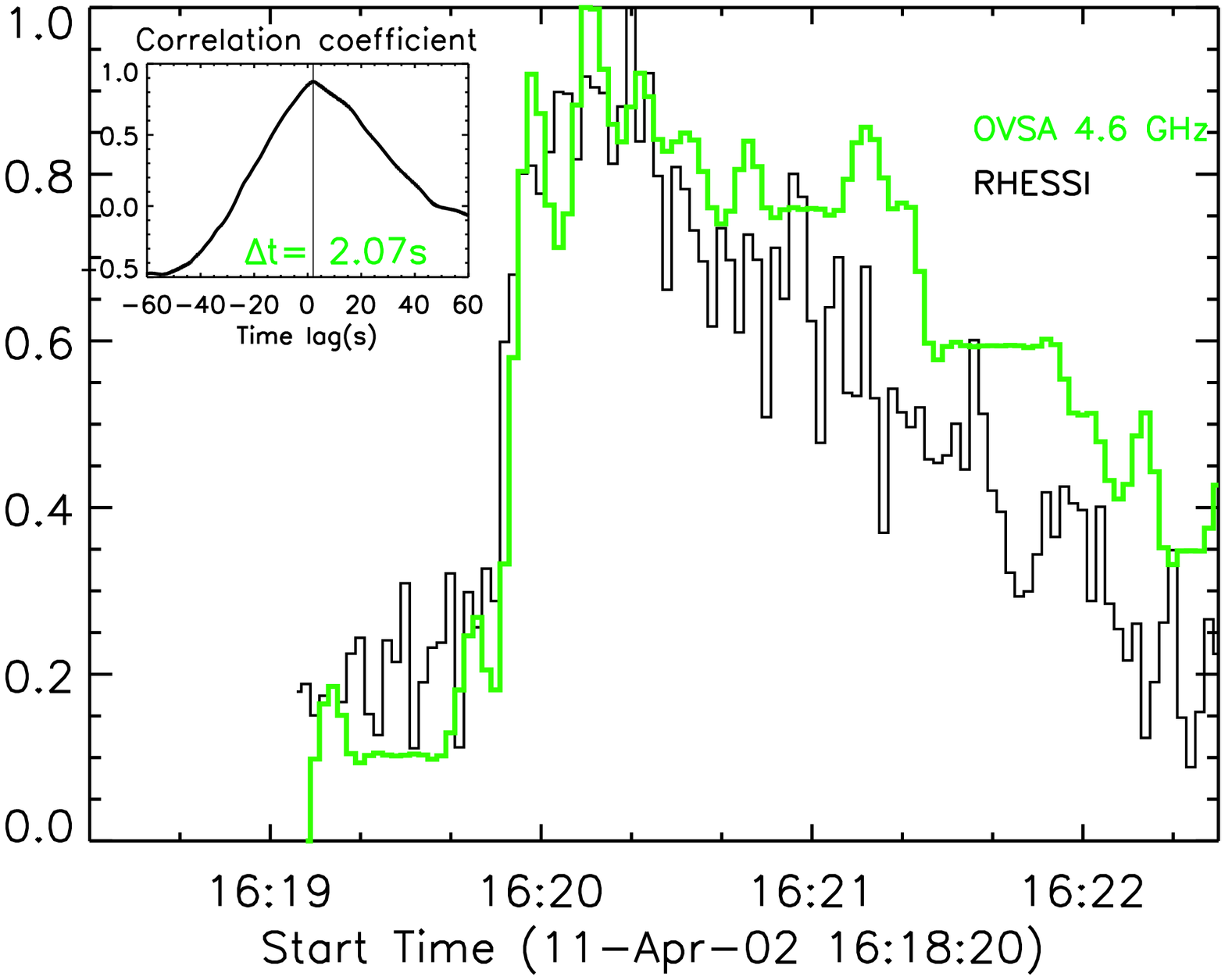}
\caption{\label{fig:Rtiming_2} Radio to HXR timing; continued.}
\end{figure}

\begin{figure}\centering
\includegraphics[width=5.3cm,angle=90]{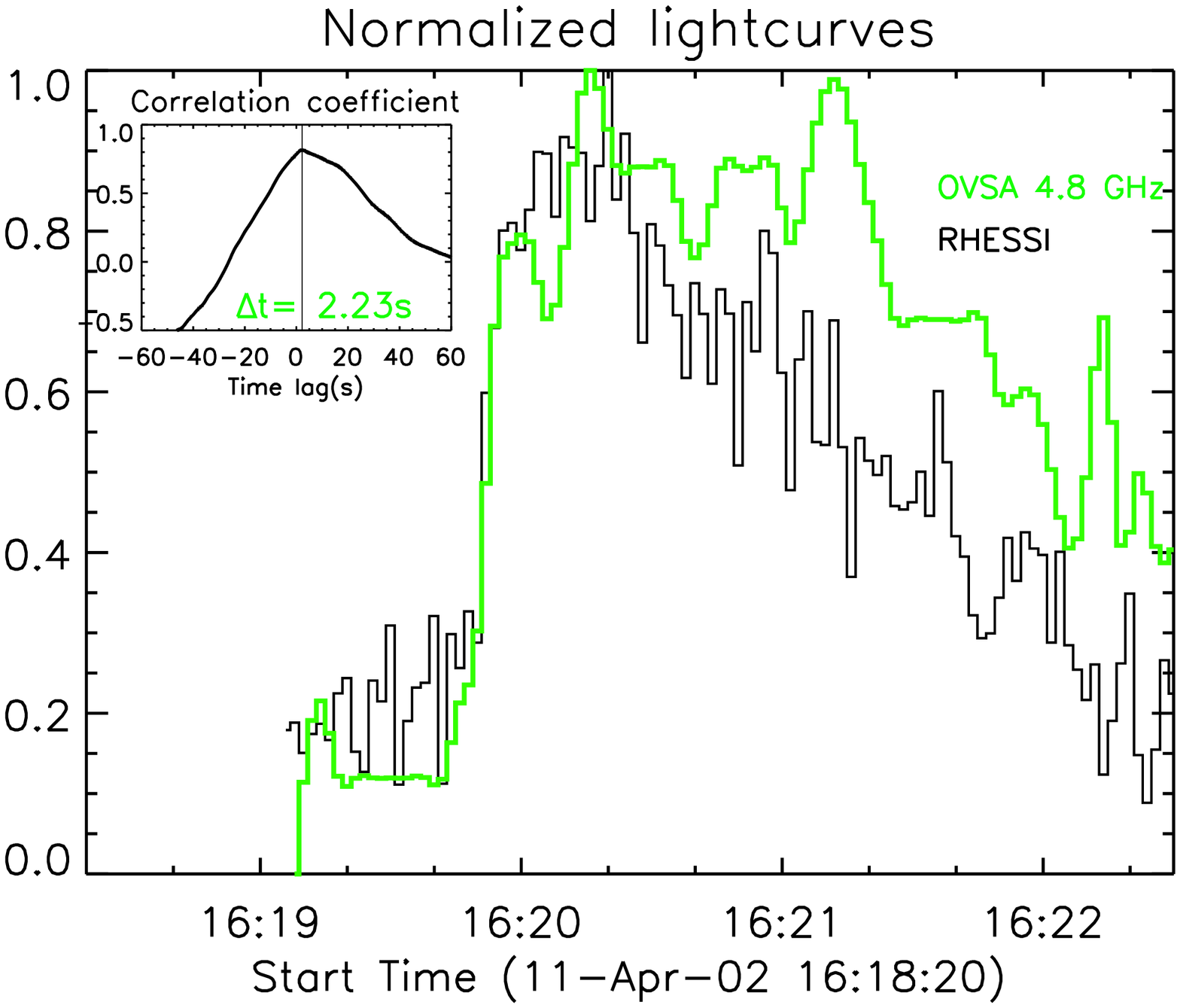}
\includegraphics[width=5.3cm,angle=90]{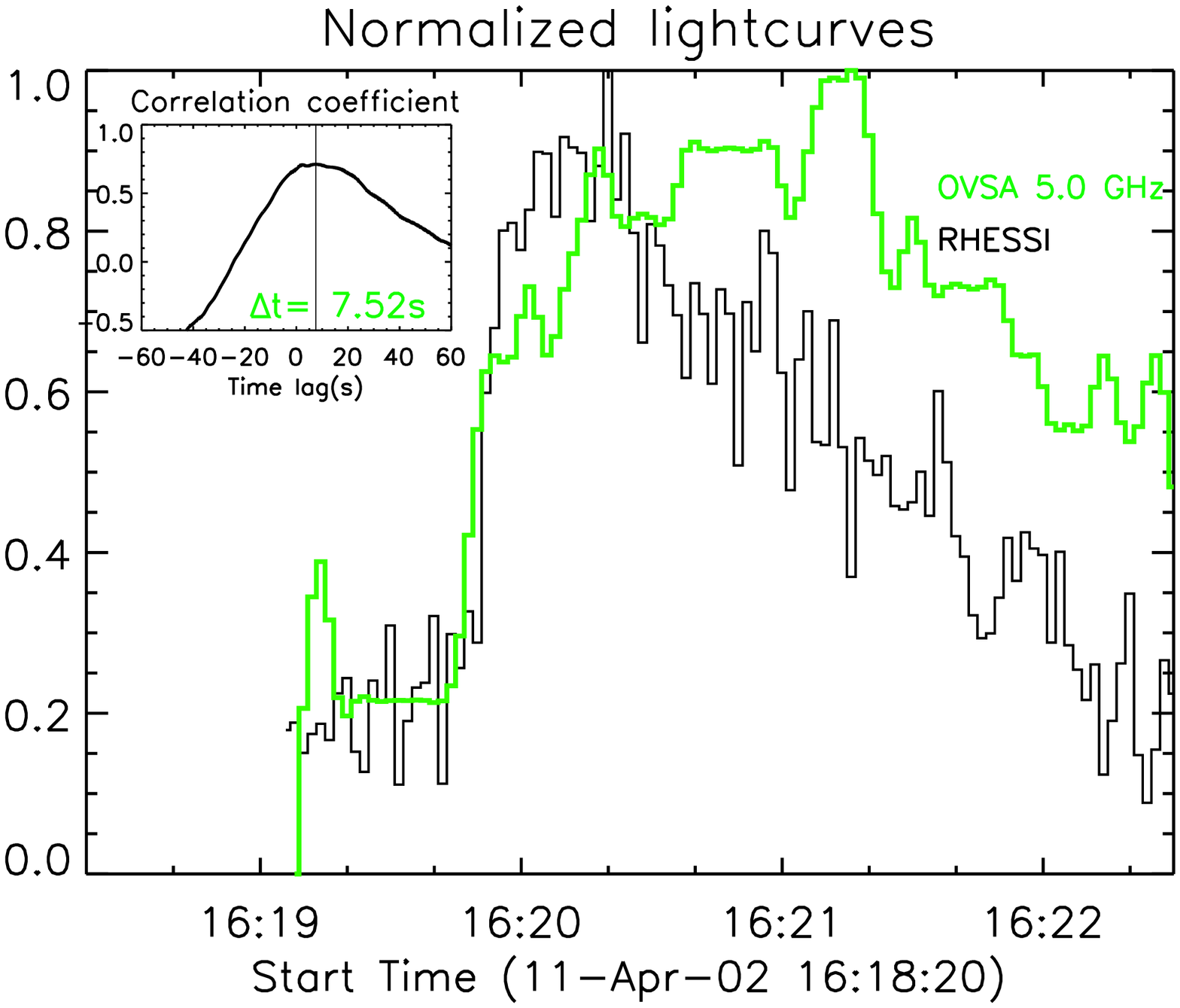}
\includegraphics[width=5cm,angle=90]{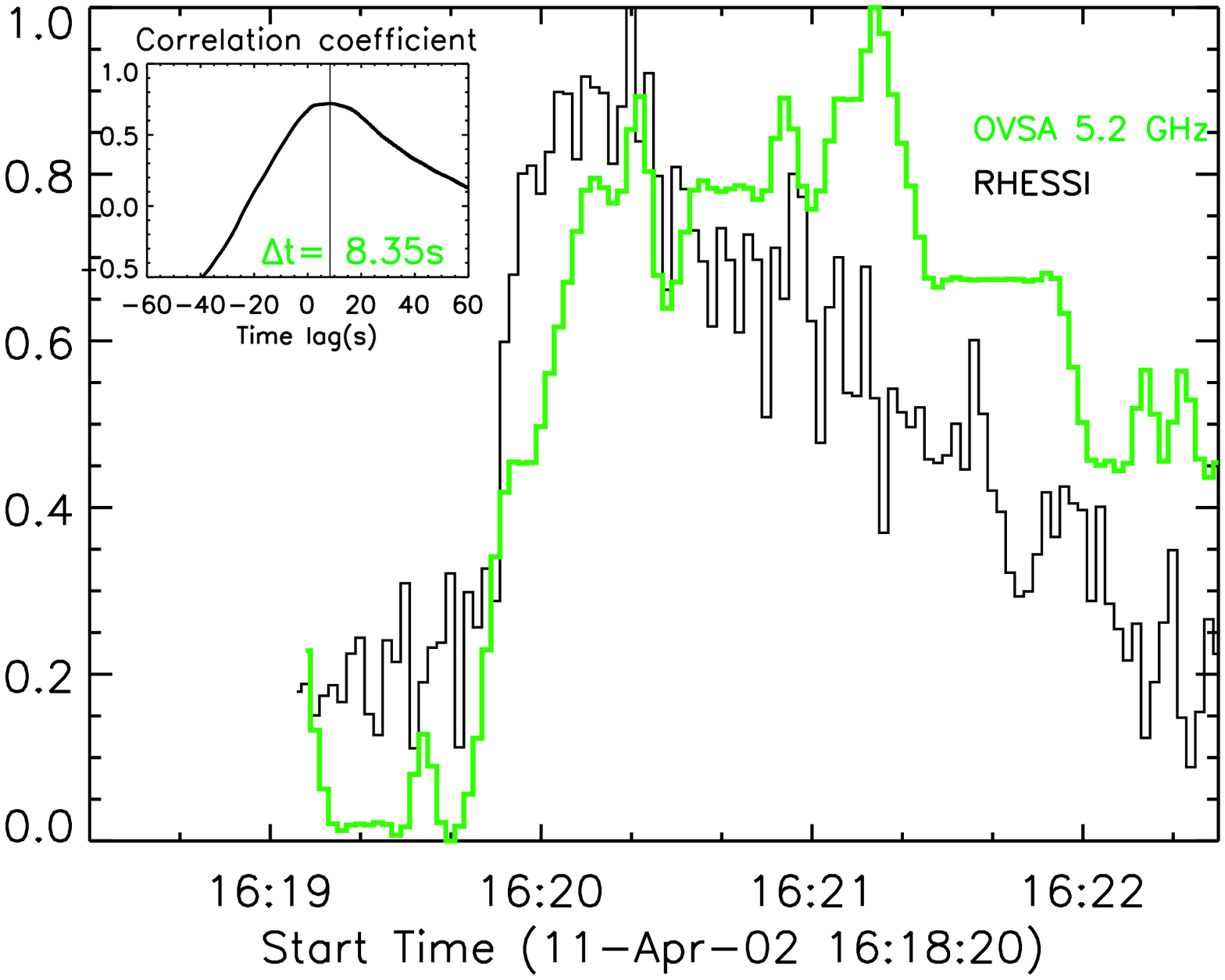}
\includegraphics[width=5cm,angle=90]{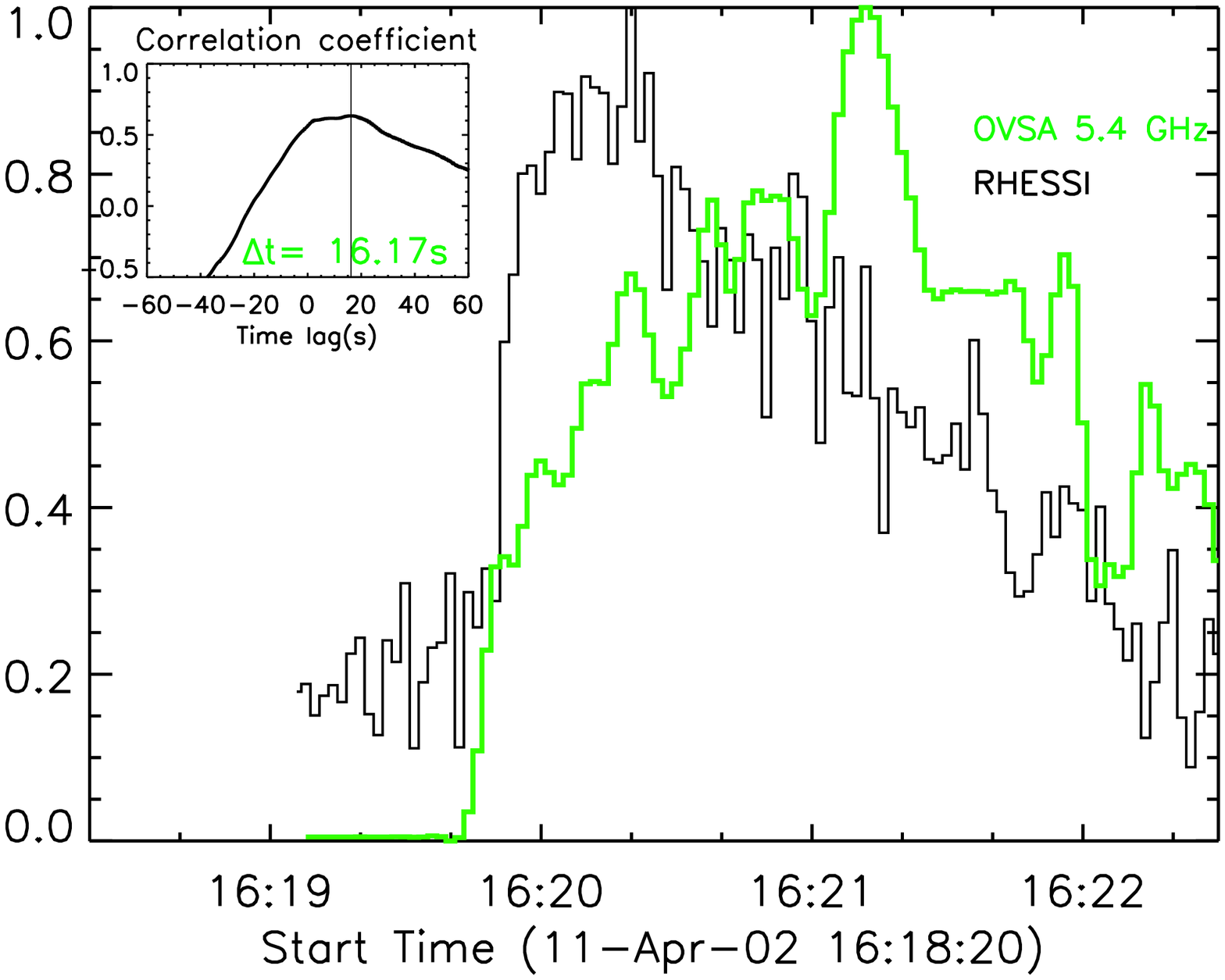}
\includegraphics[width=5cm,angle=90]{rhessi-radio_56.eps}
\includegraphics[width=5cm,angle=90]{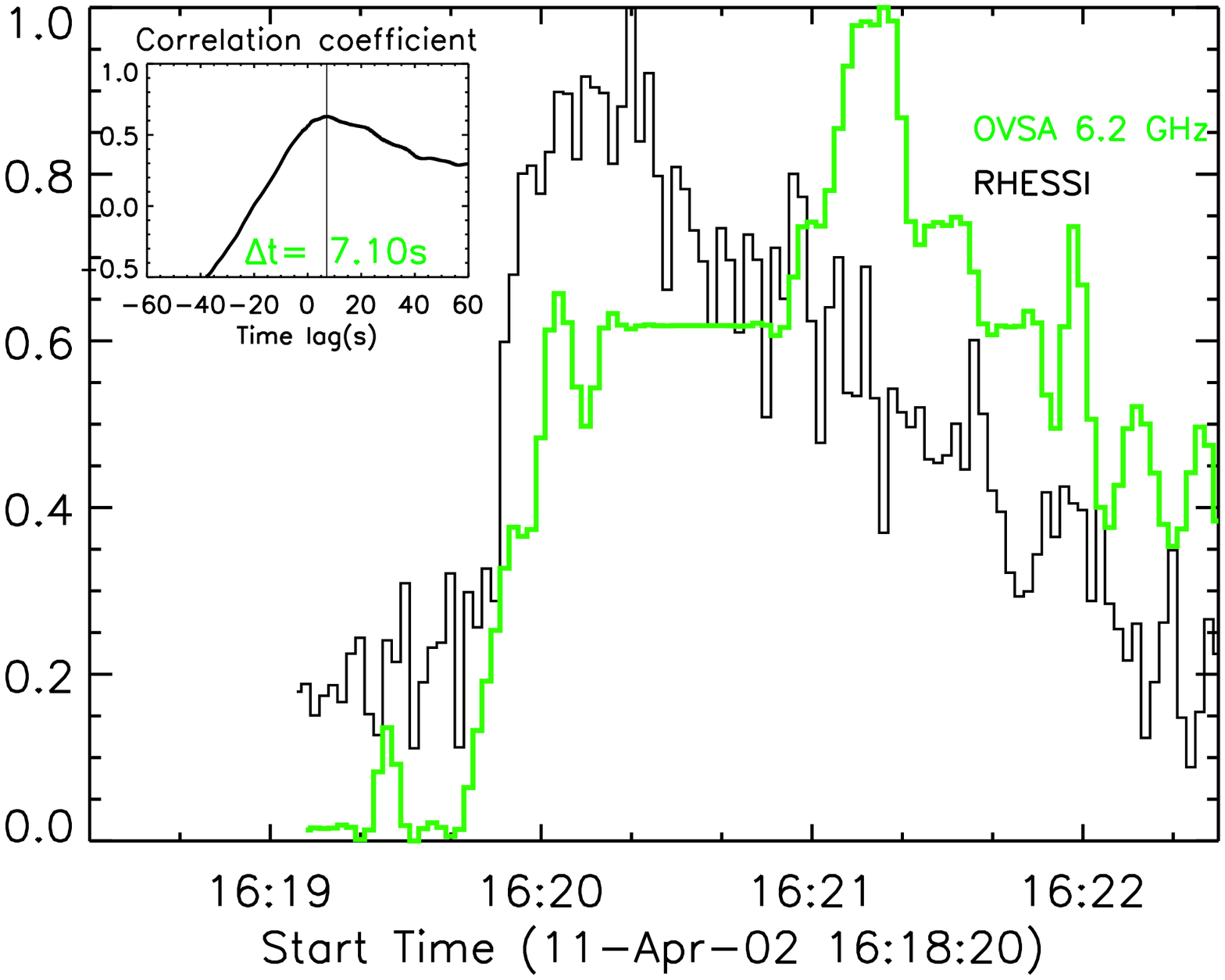}
\includegraphics[width=5cm,angle=90]{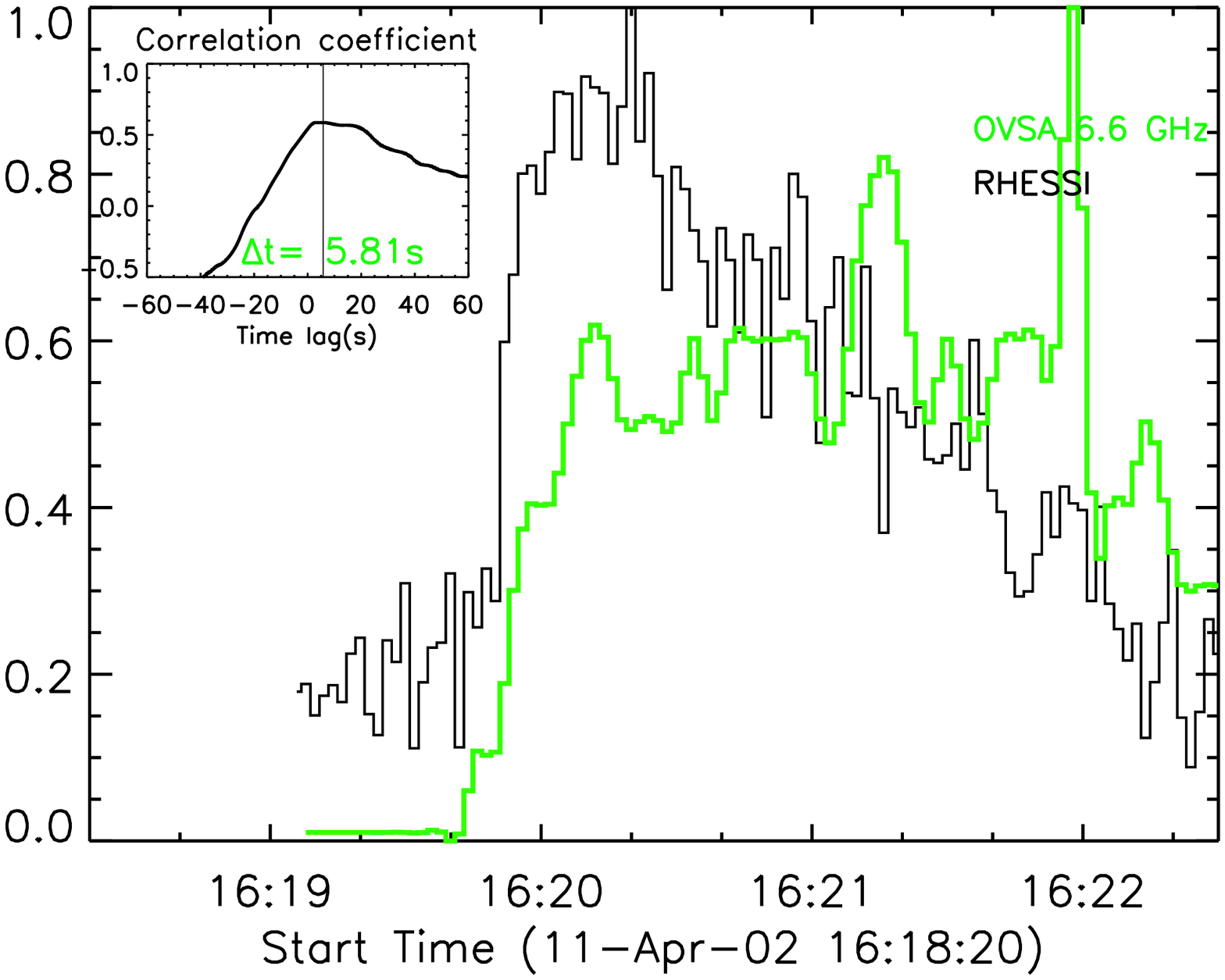}
\includegraphics[width=5cm,angle=90]{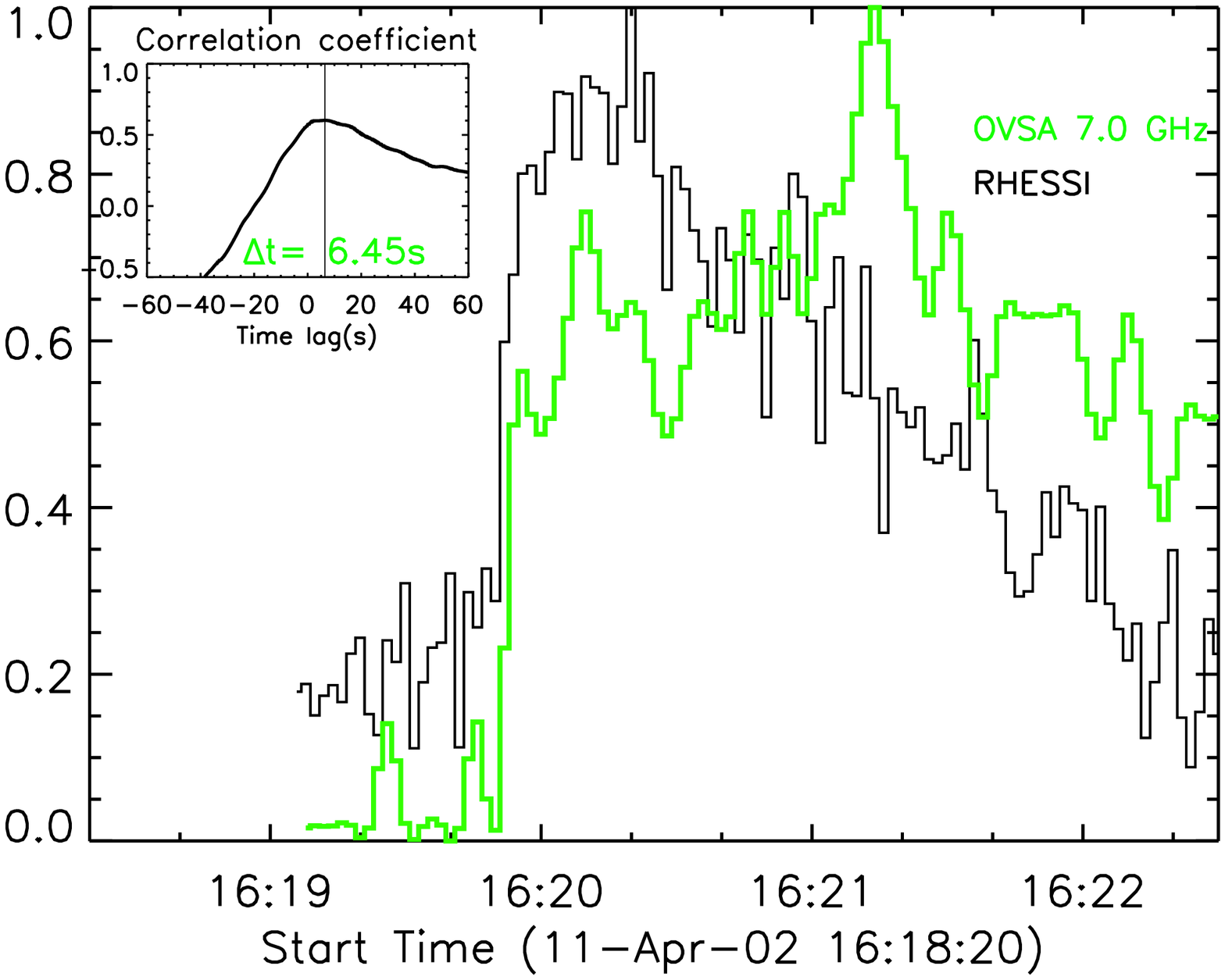}
\caption{\label{fig:Rtiming_3} Radio to HXR timing; continued.}
\end{figure}

\begin{figure}\centering
\includegraphics[width=5.3cm,angle=90]{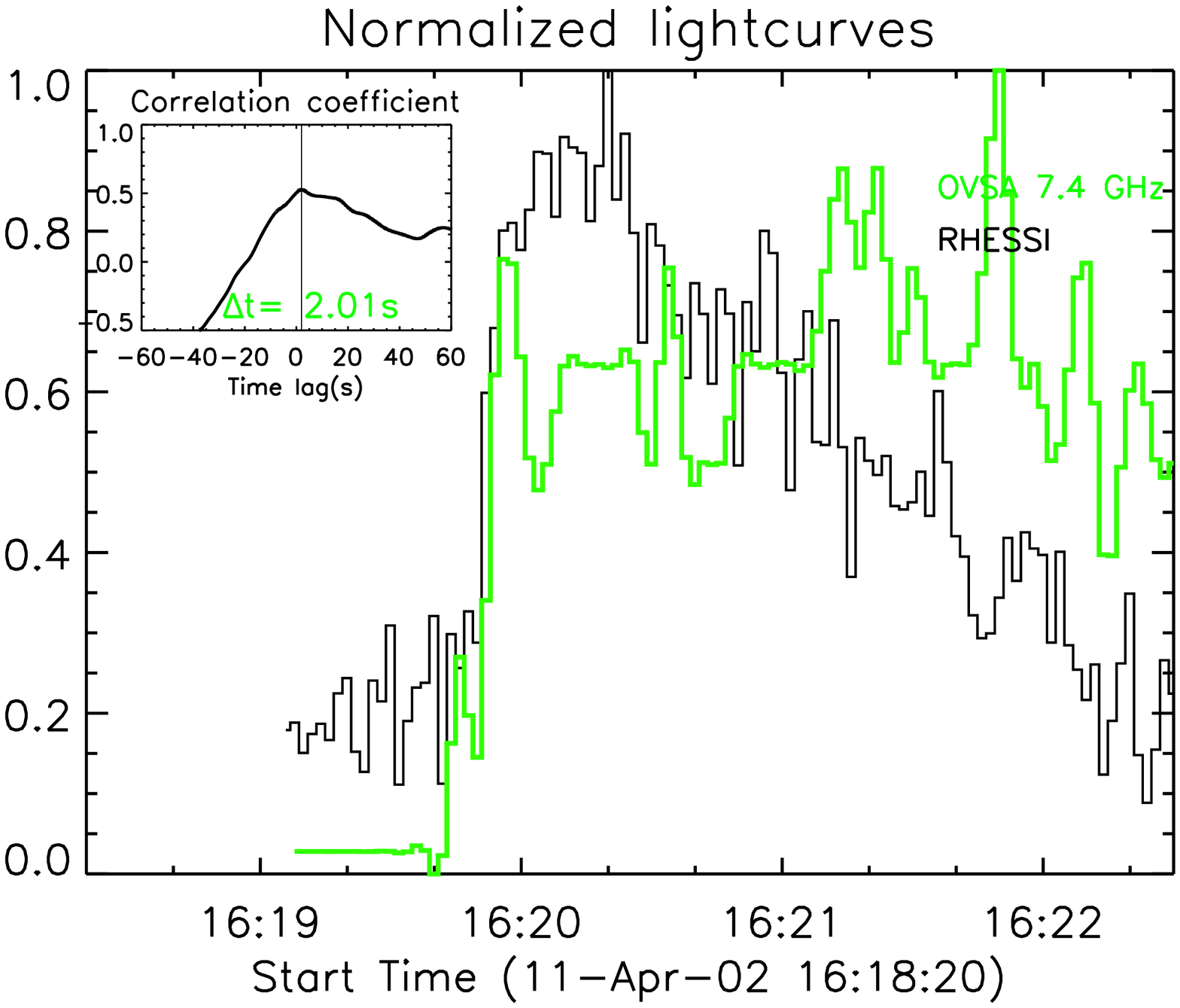}
\includegraphics[width=5.3cm,angle=90]{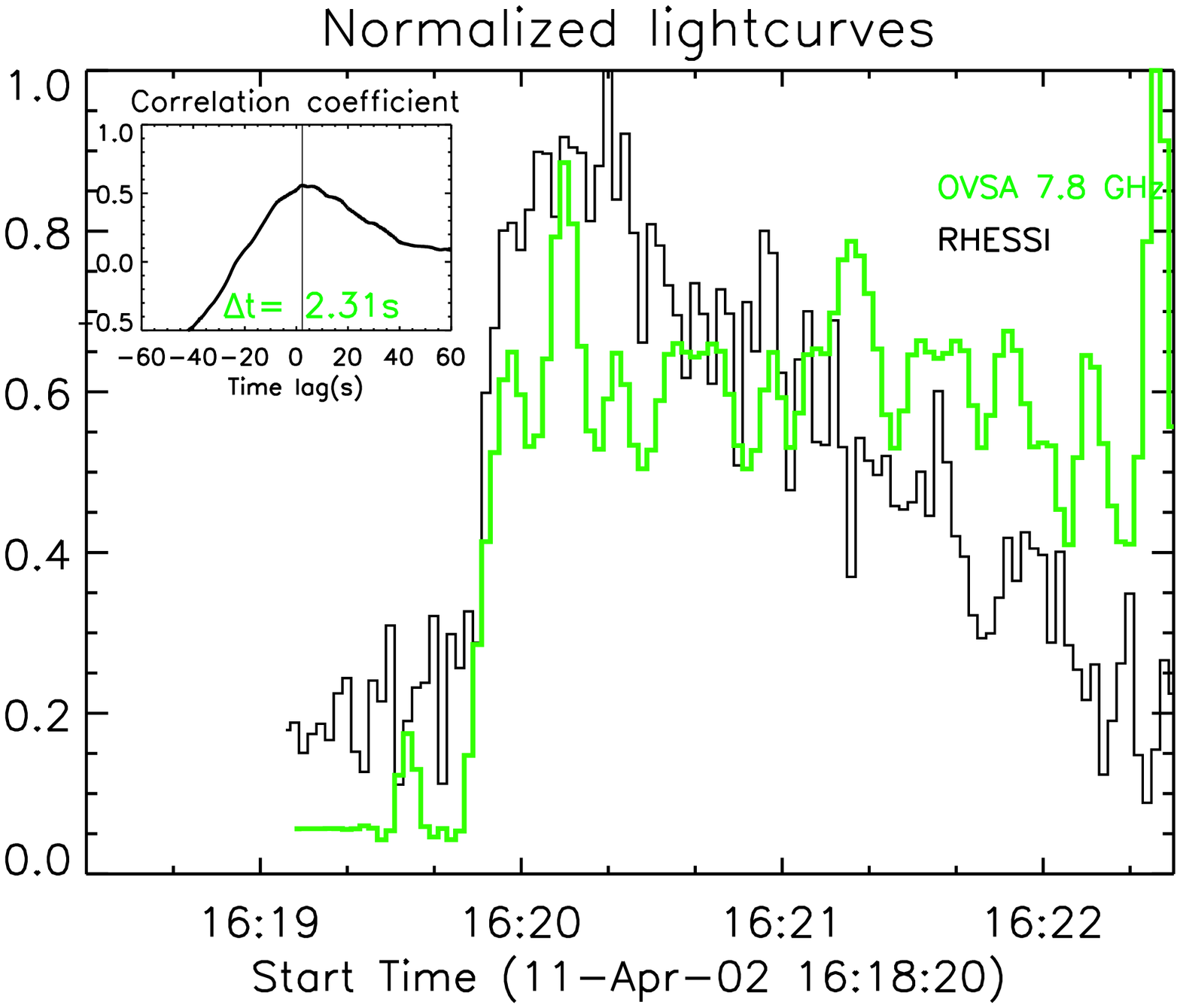}
\includegraphics[width=5cm,angle=90]{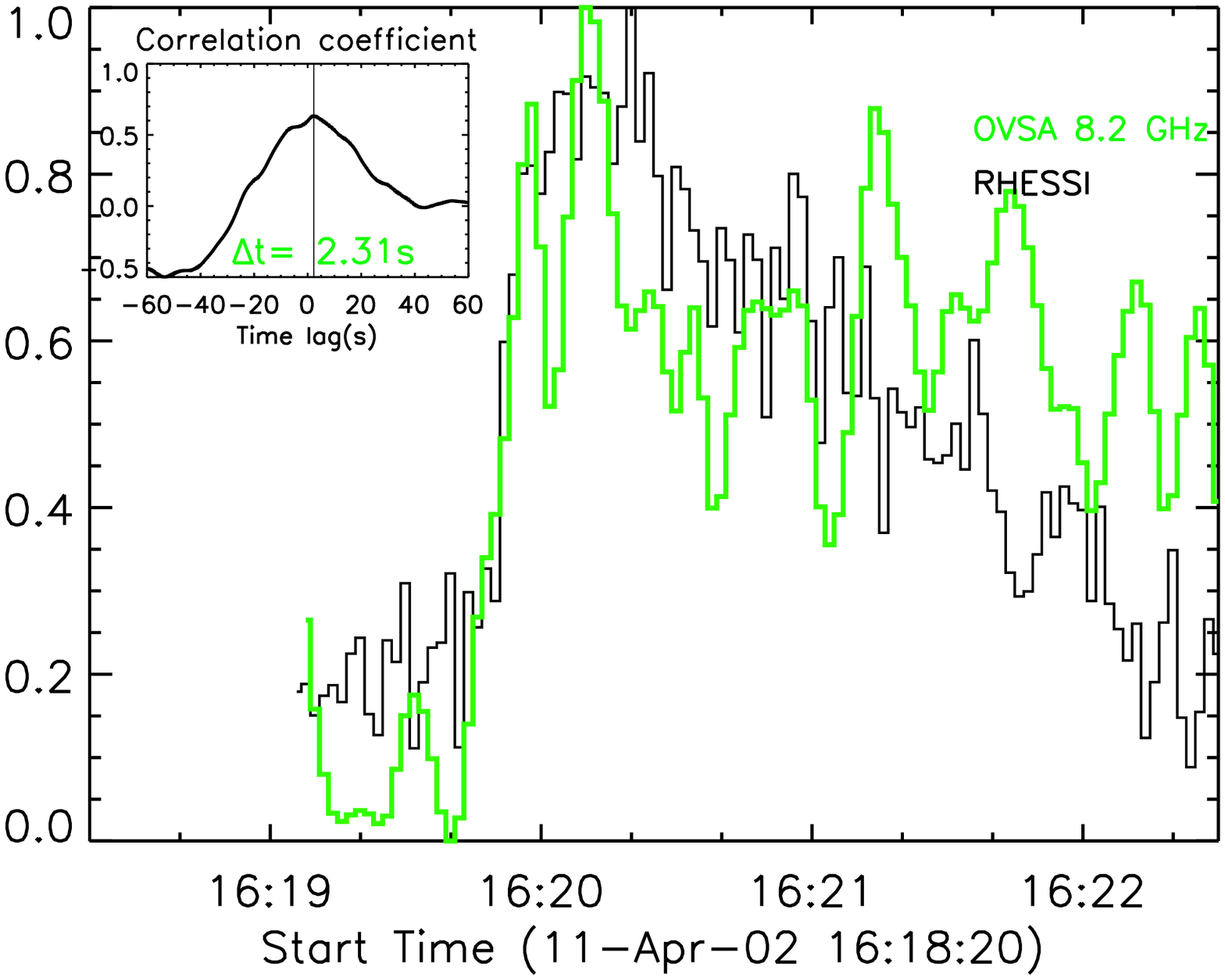}
\includegraphics[width=5cm,angle=90]{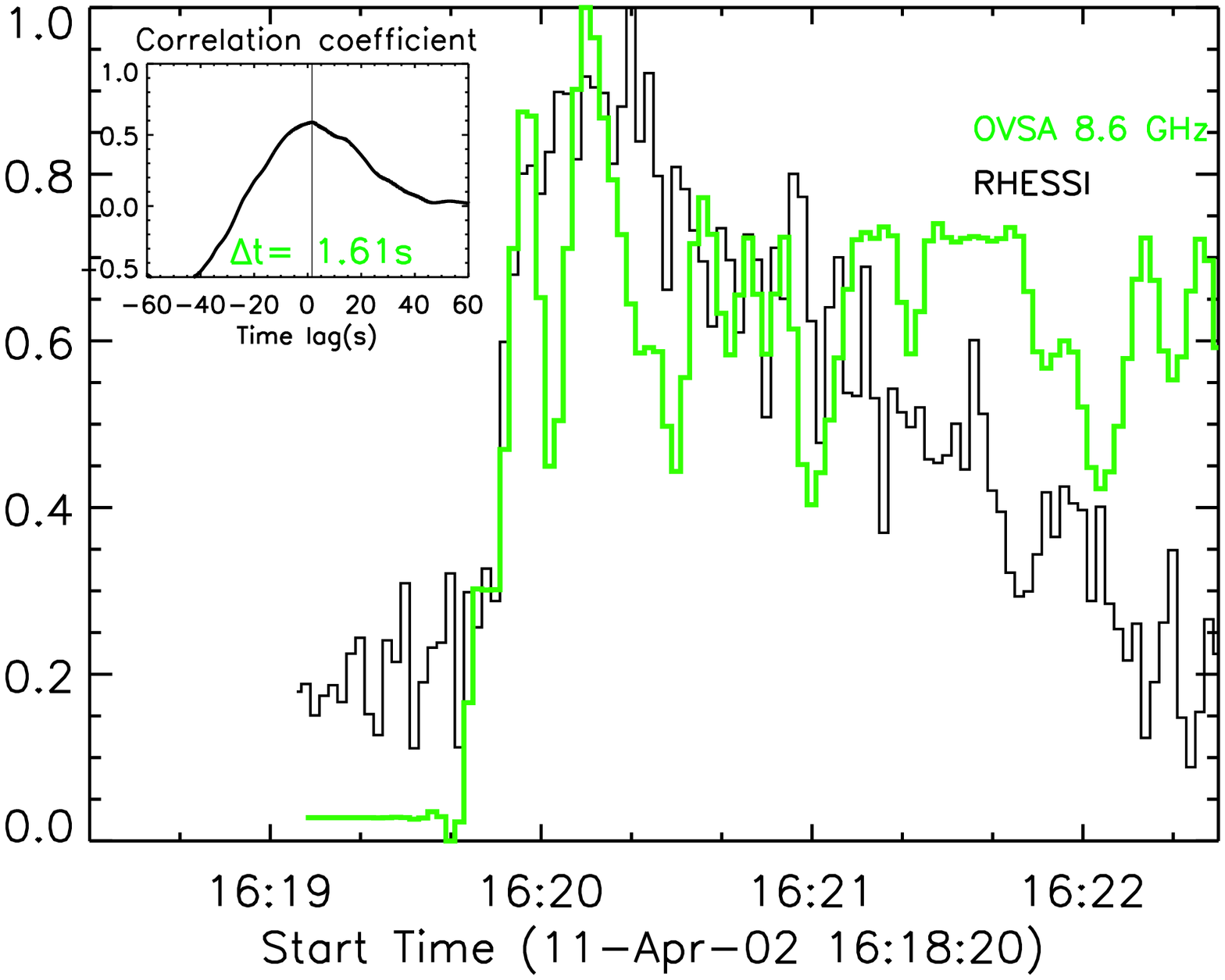}
\includegraphics[width=5cm,angle=90]{rhessi-radio_90.eps}
\includegraphics[width=5cm,angle=90]{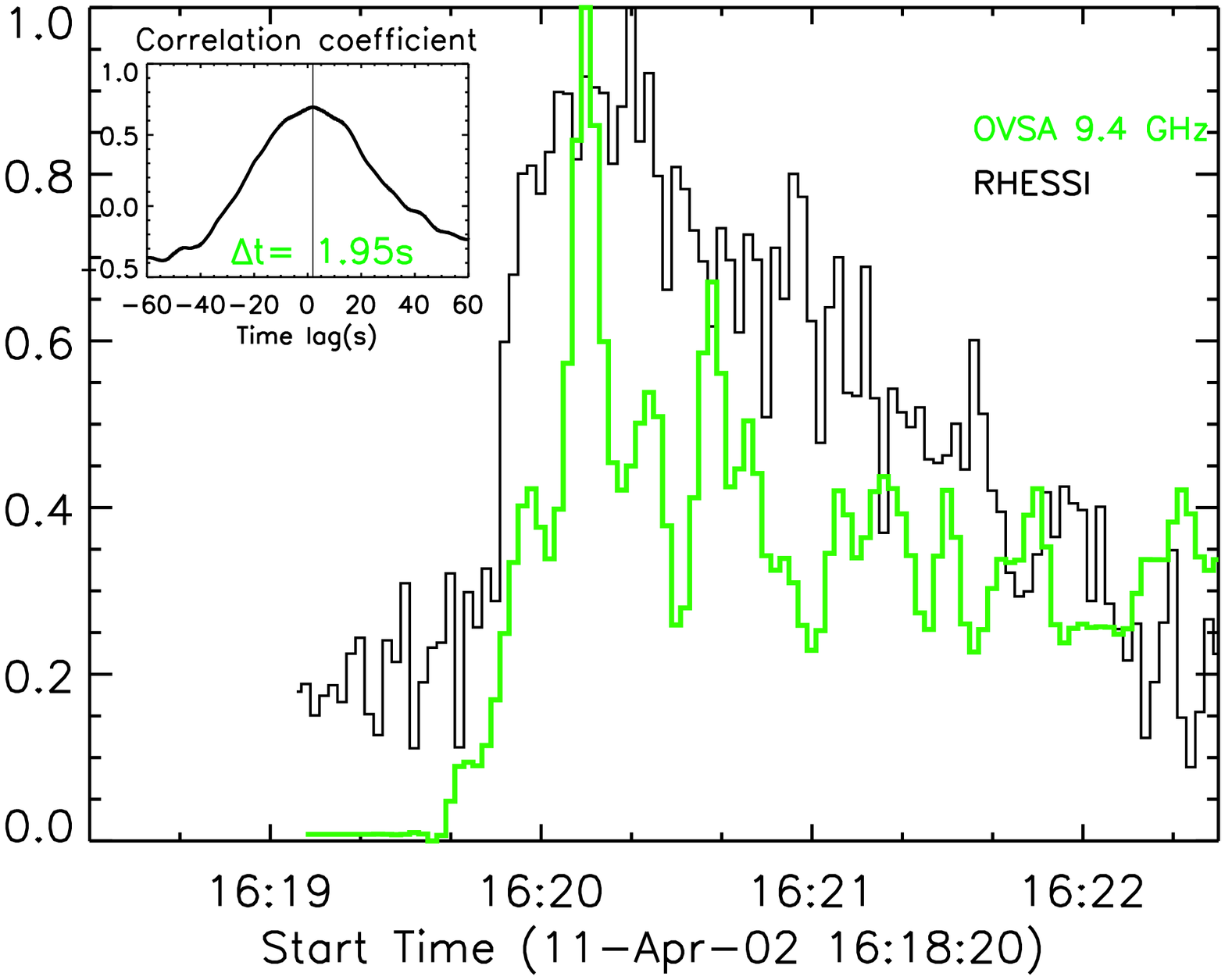}
\includegraphics[width=5cm,angle=90]{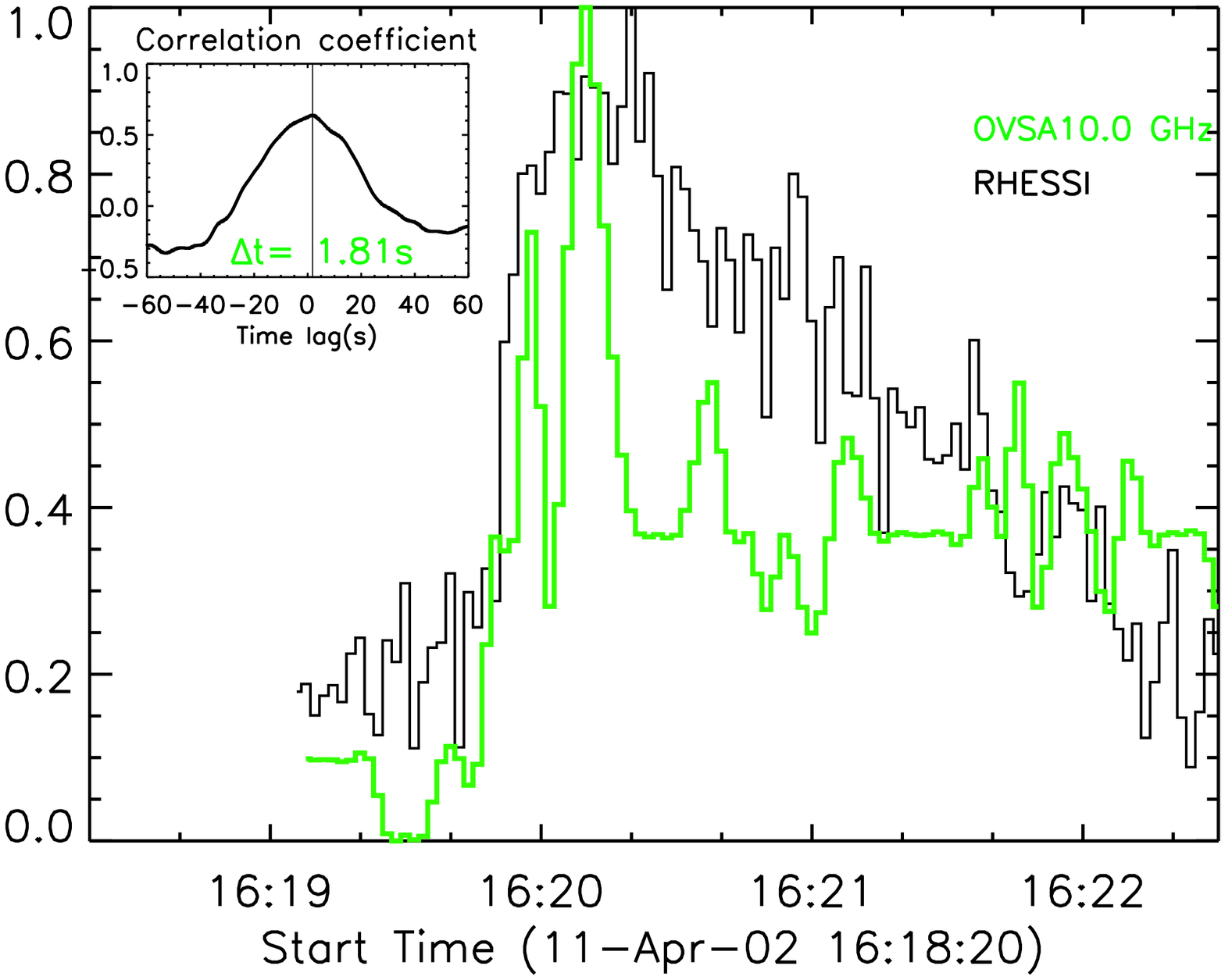}
\includegraphics[width=5cm,angle=90]{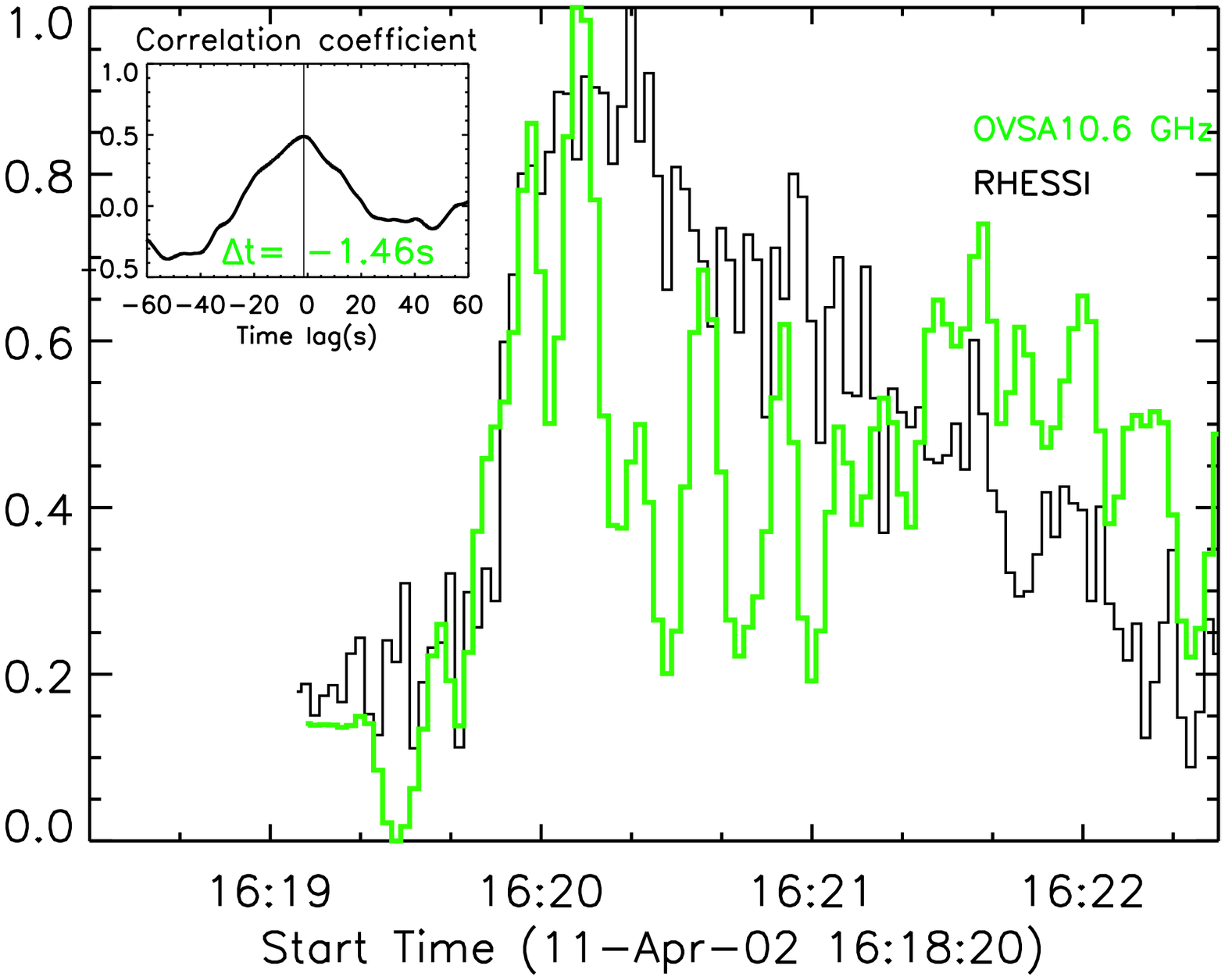}
\caption{\label{fig:Rtiming_4} Radio to HXR timing; continued.}
\end{figure}




\begin{figure}\centering
\includegraphics[width=0.9\columnwidth]{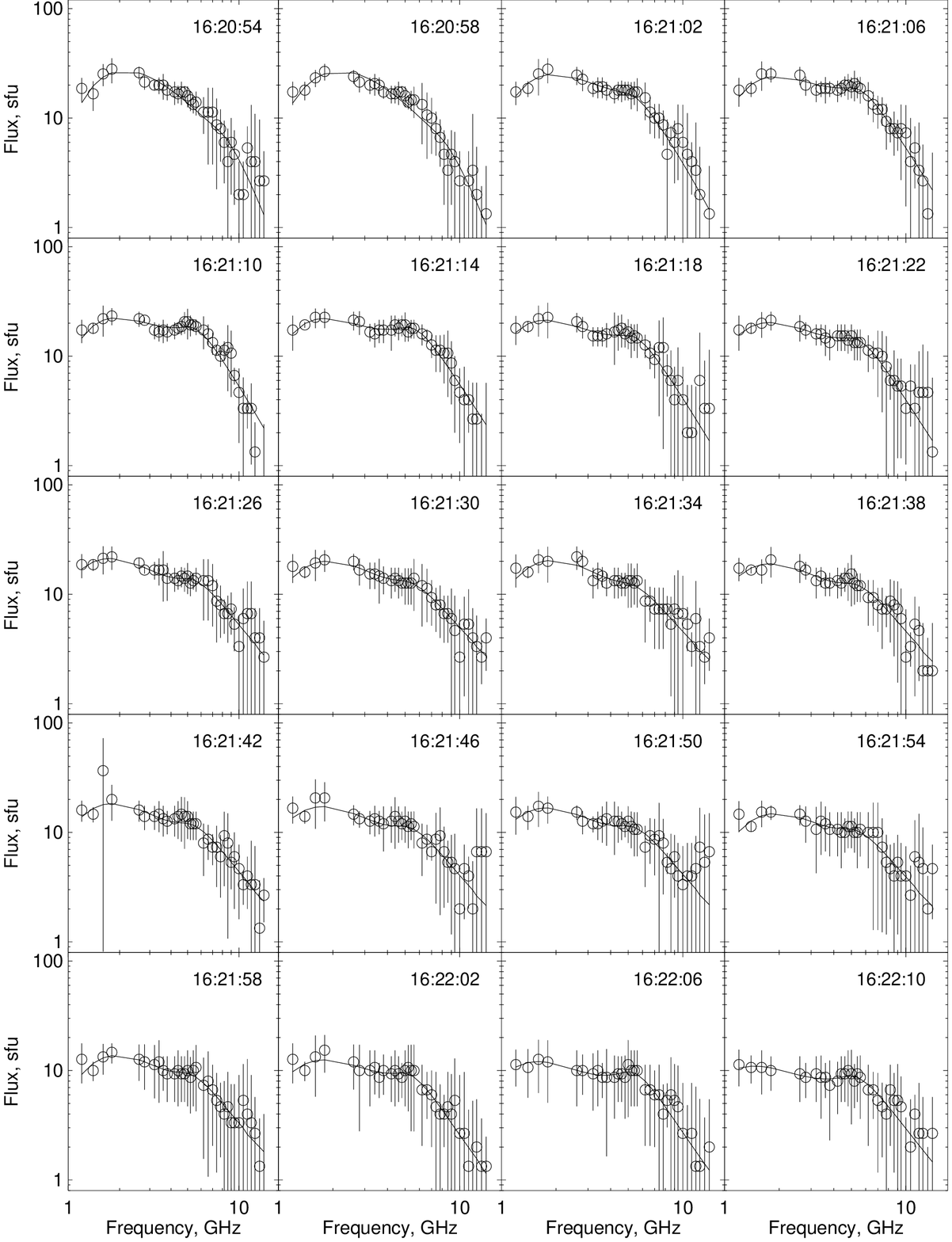}
\caption{\label{fig_OVSA_fit_Spectra_1} Radio  spectral fit of the OVSA spectra; Part II. 
}
\end{figure}

\begin{figure}\centering
\includegraphics[width=0.9\columnwidth]{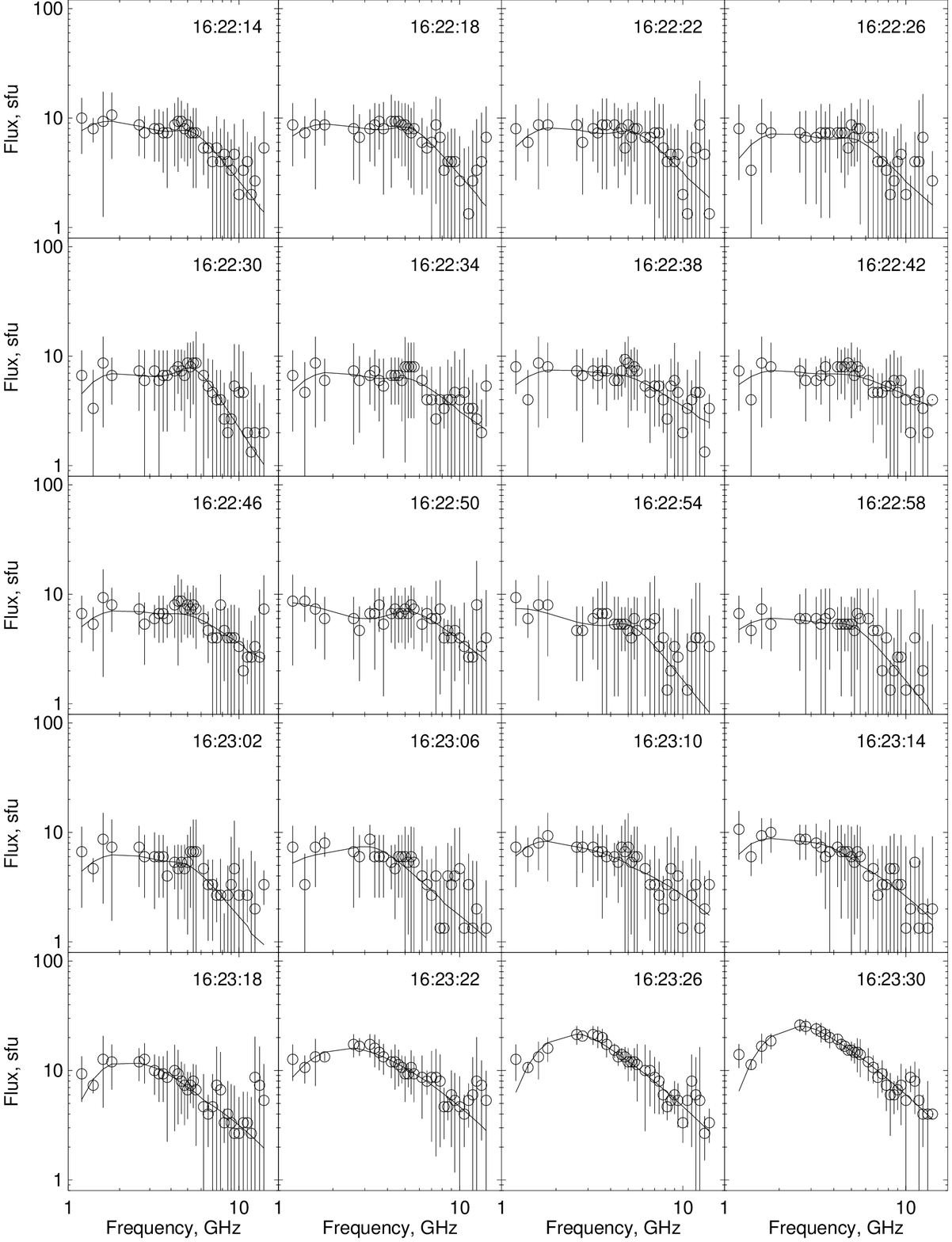}
\caption{\label{fig_OVSA_fit_Spectra_2} Radio  spectral fit of the OVSA spectra; Part III.}
\end{figure}

\begin{figure}\centering
\includegraphics[width=0.9\columnwidth]{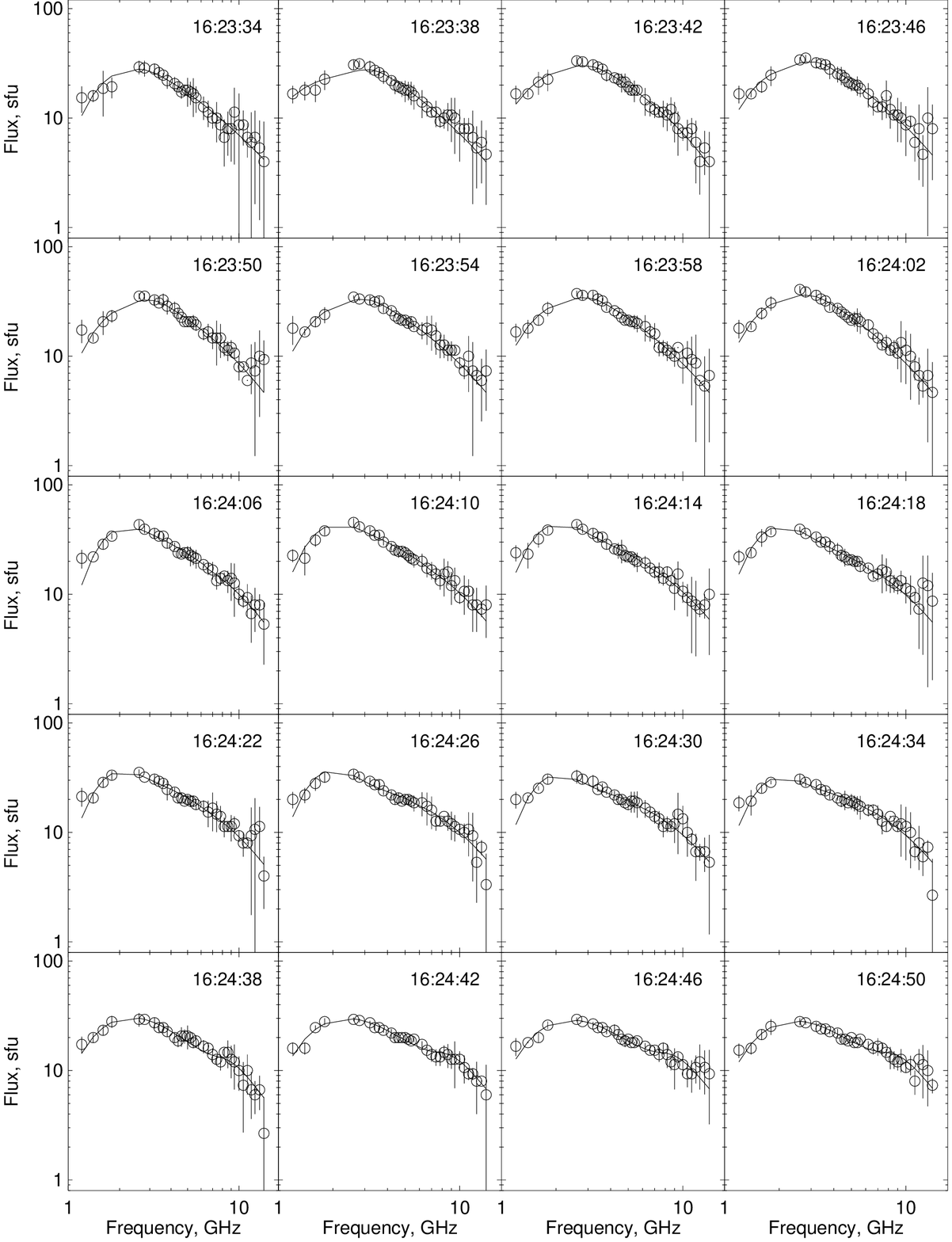}
\caption{\label{fig_OVSA_fit_Spectra_3} Radio  spectral fit of the OVSA spectra; Part IV. }
\end{figure}

\begin{figure}\centering
\includegraphics[width=0.9\columnwidth]{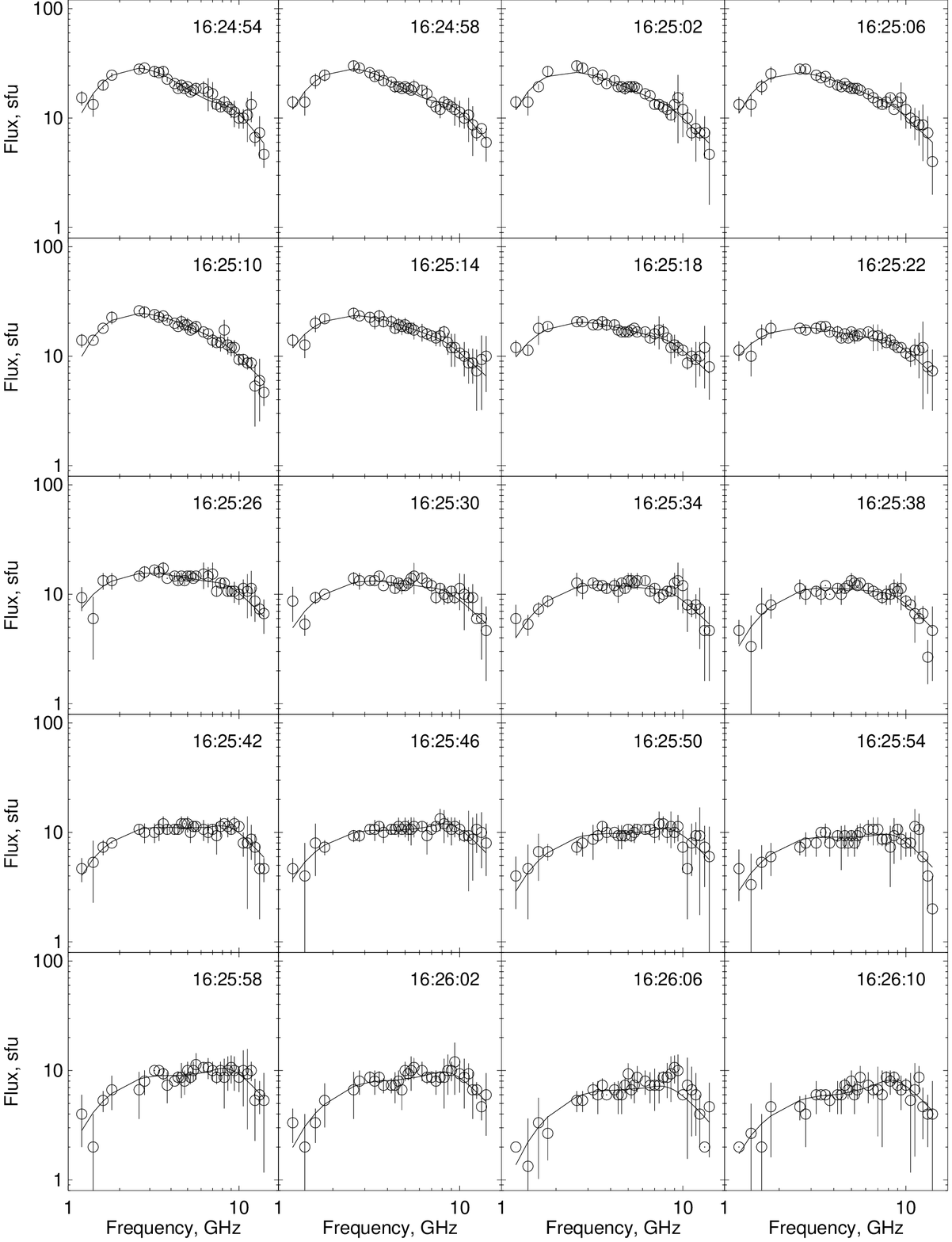}
\caption{\label{fig_OVSA_fit_Spectra_4} Radio  spectral fit of the OVSA spectra; Part V. }
\end{figure}



\end{document}